\documentclass[a4paper, twoside, 12pt]{report}
\usepackage[serbianc,english]{babel}
\usepackage{amsmath}
\numberwithin{equation}{section} 
\usepackage{enumerate}   
\usepackage{amsthm}
\usepackage{amssymb}
\usepackage{graphicx}
\usepackage{verbatim}
\graphicspath{{Slike/}}
\usepackage{wrapfig}
\title{ $\text{Поасонова заграда} \quad \{  \phi_{ij}, \mathcal{H^{'}}_\perp\}$ }
\newcommand{\vect}[1]{\boldsymbol{#1}} 
\newtheorem{Napomena}{Remark}[section]
\newtheorem{Lema}{Lemma}[section]

\newtheorem{Teorema}{Theorem}[section]

\newtheorem{Primer}{Example}[section]
\newtheorem*{Dokaz}{Proof}
\newtheorem{Definicija}{Definition}[section]
\newcommand{\mc}[1]{\mathcal{#1}}
\newcommand{\p}{\partial}
\newcommand{\fr}[2]{\frac{#1}{#2}}
\newcommand{\B}[1]{{\Bar{#1}}}

\newcommand{\KP}{{\hat{\pi}}}
\newcommand{\BB}[2]{{\Bar{#1} \Bar{#2}}}
\newcommand{\BBB}[3]{{\Bar{#1} \Bar{#2} \Bar{#3}}}
\newcommand{\BBS}[2]{{(\Bar{#1} \Bar{#2})}}
\newcommand{\NN}{\nonumber}
\newcommand{\NB}{\nonumber \\}
\usepackage[inner=3.5cm,outer=2.5cm,top=2.5cm,bottom=3.5cm]{geometry}
\usepackage{nicefrac} 
\usepackage{empheq} 
\usepackage[makeroom]{cancel}
\newcommand*\widefbox[1]{\fbox{\hspace{2em}#1\hspace{2em}}}
\usepackage{remreset}
\makeatletter
  \@removefromreset{section}{chapter}
\makeatother
\renewcommand{\thesection}{\arabic{section}}
\renewcommand{\thechapter}{\Roman{chapter}}
\usepackage{longtable}
\usepackage{hyperref}
\usepackage{caption}
\usepackage{setspace}
\usepackage{fancyhdr}
\usepackage{cite}
\usepackage[notlof,notlot,nottoc]{tocbibind}
\title{Canonical Structure of the Teleparallel Equivalent of General Relativity}

\newcommand{\thesisauthor}{Petar Mitrić}

\usepackage{titlesec}
\titleformat{\section}
  {\centering\normalfont\Large\bfseries}{\S\,\thesection.}{0.7em}{}
\usepackage{cleveref}

\newcommand{\Kristofel}[2]{\genfrac{\{}{\}}{0pt}{}{#1}{#2}}
\allowdisplaybreaks 
\begin{document}
\pagenumbering{roman}
\makeatletter
\begin{titlepage}
    \begin{center}
        \vspace*{1cm}

        \textbf{\huge \line(1,0){350}\\ Canonical Structure of the \\ Teleparallel Equivalent of \\ General Relativity\footnotemark \footnotetext{This work is based on Master's thesis, submitted to Faculty of Physics, University of Belgrade,  in September 2019.}\\ \line(1,0){350}}

        \vspace{2.5cm}
        
        



        \begin{large}
        \thesisauthor{}\phantom{\footnotemark}\footnotetext{Email address:  mitricp95@gmail.com}\\
        \vspace{0.5cm}
         \end{large}
        \textit{University of Belgrade, Faculty of Physics}\\
        \textit{Studentski trg 12, 11000 Beograd, Serbia}\\
        \vspace{1.5cm}
       
        \textbf{Abstract}: \\
        \vspace{0.3cm}
        \begin{flushleft}
         In this paper we review the canonical analysis of constrained systems and apply it in the case of teleparallel equivalent of general relativity and teleparallel gravity. For each of them we find all the Poisson brackets, generators of gauge symmetries, as well as the number of degrees of freedom. \\
        \end{flushleft}

        \vspace{1cm}
        \large

    \end{center}
\end{titlepage}
\makeatother
\clearpage
\thispagestyle{empty}
\clearpage

$\quad$
\clearpage
\pagenumbering{Roman}
\tableofcontents
\clearpage
\addcontentsline{toc}{chapter}{Acronyms and Conventions }
\clearpage
\thispagestyle{empty}
\section*{Abbreviations and Conventions }
\label{sec:SOOA}
\vspace{1.5cm}
\doublespacing
\subsection*{Abbreviations}
\vspace{0.5cm}
\begin{flushleft}
\begin{tabular}{l l }
\textbf{GR}     &  \hspace{1cm}     General Relativity  \\
{$\mathbf{C_{PFC}}$}     &  \hspace{1cm}     Primary First Class Constraint  \\
\textbf{TG}     &  \hspace{1cm}     Teleparallel Gravity  \\
\textbf{TEGR}     &  \hspace{1cm}     Teleparallel Equivalent of General Relativity  \\

\end{tabular}
\end{flushleft}
\vspace{0.5cm}
\onehalfspacing
\subsection*{Conventions}
\vspace{0.5cm}
\begin{flushleft}
\begin{itemize}
    \item Greek indices are reserved for the coordinate basis
    \item Latin indices are reserved for the Lorentz basis
    \item Signature of the metric is $\eta_{ij} = diag[1, -1, -1, -1]$
    \item The first letters of both alphabets $(\alpha, \beta, \gamma, \delta ,\epsilon...; a, b, c, d...)$ run over $1,2,3$, whereas the rest of them run over $0, 1, 2, 3$ 
    \item Antisymmetrization $A_{[imk]}$ is performed only over the first and last index in the square brackets: $A_{[imk]} = \frac{1}{2} \left(A_{imk} - A_{kmi} \right)$
    \item In the case of vanishing connection, the torsion is defined as: $T^i{}_{\mu\nu} := \p_\mu b^i{}_\nu - \p_\nu b^i{}_\mu$
    \item $T_k := T^m{}_{mk}$
\end{itemize}
\end{flushleft}

\vspace{0.5cm}
\subsection*{Labels}
\vspace{0.5cm}
\begin{flushleft}
\begin{tabular}{l l }
\hspace{0.5cm} $b^k{}_\mu$   &  \hspace{0.5cm}     Tetrad field  \\
\hspace{0.5cm} $h_k{}^\mu$   &  \hspace{0.5cm}     Inverse tetrad field  \\
\end{tabular}
\end{flushleft}

\singlespacing
\vspace{0.5cm}
\noindent

\chapter*{Introduction}
\addcontentsline{toc}{chapter}{Introduction}
\markboth{INTRODUCTION}{}

The main goal of this paper is to examine the canonical structure of the teleparallel equivalent of general relativity (TEGR)\cite{BlagojevicNikolic} and teleparallel gravity (TG). We will see that TEGR is equivalent to GR because their actions are the same up to a total divergence term. This total divergence term plays a significant role in some generalizations of GR. Its presence is the reason why $f(R)$ theories are not equivalent to  $f(\mc{L}_T)$ theories\footnote{$\mc{L}_T$ is the Lagrangian of TEGR}. The reason why we want to examine any generalization of GR  \cite{Izumi} is that we are trying to avoid some characteristics of GR such as singular solutions, nonrenormalizability... Yet, TEGR is significant by itself, regardless of its generalizations. Certain problems, such as proof of the positivity of energy \cite{Nester} in GR, are simpler in this equivalent formulation.\par\null\par

Even though it is not necessary to analyze TEGR and TG in the canonical formalism, there are some advantages if we decide to do so:
\begin{itemize}
    \item there is a straightforward way to calculate the number of degrees of freedom and conserved charges \cite{Blagojevic}
    \item there exists an algorithm for constructing generators of gauge  symmetries\cite{Kastelani}
    \item a large number of numerical methods are suited for the canonical formalism\cite{NumerickeMetode1,NumerickeMetode2,NumerickeMetode3}
    \item  there is a well-known way to quantize theory from the canonical formalism \cite{Teitelboim}, although in this paper the quantization of gravity will not be discussed
\end{itemize}
\noindent
Regardless of its specific uses, the canonical approach is equivalent to Lagrange's\cite{Mukunda} and as such, it is important to examine it carefully.\par\null\par

In chapter \ref{Teorija gravitacije sa torzijom}, we motivate the reason for introducing the notion of torsion in GR. We construct the Lagrangian of TEGR and then slightly generalize it, thus acquiring the Lagrangian of TG. Further, we introduce the so-called ADM basis\cite{ADMoriginal,ADMPregled}, which turns out to be very useful for the irreducible decomposition and calculation of the Poisson brackets. \par\null\par \newpage

In chapter \ref{Hamiltonov formalizam sa vezama}, we give a short review of the Hamiltonian formalism for constrained systems\cite{Dirak}. The notions of first and second class constraints are introduced and then used to express the number of degrees of freedom in terms of them. Also, we formulate the so-called Castellani's algorithm, for constructing the generators of gauge symmetries. Then, we consider electrodynamics as an example.  \par\null\par

In chapters \ref{genericki slucaj teleparalelne teorije} and \ref{specijalni slucaj teleparalelne teorije}, we examine the canonical structures of TG and TEGR respectively, by constructing  their canonical Hamiltonians and calculating the corresponding constraint algebras. The obtained results are then used to classify the constraints. \par\null\par

%

In chapter \ref{glava konstrukcija generatora}, we find the generators of gauge symmetries. It turns out that one of them generates local translations. That means that TG can be derived by imposing the condition of invariance under local translations. In addition, we will see that there exists an extra symmetry in TEGR that could remain unnoticed in Lagrange's formalism.\par\null\par


Our work gives an essentially equivalent description of the results that have been obtained earlier\cite{BlagojevicNikolic,BlagojevicHehl}. Nevertheless, we believe that certain aspects of this paper represents a useful contribution to the subject. For example, we present a systematic way to calculate all nontrivial Poisson brackets in TG and TEGR. Moreover, all the calculations are presented with details that cannot be found in the literature. This work can be useful in studying certain generalizations of TEGR, such as $f(\mc{L}_T)$ gravity.

\newpage
\chapter*{}
\clearpage
\pagenumbering{arabic}
\chapter{ Gravitation with Torsion}
\label{Teorija gravitacije sa torzijom}

In this chapter, we will present a natural way to construct the action for the so-called \textit{teleparallel equivalent of general relativity} (TEGR), which is the central theory we wish to examine in this paper. The starting point will be considering certain aspects of theories of gravity with \textit{torsion} \cite{Blagojevic,Ivanenko,BlagojevicHehl}, which represent one of the most natural generalizations of general relativity (GR). We assume that the reader is familiar with GR at the level of \cite{Kerol} even though, in the following section, we will give a short review of specific elements of GR that we wish to modify. Results from differential geometry, used in this chapter, can be found in \cite{Novikov}.

\section{The Connection and Metric in GR}

Very accurate measurements in the Solar System showed that in the fixed gravitational field, bodies with given initial conditions, despite their individual characteristics, have the same trajectories. This indicates that the effect of gravitational interaction can be replaced by the introduction of nontrivial geometry.\par\null\par

Notions of metric and connection are introduced independently in differential geometry. The connection can be used to define parallel transport of vectors, while the metric enables us to define lengths of curves.

\begin{Definicija}
\normalfont
Let $M$ be a manifold and let $\gamma : (0,1) \rightarrow M$ and $v_\gamma$ be a smooth curve and its tangent vector, respectively. Then

\begin{enumerate}
    \item we say that $\gamma$ is autoparallel curve if $\nabla_{v_\gamma} v_\gamma = 0$.
    \item we say that $\gamma$ is geodesic curve if it is stationary with respect to \newline {${L[\gamma] := \int_0^1 d\lambda \left( \sqrt{g(v_\gamma, v_\gamma)} \right)_{\gamma(\lambda)}}$}. Geodesic curve which connects points $A$ and $B$ represents the shortest curve between those two points.
\end{enumerate}

\end{Definicija}

In Euclidean space, notions of autoparallel curve and geodesic curve are the same. However, not every manifold has this property. On an arbitrary manifold, a relation between the metric and connection, which is mathematically simple,  could be geometrically illustrated by examining the link between the geodesic and autoparallel curves. GR imposes these relations by postulating that:

\begin{enumerate}
    \item connection is compatible with the metric: $\nabla g = 0$
    \item connection is symmetric:
    $\Gamma^\mu{}_{\sigma \rho} - \Gamma^\mu{}_{\rho \sigma} = 0$
\end{enumerate}

These conditions define the so-called \textit{Levi-Civita} connection:
\begin{equation} \label{Koneksija Levi Civita}
    \Gamma^\lambda{}_{\mu \nu} = \fr{1}{2} g^{\lambda \sigma} 
            \left(
                    \p_\mu g_{\nu \sigma} + 
                    \p_\nu g_{\mu \sigma} - 
                    \p_\sigma g_{\mu \nu}
            \right)
\end{equation}
$\Gamma^\lambda{}_{\mu \nu}$ are called \textit{Christoffel symbols}. 

Conditions 1. and 2. are equivalent to the fact that geodesic and autoparallel curves are the same. This, very intuitive geometrical picture, can lead us to believe that these conditions are fulfilled in nature.\par\null\par

Since we want to replace the effect of the gravitational interaction with nontrivial geometrical properties, it is important to find the quantities that will measure the discrepancy between the physical space-time manifold and Euclidean space. One of the many properties of Euclidean space is Schwarz's theorem: $\p_i \p_j f = \p_j \p_i f$. If space-time is actually Euclidean, we can always find coordinates in which ordinary and covariant derivatives are the same, so we can conclude that $\left[ \nabla_\mu, \nabla_\nu   \right] A= 0$. However, on the arbitrary smooth manifold, it can be shown that:
 \begin{equation}
   \left[ \nabla_\mu, \nabla_\nu   \right] A^\sigma = R^\sigma{}_{\rho \mu \nu}                                                          A^\rho
                                                    +T^\rho{}_{\mu\nu}\nabla_\rho
                                                    A^\sigma
 \end{equation}
 We can see that $R^\sigma{}_{\rho \mu \nu}$ and $T^\rho{}_{\mu\nu}$ are measuring the discrepancy between the space-time manifold and Euclidean space. They are called \textit{curvature tensor} and \textit{torsion tensor}, respectively. We can express them in terms of connection as:

\begin{subequations}
\begin{eqnarray}
R^\sigma{}_{\rho \mu \nu}
            &=& \p_\mu \Gamma^\sigma{}_{\rho \nu} -
                \p_\nu \Gamma^\sigma{}_{\rho \mu} +
                \Gamma^\sigma{}_{\lambda \mu} \Gamma^\lambda{}_{\rho \nu} -
                \Gamma^\sigma{}_{\lambda \nu} \Gamma^\lambda{}_{\rho \mu} \\
T^\rho{}_{\mu\nu}
            &=& \Gamma^\rho{}_{\nu \mu} -\Gamma^\rho{}_{\mu \nu}
\end{eqnarray}
\end{subequations}
It is now obvious that one of the imposed conditions of GR is that space-time manifold is torsion-free.\par\null\par 

Even though GR has simplicity, beauty, and intuitive geometrical picture, there are at least a few reasons why theories with torsion should be considered:

\begin{itemize}
     \item It is a well-known fact that the electromagnetic, weak and strong interactions, can all be obtained by localization of certain symmetry groups. Gravitation can also be obtained in the same way\cite{BlagojevicHehl}. Thus obtained \textit{gauge theory of gravity} has to have torsion. This approach is very tempting because it would describe all fundamental interactions from the unique principle.
     \item Both mass and spin of matter are contributing to the gravitational field. On the other hand, GR does not take into consideration the contribution of spin to the gravitational field.
     \item The paper~\cite{Birkof} shows that there exist Lagrangians of $R+R^2$ type with torsion, that satisfy strong Birkhoff's theorem which says that for $SO(3)$ spherically symmetric space-time, the unique solution to the vacuum field equations is Schwarzschild solution with vanishing torsion. That means that such theories have the same classical predictions as GR in the case of spherically symmetric solutions, which happen to be almost all the cases in which GR was tested. That's why those theories with torsion can be interpreted as generalizations of GR. They agree with GR in all nonextremal cases, whereas at the very small scale or in the early stadiums of the universe, their predictions can be different.\cite{shie}. 
\end{itemize}

The listed facts provide sufficient reasons for us to study theories of gravitation with torsion, thoroughly. 
In the next section, we will give a short review of Riemann-Cartan geometry~\cite{Blagojevic,Ivanenko,BlagojevicHehl}. 
\section{Riemann-Cartan Geometry}
\label{Riman Kartanova geometrija}

In 1922 Élie Cartan started developing theories of gravitation with torsion  \cite{Kartan}. He supposed that space-time is a four-dimensional manifold with a nonsymmetric, metric-compatible connection. The geometry of such space-time is called Riemann-Cartan geometry. Cartan coined the term torsion because he found out that a body in space-time, with metric-compatible connection and Christoffel symbol part equals to zero \eqref{Vrednost Koneksije},  has a helicoid trajectory in the three-dimensional space section of space-time\cite{BlagojevicHehl}. Later, he proposed that the torsion tensor is related to some kind of an internal angular momentum of matter. However, his work has attracted more attention much later. 

\subsection*{Lorentz Basis}

In order to examine theories with torsion, we need to define the underlying geometry of space-time. As in the case of GR, we demand that space-time is a four-dimensional manifold with smooth atlas and Lorentzian metric\footnote{A metric is called Lorentzian if it's signature is $\left(+, -, -, - \right)$}. On this manifold, at every point, it is always possible to find the so-called \textit{Lorentz basis}, for which: 
\begin{equation}
    g\left( \vect{e_i}, \vect{e_j} \right) = \eta_{ij}
\end{equation}
where $\eta_{ij}$ is metric of the Minkowski space-time $\eta_{ij} = diag[1, -1 , -1, -1]$. We can use the Lorentz basis to impose all the additional conditions that we want our theory to satisfy, such as the relation between the connection and metric. Afterward, we can derive what kind of consequences do those conditions imply in the coordinate basis.\par\null\par 

In order to change the basis from Lorentz to the coordinate basis (or other way around), we need to define \textit{tetrad} fields $b^i{}_\mu$ (inverse tetrad fields $h_i{}^\mu$):
\begin{equation} \label{Prelazak iz L u K bazu}
    \vect{e_\mu} = b^i{}_\mu \vect{e_i}, \qquad \vect{e_i} = h_i{}^\mu \vect{e_\mu}
\end{equation}
We can express components of the metric tensor in terms of tetrads:
\begin{equation}
    g_{\mu \nu} = \eta_{ij} b^i{}_\mu b^j{}_\nu = \Vec{e}_\mu \cdot \Vec{e}_\nu
    \label{metrika preko tetrada}
\end{equation}
Now, we can easily prove a couple of useful identities:
\begin{subequations}
\begin{align}
  \eta_{ij} &= g_{\mu\nu} h_i{}^\mu h_j{}^\nu=\Vec{e}_i \cdot \Vec{e}_j,   &  \eta^{ij}  &= g^{\mu\nu} b^i{}_\mu b^j{}_\nu  \\
    g^{\mu\nu} &= \eta^{ij} h_i{}^\mu h_j^{\nu},   &  b &= \sqrt{|g|} \\
    b^i{}_\mu h_i{}^\nu &= \delta^\nu_\mu,   &  b^i{}_\mu h_j{}^\mu &= \delta^i_j 
\end{align}
\end{subequations}

Covariant derivative of vector $v^i$ is defined\footnote{Covariant derivative of $v^\nu$ is defined in the standard way: $\nabla_\mu v^\nu = \p_\mu v^\nu + \Gamma^\nu{}_{\lambda \mu}v^\lambda$, where $\Gamma^\nu{}_{\lambda \mu}$ is coordinate connection} as:

\begin{equation}
    \nabla_\mu v^i = \p_\mu v^i + \omega^i{}_{j \mu} v^j
\end{equation}
where we also defined the \textit{spin connection} $ \omega^i{}_{j \mu}$. As in GR, we will now impose the condition that the connection is compatible with metric:

\begin{equation}
    \nabla_\mu \eta_{ij} = 0 \label{uslov metricnosti Minkovski}
\end{equation}
This condition is equivalent to the requirement that the spin connection is antisymmetric with respect to Lorentz indices:
\begin{equation}
   0 =  \nabla_\mu \eta_{ij} = - \omega^k{}_{i \mu} \eta_{kj}
                               - \omega^k{}_{j\mu}\eta_{ik}
                               = -\omega_{ji\mu} - \omega_{ij\mu}
\end{equation}
\subsection*{Consequences in the Coordinate Basis}

There exists a relation between the coordinate and spin connection. This relation can be easily derived from the fact that parallel transport is a geometric notion, independent of the choice of basis:

\begin{equation}
    u^i + \delta u^i = b^i{}_\mu (x+dx) \cdot \left( u^\mu + \delta u^\mu \right)
\end{equation}
where:
\begin{subequations}
\begin{eqnarray}
\delta u^i &=& -\omega^i{}_{j\mu} u^j dx^\mu \\ 
\delta u^\mu &=& - \Gamma^\mu{}_{\lambda \sigma} u^\lambda dx^\sigma
\end{eqnarray}
\end{subequations}
Now, we can easily obtain relations  $\Gamma(\omega)$ and $\omega(\Gamma)$:

\begin{subequations}
\begin{eqnarray}
\Gamma^\sigma{}_{\nu\mu} &=& \omega^i{}_{s\mu} b^s{}_\nu h_i{}^\sigma
                        + h_s{}^\sigma \p_\mu b^s{}_\nu \label{koneksija preko spinske koneksije} \\
\omega^i{}_{k\sigma} &=& \Gamma^\mu{}_{\lambda\sigma} b^i{}_\mu h_k{}^\lambda
                        -h_k{}^\lambda \p_\sigma b^i{}_\lambda
\end{eqnarray}
\end{subequations}
Previous two results, as well as~\eqref{metrika preko tetrada} imply that the metricity condition is also satisfied in the coordinate basis:

\begin{equation}  \label{Uslov metricnosti}
    \nabla_\mu g_{\nu \lambda} =\p_\mu g_{\nu \lambda}
                            -\Gamma^{\rho}{}_{\nu\mu} g_{\rho \lambda}
                            -\Gamma^{\rho}{}_{\lambda \mu} g_{\nu\rho} = 0
\end{equation}
Geometrically, that means that parallel transport preserves lengths and angles, which is easily seen from: $\nabla \left[ g(X,Y)\right] = (\nabla g)(X,Y) + g(\nabla X,Y) + g(X,\nabla Y)$. \par\null\par

Even though the metricity condition is fulfilled, torsion can still be nonvanishing, thus our newly defined theory is more general than GR. That means that the connection does not have to be Levi-Civita, so it is necessary to find the correct form of the equation~\eqref{Koneksija Levi Civita} in the Riemann-Cartan case. Such an equation can easily be obtained from~\eqref{Uslov metricnosti}:

\begin{equation} \label{Vrednost Koneksije}
    \Gamma^{\mu}{}_{\lambda \nu} =
            \Kristofel{\mu}{\lambda \nu} + K^\mu{}_{\lambda \nu} \;,
\end{equation}
where $\Kristofel{\mu}{\lambda \nu}$ is Christoffel symbol, while $K^\mu{}_{\lambda \nu}$ is the \textit{contorsion} tensor:

\begin{subequations}
\begin{eqnarray} 
\Kristofel{\mu}{\lambda \nu} &=& \fr{1}{2} g^{\mu \sigma} \left(
                                    \p_\lambda g_{\nu \sigma} +
                                    \p_\nu g_{\lambda \sigma} -
                                    \p_\sigma g_{\lambda \nu}
                                \right) \\
 K^\mu{}_{\lambda \nu}      &=& -\fr{1}{2} \left(
                                T^{\mu}{}_{\lambda \nu} -
                                T_{\nu}{}^{\mu}{}_\lambda +
                                T_{\lambda \nu}{}^\mu
                                \right)
\end{eqnarray}
\end{subequations}
From these equations, we can conclude that in the Riemann-Cartan geometry, it is not sufficient to know the metric tensor to calculate connection coefficients. We also need to know the value of the torsion tensor. That's one of the reasons why we need both curvature and torsion to characterize Riemann-Cartan space.

\begin{Napomena}
\normalfont
Dynamical variables in this theory are tetrads and spin connection coefficients. They will represent generalized coordinates in our physical theory, although we will constrain ourselves to the case of vanishing spin connection.

\end{Napomena}

We should now define torsion and curvature as functions of the generalized coordinates:
\begin{subequations}
\begin{eqnarray}
R^{ij}{}_{\mu \nu} &:=& \p_\mu \omega^{ij}{}_\nu -\p_\nu \omega^{ij}{}_\mu
                    +   \omega^i{}_{s\mu} \omega^{sj}{}_\nu 
                    -   \omega^i{}_{s\nu} \omega^{sj}{}_\mu \\
T^i{}_{\mu \nu}     &:=& \p_\mu b^i{}_\nu - \p_\nu b^i{}_\mu 
                        + \omega^i{}_{k\mu} b^k{}_\nu - \omega^i{}_{k\nu}b^k{}_\mu
\end{eqnarray}
\end{subequations}
It turns out that it is also useful to define:
\begin{subequations}
\begin{eqnarray}
c^i{}_{\mu\nu} &:=& \p_\mu b^i{}_\nu - \p_\nu b^i{}_\mu \\
\Delta_{ij\mu} &:=& \fr{1}{2} (c_{ijm} - c_{mij} + c_{jmi} ) b^m{}_\mu
\end{eqnarray}
\end{subequations}
It would be useful if we would know the relation between the old and new definitions of torsion and curvature. The next lemma specifies such relations:\newline
\vspace{1cm}
{
\begin{Lema} \label{Prva Lema} $\quad$
{
\begin{enumerate}[(i)]
\item $R^\mu{}_{\nu \rho \sigma} = h_i{}^\mu b_{j\nu} R^{ij}{}_{\rho \sigma}$
\item $b^i{}_\sigma T^\sigma{}_{\mu\nu} = T^i{}_{\mu\nu}$
\item $\omega_{ij\mu} = \Delta_{ij\mu} + K_{ij\mu}$
\end{enumerate}
}
\end{Lema}
}
\noindent
First two items can be shown using the equation~\eqref{koneksija preko spinske koneksije}, while the last item can be shown using $T_{ij \mu}= c_{ij\mu}+\omega_{i\mu j} - \omega_{ij\mu}$.\par\null\par

In this section, we defined curvature and torsion with Lorentz and coordinate indices independently. Afterward, we showed that there is a straightforward way to relate these definitions using tetrads and inverse tetrads. For future reference, it is important to emphasize that the change of indices (from Lorentz to coordinate and the other way around) will always be performed using tetrads and inverse tetrads because they give correct tensor law of transformation, which can be seen from~\eqref{Prelazak iz L u K bazu}. For example:
\begin{equation*}
    b^i{}_\mu v_i = v_\mu, \quad h_i{}^\mu v_\mu = v_i
\end{equation*}

\section{Teleparallel Theory}
\label{Teleparalelna}
In Riemann-Cartan geometry, there is one very important theorem:
\begin{Teorema} \normalfont
Let $R(\omega)$ and $R(\Delta)$ be Ricci scalar curvature with respect to spin connection and Levi-Civita connection respectively. Also, let $K^\mu = K^{\mu n}{}_n$ and $T^k=T^m{}_{mk}$. Then

\begin{equation} \label{relacija TEOTR}
    bR(\omega) = bR(\Delta) + 
                b\left( \frac{1}{4} T_{ijk}T^{ijk} + 
                \frac{1}{2} T_{ijk}T^{jik} - T^k T_k \right)
                +2\p_\mu \left( b K^\mu \right)
\end{equation}
\end{Teorema}
\noindent
Proof of this theorem will be given in \S~\ref{Ekvivalentnost}. \newline
If we now limit ourselves to the case when $R^{ij}{}_{\mu\nu}(\omega)=0$, then the previous identity becomes:
\begin{equation} \label{ekv rel 1}
   bR(\Delta)= - b\left( \frac{1}{4} T_{ijk}T^{ijk} + 
                \frac{1}{2} T_{ijk}T^{jik} - T^k T_k \right) -2\p_\mu \left( b K^\mu \right)
\end{equation}
The left-hand side of the previous equation represents the Einstein-Hilbert Lagrangian, while one of the terms on the right-hand side is total divergence, which can be ignored. We can conclude that the theory with vanishing curvature, defined by the Lagrangian:
\begin{equation} \label{lagranzijan TFOTR}
    \Tilde{\mc{L}} = b \mc{L}_T =a b\left( \frac{1}{4} T_{ijk}T^{ijk} + 
                \frac{1}{2} T_{ijk}T^{jik} - T^k T_k \right)
\end{equation}
is equivalent to GR. That is why this theory is called \textit{teleparallel equivalent of general relativity}(TEGR).
Factor $a$ in \eqref{lagranzijan TFOTR} represents some dimensional constant which is relevant only when Lagrangian has more terms, except \eqref{lagranzijan TFOTR} (for example if we add matter part of Lagrangian). Even though we will examine only the case of the free gravitational field, we will nevertheless keep the constant $a$.

\begin{Napomena}
\normalfont
If we naively look at \eqref{relacija TEOTR} we could conclude that theory with vanishing Ricci scalar $R(\omega)=0$  and Lagrangian \eqref{lagranzijan TFOTR} is also equivalent to GR. However that's not the case, because even though Lagrangians are numerically equal, up to a total divergence, those two theories have different dynamical variables - GR has tetrads\footnote{It is possible to take tetrads as generalized coordinates instead of metric in GR}, while the theory with $R(\omega)=0$ has both tetrads and spin connection. However, if we impose a more restrictive condition $R^{ij}{}_{\mu\nu}(\omega)=0$, the connection becomes trivial and thus irrelevant. In order to further simplify our theory, without losing any physical significance, we will impose additional conditions that the spin connection coefficients are vanishing. It is important to note that it is not necessary to do that, but nevertheless, we choose to do so, because of the simplicity.
\end{Napomena}

It turns out that it is also important to examine a theory defined by the Lagrangian:
\begin{equation} \label{Lagranzijan teleparalelne gr}
    \mc{L}_T =a \left( A T_{ijk}T^{ijk} + 
                B T_{ijk}T^{jik} +C T^k T_k \right)
\end{equation} 
which looks like a generalization of TEGR because we can get TEGR Lagrangian \eqref{lagranzijan TFOTR} by choosing coefficients $A, B, C$ as:
\begin{equation}\label{spec koef}
    A=\fr{1}{4},\qquad B=\fr{1}{2}, \qquad C=-1 
\end{equation}
It turns out that torsion tensor has only three irreducible parts. That means that \eqref{Lagranzijan teleparalelne gr} is the most general Lagrangian of teleparallel theory. However, we will restrict ourselves to the case in which coefficients $A, B, C$ are chosen in such a way that we do not have additional constraints in our theory\footnote{The constraints are defined in chapter  \ref{Hamiltonov formalizam sa vezama}}. That theory will be referred to as \textit{Teleparallel gravity} (TG). \par\null\par

Let us now take one step back and ask ourselves why do we even want to examine a theory equivalent to GR? Didn't we introduce the torsion in our theory in order to fix the problems that GR has? Well, there are two main reasons why we are interested in an equivalent formulation of any theory, especially TEGR:
\begin{itemize}
    \item Some problems can be much easier in the new formulation of the theory. For example, Nester  \cite{Nester} proved the positivity of energy in GR using the tetrad formalism.
    \item Sometimes, generalizing a theory can be very easy from one formulation, and very hard from the other. For example, even though GR agrees with the experiment, it still has theoretical problems such as singular solution and nonrenormalizability that we wish to fix. That is why generalizations of GR are especially attractive. One way to generalize GR is to consider theories with $f(R)$ Lagrangians. Similarly, we can also consider  $f(\mc{L}_T)$ theories \cite{Izumi}. It is important to emphasize that new theories $f(R)$ and $f(\mc{L}_T)$ are not equivalent because the total divergence term in \eqref{ekv rel 1} now plays a significant role. The first step towards understanding $f(\mc{L}_T)$ theories properly, is analyzing TEGR.
\end{itemize}

\subsection*{Geometrical Properties}
\begin{wrapfigure}{i}{0.35\textwidth}
\vspace*{-2cm}\includegraphics[width=0.9\linewidth]{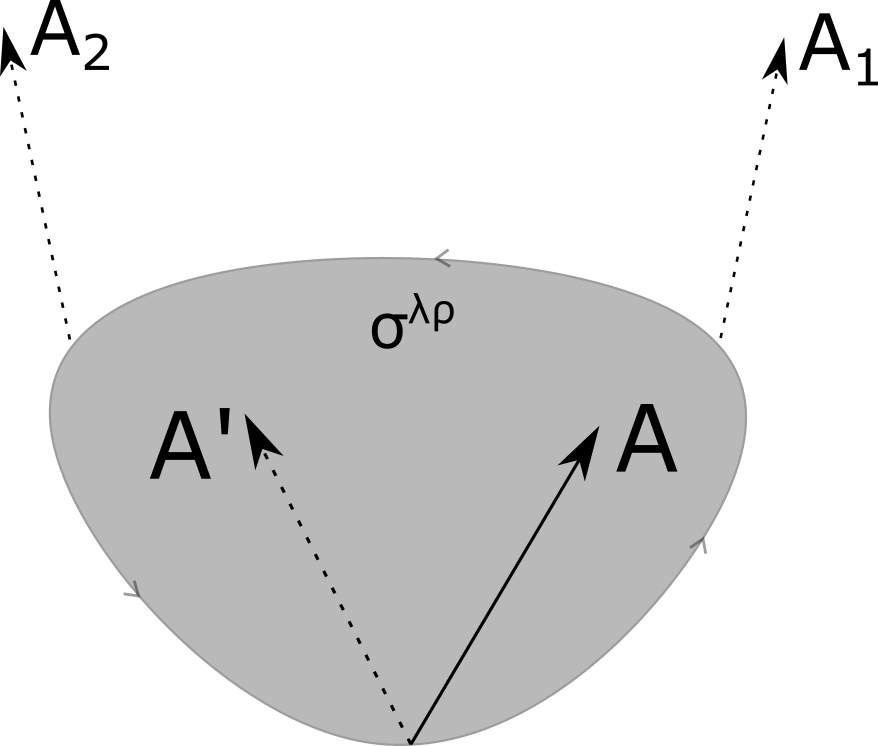} 
\caption{$A'$ is obtained by parallelly transporting  vector $A$ around the closed contour}
\label{parprenos}
\end{wrapfigure}

It is well-known that parallel transport is path-dependent. If we consider parallel transport of vector $A^\mu$ around an infinitesimally small closed loop it is easy to see that:

\begin{equation}
    \Delta A_\nu = \fr{1}{2} R^\mu{}_{\nu \lambda \rho } A_\mu \sigma^{\lambda\rho}
\end{equation}
where $\sigma^{\lambda\rho}$ is the infinitesimally small surface (pic. \ref{parprenos}) \par\null\par

In TG and TEGR curvature tensor is zero. Hence, parallel transport is path-independent, so we can define the notion of parallelism globally. This is the reason why we called them teleparallel theories - greek word $\tau \Tilde{\eta}\lambda \varepsilon$ should be pronounced 'tele' and it means distant.

\section{ADM Basis}
\subsection*{Definitions and Basic Features}
The main goal of this paper is examining the canonical structure of TEGR. Any canonical analysis requires a notion of time. It is thus useful to introduce a new basis, one in which the coordinate perpendicular to a spacelike hypersurface will play an important role. Such a basis represents 3+1 decomposition of space-time.\par\null\par 

\begin{wrapfigure}{o}{0.4\textwidth}
\includegraphics[width=0.9\linewidth]{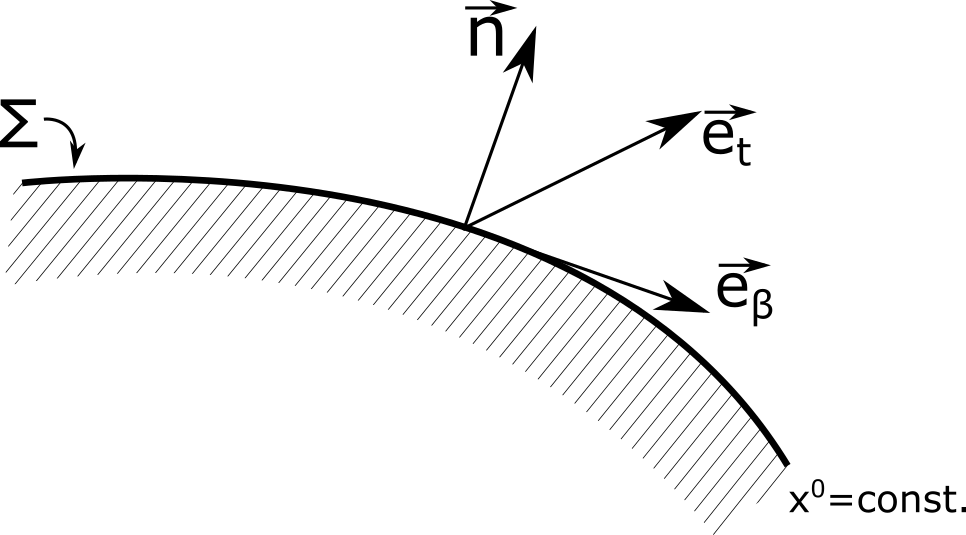} 
\caption{ADM basis}
\label{parprenos1}
\end{wrapfigure}
In this section, we will not only define the ADM basis but also state many useful identities which we will later use extensively.

Let ($\Vec{e}_t$, $\Vec{e}_\alpha$) be the coordinate basis. We define  Arnowitt Deser Misner (ADM) basis\cite{ADMoriginal, ADMPregled} ($\Vec{n}$, $\Vec{e}_\alpha$) by requiring that $\Vec{n}\cdot \Vec{e}_\alpha = 0$ and $\Vec{n}\cdot \Vec{n} =1 $. Historically, the first man to use such a basis, in a slightly different form, was Dirac\cite{DirakADM}. Let us now express  $\Vec{n}$ in terms of previously defined variables:

\begin{Lema}
\begin{equation} \label{vektor n}
    n_k =  \fr{h_k{}^0}{\sqrt{g^{00}}}
\end{equation}
\end{Lema}

\begin{Dokaz} \normalfont
It is obvious that $\Vec{n}$ is unique. We can conclude that it is sufficient to prove that~\eqref{vektor n} satisfies all the required conditions: 
\begin{equation*}
    \Vec{e}_\alpha \cdot \Vec{n} \sim b^k{}_\alpha h_k{}^0 = \delta^0_\alpha = 0, \qquad
    \Vec{n} \cdot \Vec{n} = \fr{\eta^{kl} h_k{}^0 h_l{}^0}{g^{00}} = 1
\end{equation*}
\flushright
\qedsymbol
\end{Dokaz}

Every vector  $\Vec{A}$ has a component along the spacelike hypersurface  $\Sigma$ ( ${\Vec{A}_{||}}$) and a component along  $\Vec{n}$ ($\Vec{A}_\perp$):

\begin{subequations}
\begin{eqnarray}
  \Vec{A}_\perp &=& (\Vec{n}\cdot \Vec{A})  \Vec{n},\quad\quad\;\;\;\text{тј.} \quad   (A_\perp )_k = n_k (n^l A_l)  \\
   \Vec{A}_{||} &=& \Vec{A} - (\Vec{n}\cdot\Vec{A})\Vec{n},\quad\text{тј.} \quad   (A_{||} )_k = (\delta^l_k - n^l n_k) A_l 
\end{eqnarray}
\end{subequations}
We can see that the corresponding projectors are: ${(P_\perp)^l_k = n^l n_k, \; (P_{||})^l_k = \delta^l_k - n^l n_k}$.
For calculations in ADM basis, it is very useful to introduce new notation $A_\B{k}$, which represents a component of the vector $A_k$ along hypersurface $\Sigma$. Similarly, we define  $A_\perp$ as a component of $A_k$ along $\Vec{n}$. Then, the decomposition of vector $A_k$ can be written as:

\begin{equation} \label{razlaganje na perp i n}
    A_k = A_\B{k} + n_k A_\perp,\qquad n^k A_\B{k} = 0, \qquad n^k A_k = A_\perp
\end{equation}
Having in mind that indices with bars represent components in hypersurface $\Sigma$, it is somewhat natural to define $\delta^\B{l}_\B{k}$ to be equal to projector  $(P_{||})^l_k$:
\begin{equation}
    \delta^\B{l}_\B{k} = \delta^\B{l}_k= \delta^l_\B{k}  =\delta^l_k - n^l n_k
\end{equation}

If we have tensor with more than one index, then the decomposition~\eqref{razlaganje na perp i n} is valid for each index separately. For example, if  $B_{km}$ is an antisymmetric, second rank tensor (like torsion with respect to second and third index), then:

\begin{eqnarray}
    B_{km} &=& B_{\B{k} m} + n_k B_{\perp m} = B_\BB{k}{m} + n_k B_{\perp m} + n_m B_{\B{k} \perp} \NB
    &=&  B_\BB{k}{m} + n_k B_{\perp \B{m}} + n_k n_m B_{\perp \perp } + n_m B_{\B{k} \perp} \NB
    &=& B_\BB{k}{m} + 2B_{[\B{k} \perp} n_{m]}
\end{eqnarray}
where $B_\BB{k}{m} = \delta^l_\B{k} \delta^n_\B{m} B_{ln}, \quad B_{\B{k}\perp} = \delta^n_\B{k} n^m B_{nm}$.
It is now easy to represent any tensor in a similar form. 

We noted that the ADM basis represents 3+1 decomposition of space-time. It should be natural then to define the notion of three-dimensional metric ${}^3 g^{\alpha\beta}$. We define it as the inverse of  $g_{\alpha\beta}$:
\begin{Lema} \normalfont
 ${}^3 g^{\alpha\beta} = g^{\alpha\beta} - \fr{g^{\alpha 0}g^{\beta 0}}{g^{00}}$
\end{Lema}

\begin{Dokaz}
\begin{equation}
  {}^3 g^{\alpha\beta}   g_{\beta\gamma} = g^{\alpha \mu} g_{\mu \gamma} - g^{\alpha 0} g_{0 \gamma} - \fr{g^{\alpha 0}}{g^{00}} \left( g^{0\mu} g_{\mu\gamma} -g^{00}g_{0\gamma} \right) = \delta^\alpha_\gamma \NN
\end{equation}
\flushright
\qedsymbol
\end{Dokaz}
\noindent
It should be noted that ${}^3 g^{\alpha\beta}$ is defined for $\alpha, \beta = 1,2,3$. However, we will formally extend the definition to the case when either one of indices is zero: ${}^3 g^{\alpha0}={}^3 g^{0\alpha}= {}^3 g^{00}=0 $. \par\null\par

It turns out that it is useful to also define $N$ и $N^\alpha$ as:

\begin{equation} 
    \Vec{e}_t = N \Vec{n} + N^\alpha \Vec{e}_\alpha \label{uvodjenje N i N alfa}
\end{equation}

\begin{Lema}\label{ N i N alfa}
\begin{equation}
    N = n_k b^k{}_0 = \fr{1}{\sqrt{g^{00}}}, \quad N^\alpha = h_\B{k}{}^\alpha b^k{}_0 = -\fr{g^{0\alpha}}{g^{00}} \label{lema 4 aaa}
\end{equation}
\end{Lema}

\begin{Dokaz}\normalfont
If we multiply ~\eqref{uvodjenje N i N alfa} by  $\Vec{n}$, using the fact that $\Vec{e}_\alpha \cdot \Vec{n} = 0$, we get:
\begin{equation*}
    N= \Vec{e}_t \cdot \Vec{n} = n_k b^k{}_0 = \fr{1}{\sqrt{g^{00}}}
\end{equation*}
We can just as easily prove the second relation in \eqref{lema 4 aaa}. The only difference is that we should multiply~\eqref{uvodjenje N i N alfa} by  $\Vec{e}_\beta$ and then use that $\Vec{e}_t \cdot \Vec{e}_\beta = g_{0\beta},\; \;$  $\Vec{e}_\alpha \cdot \Vec{e}_\beta = g_{\alpha \beta}$  $\Vec{e}_\beta \cdot \Vec{n} = 0$:

\begin{equation*}
    g_{0 \beta} = N^\alpha g_{\alpha \beta} 
\end{equation*}
If we now multiply both sides by ${}^3 g^{\gamma\beta}$:
\begin{equation*}
    N^\alpha = {}^3 g^{\alpha \beta} g_{0 \beta} = {}^3 g^{\alpha \mu} g_{0 \mu} = -\fr{g^{\alpha 0}}{g^{00}}
\end{equation*}
In order to complete the proof, we also need:
\begin{equation*}
    h_\B{k}{}^\alpha b^k{}_0 = (\delta^m_k - n^m n_k) h_m{}^\alpha b^k{}_0 = -\fr{h^{m0} h_k{}^0}{g^{00}}h_m{}^\alpha b^k{}_0 = -\fr{g^{\alpha 0}}{g^{00}}
\end{equation*}

\flushright
\qedsymbol
\end{Dokaz}

\subsection*{Useful Identities}
There are quite a few identities that we now want to state. We will divide them into groups:
\begin{itemize}
    \item Group I will consist of identities useful for transition from the coordinate to ADM basis (and other way around). It turns out that group I will be particularly useful for constructing canonical Hamiltonian and generalized momenta.
    \item Group II consists of identities in the ADM basis. These identities will not have derivatives in them. They will be used to bring some expressions into suitable forms. 
    \item Group III consists of identities in the ADM basis that have derivatives with respect to tetrads. They are very useful for calculating Poisson brackets, which are inevitable in the canonical analysis.
\end{itemize}
\subsubsection*{Group I}
\vspace{0.4cm}
\begin{subequations}
\begin{empheq}[box=\widefbox]{align}
  A_\perp &= NA^0   &  A_0  &= N A_\perp + N^\alpha A_\alpha \label{ADM grupa I a} \\
   A_\B{m} &= A_\alpha \left[ h_m{}^\alpha + N^\alpha h_m{}^0 \right]   &  A^\B{m} &= b^m{}_\alpha A^\alpha + N^\alpha b^m{}_\alpha A^0  \label{ADM grupa I b}\\
   A^\B{m}B_\B{m} &= \left[ A^\beta +N^\beta A^0 \right] B_\beta    \label{ADM grupa I c}
\end{empheq}
\end{subequations}
We can see that there is an analogy between the perpendicular component ($A_\perp$) in the ADM basis and the zeroth component in the coordinate basis ($A_0$). A similar analogy can be drawn between $A_{\B{m}}$ and $A_\alpha$. They are not interchangeable in a straightforward way, but it is useful to have such  analogy in mind.
\vspace{1cm}
\subsubsection*{Group II}
\vspace{0.4cm}
\begin{subequations}
\begin{empheq}[box=\widefbox]{align}
  A_\B{k} B^{\B{k}} &= A_k B^{\B{k}} = A_\B{k} B^k   &  \eta_{kl} \eta^\BB{l}{m}  &= \eta_\BB{k}{l} \eta^\BB{l}{m} = \delta^\B{m}_\B{k}  \label{ADM grupa II a}\\
   \delta^\B{k}_\B{k} &= 3   &  \epsilon_{\BBB{k}{l}{m}\perp} \epsilon^{\BBB{k}{l}{m}\perp} &= -6 \label{ADM grupa II b}\\
   h_\B{k}{}^0 &= 0   &   n_\B{k} &= 0 \label{ADM grupa II c}\\
   b^k{}_\perp &= n^k   &   {}^3 g^{\alpha \beta} &= h_\B{k}{}^\alpha h^{\B{k}\beta} \label{ADM grupa II d}\\
   h_\B{k}{}^\alpha A^k &= {}^3 g^{\alpha \beta} A_\beta   &   {}^3 g^{\alpha \beta}b_{m\alpha} &= h_\B{m}{}^\beta \label{ADM grupa II e}\\
   h_\B{k}{}^\alpha b^k{}_\beta &= \delta^\alpha_\beta   &   h_\B{k}{}^\alpha b^m{}_\alpha &= \delta^\B{m}_\B{k} \label{ADM grupa II f} 
\end{empheq}
\end{subequations}
Some of these relations are equivalent, but from a practical point of view, it is useful to list all of them. Let us state one more relation that belongs to this group, but we intentionally separate it from the rest of the group:
\begin{equation}
    b:= det(b^k{}_\mu) = \fr{N}{n_0} det(b^a{}_\alpha) = JN
\end{equation}
where  $J=\fr{1}{n_0} det(b^a{}_\alpha) $. Proof of this relation is given in lemma  \ref{B: lema 4}.\newline
\noindent
It turns out that the calculation of Poisson brackets can be somewhat simplified using this relation because it expresses determinant $b$ in terms of $J$, which does not depend on  $b^k{}_0$, and $N$, which linearly depends on $b^k{}_0$.
\subsubsection*{Group III}
\begin{subequations}
\begin{empheq}[box=\fbox]{align}
  \fr{\p h_i{}^\mu}{\p b^j{}_\nu} &= -h_i{}^\nu h_j{}^\mu   &  \fr{\p g^{\mu\nu}}{\p b^i{}_\rho} &= -h_i{}^\mu g^{\nu\rho} - h_i{}^\nu g^{\mu\rho} \label{Grupa III a}\\
   \fr{\p n_k}{\p b^i{}_\mu} &= -n_i h_\B{k}{}^\mu = -n_i b^k{}_\alpha {}^3g^{\alpha \mu}   &  \fr{\p N}{\p b^k{}_\alpha} &= -N^\alpha n_k \label{Grupa III b}\\
   \fr{\p J}{\p b^i{}_\mu} &= J b_{i\alpha} {}^3g^{\alpha\mu}=J h_\B{i}{}^\mu   &   \fr{\p b}{\p b^k{}_\mu} &=b h_k{}^\mu \label{Grupa III c}\\
   n^k \fr{\p N^\beta}{\p b^k{}_\alpha} &= N {}^3 g^{\alpha\beta}   &   n^k \fr{\p h_\B{m}{}^\beta}{\p b^k{}_\alpha} &= n_m {}^3 g^{\alpha\beta}   \label{Grupa III d}\\
   b^s{}_\alpha \fr{\p h_\B{l}{}^\gamma}{\p b^s{}_\mu} &= -\delta^\gamma_\alpha h_\B{l}{}^\mu  &   b^l{}_\alpha \fr{\p \delta^\B{j}_\B{p}}{\p b^l{}_\mu} &= 0
   \label{Grupa III e} \\
    n^k \frac{\partial n_m}{\partial b^k{}_{\alpha }}&=- b_{m\rho } \;^3g^{\rho \alpha } \label{Grupa III f}
\end{empheq}
\end{subequations}
Many of those relations, from all three groups, can be easily derived from~\eqref{razlaganje na perp i n}. We will not prove all of them because we do not want to wander off too far from our actual work. However, we think that it is important to prove at least some of them to illustrate methods for calculation in the ADM basis. That's why we will prove a few relations from each group:
\newline
Group I:
\begin{eqnarray*}
   A_\perp
        &=& n^m A_m = n^m h_m{}^\mu A_\mu = \fr{g^{0\mu}}{\sqrt{g^{00}}}A_\mu = N A^0 \\
   A_0 
        &=& \fr{1}{g^{00}} \left[ g^{0\mu} A_\mu - g^{0\alpha} A_\alpha \right] = \fr{1}{g^{00}} A^0 - \fr{g^{0\alpha}}{g^{00}}A_\alpha = 
   NA_\perp + N^\alpha A_\alpha \\
   A_\B{m} 
        &=& A_m - n_m A_\perp = h_m{}^\mu A_\mu - n_m N A^0 = h_m{}^0 A_0 -\fr{h_m{}^0}{g^{00}} A^0 + h_m{}^\alpha A_\alpha \\
        &=& h_m{}^0 A_0 -\fr{h_m{}^0}{g^{00}}\left[ g^{00}A_0 + g^{0\alpha}A_\alpha \right] + h_m{}^\alpha A_\alpha =
        A_\alpha \left[ h_m{}^\alpha + N^\alpha h_m{}^0 \right]
\end{eqnarray*}

Group II:
\begin{eqnarray*}
    A_\B{k} B^\B{k} 
        &=& \delta^m_\B{k} A_m B^\B{k} = \left( \delta^m_k - n^m n_k \right) A_m B^\B{k} = A_k B^\B{k} \\ 
    {}^3 g^{\alpha \beta} A_\beta
        &=& {}^3 g^{\alpha \rho} A_\rho = A^\alpha - A^0 \fr{g^{\alpha 0}}{g^{00}} = A^\alpha - \fr{g^{\alpha 0}}{\sqrt{g^{00}}}A_\perp = h_\B{k}{}^\alpha A^k \\
        {}^3g^{\rho \beta }b_{m\rho }&=&(g^{\rho \beta }-\frac{g^{\rho 0}g^{\beta 0}}{g^{00}})b_{m\rho }=(\delta^n_m-n^n n_m)h_n{}^{\beta }=h_{\Bar{m}}{}^{\beta }
\end{eqnarray*}

Group III: \par\null\par

Relation $\frac{\partial h_i{}^{\mu}}{\partial b^j{}_{ \nu}} = 
-h_j{}^{ \mu} h_i{}^{ \nu}$ can be obtained by multiplying each side of $\delta^k_i = b^k{}_{\rho} h_i{}^{\rho}$ with  $h_k{}^{ \mu} \partial b^j{}_{ \nu}$.

\begin{equation*}
    0=  h_k{}^{\mu} \frac{\partial  b^k{}_{\rho}}{\partial b^j{}_\nu}h_i{}^{\rho}+ h_k{}^{\mu}
 b^k{}_{\rho} \frac{\partial h_i{}^{\rho}}{\partial b^j{}_{ \nu}} =
 h_i{}^{ \nu}h_j{}^{ \mu}+\frac{\partial h_i{}^{ \mu}}{\partial b^j{}_{ \nu}}
\end{equation*}
The second relation in  \eqref{Grupa III a} is also easy to prove, using \eqref{metrika preko tetrada}. Further, using previously obtained results, as well as chain rule:
\begin{eqnarray*}
   \frac{\partial n_k}{\partial b^i{}_{\mu}} &=& \frac{1}{\sqrt{g^{00}}}\frac{\partial h_k{}^{0}}{\partial b^i{}_{\mu}}-\frac{h_k{}^{0}}{2(g^{00})^{\frac{3}{2}}} \frac{\partial g^{00}}{\partial b^i{}_{ \mu}} 
= -\frac{h_i{}^{0}h_k{}^{ \mu}}{\sqrt{g^{00}}}-\frac{h_k{}^{0}}{2(g^{00})^{\frac{3}{2}}}\left[ -h_i{}^{0}g^{0\mu}-h_i{}^{0}g^{0\mu} \right]\\
&=& -n_i h_k{}^{\mu}+n_i n_k \frac{g^{\mu 0}}{\sqrt{g^{00}}} 
\end{eqnarray*}
Let us also prove a bit more interesting result $\frac{\partial b}{\partial b^k_{\;\mu}}=b \cdot h_k^{\;\mu}$: Using Jacobi's formula $\frac{\partial \mathrm{det} A}{\partial A_{ij}}=\mathrm{adj}^T (A)_{ij}$, as well as the formula for inverse of matrix $\mathrm{adj}(A)=\mathrm{det}A \cdot A^{-1}$, we get:

\begin{equation}
    \frac{\partial \mathrm{det} A}{\partial A_{ij}}=\mathrm{adj}^T (A)_{ij}=\mathrm{det}A \cdot (A^{-1})^T_{ij}
\end{equation}
In our case, matrix $A$ is the tetrad matrix $b^k{}_{\mu}$, while $A^{-1})^T$ is the inverse tetrad matrix $h_k{}^{\mu}$. That proves our relation.\par\null\par

We could also prove all the other relations similarly. These results are the starting point for proving much more complicated results, as we will see in the following sections.

\section{Proof of the Equivalence Between GR and TEGR}
\label{Ekvivalentnost}
In this section, we will prove~\eqref{relacija TEOTR} because in~\S~\ref{Teleparalelna} we concluded that the equivalence between GR and TEGR is a consequence of this relation. We believe that the proof will be clearer if we divide it into a couple of simpler lemmas.\par\null\par

It is important to note that~\eqref{relacija TEOTR} is defined in Riemann-Cartan geometry, so it is necessary to temporarily leave our teleparallel case. It turns out that it is useful to define $H_{ij}^{\mu\nu} := b \left( h_i{}^\mu h_j{}^\nu - h_i{}^\nu h_j{}^\mu \right)$. In Appendix~\ref{DodatakD} in order to prove a couple of important identities, we used the fact that $H_{ij}^{\mu\nu}$ is defined as a product of two determinants\footnote{$b$ is the determinant of the tetrad matrix, while $h_i{}^\mu h_j{}^\nu - h_i{}^\nu h_j{}^\mu$ is the determinant of a submatrix}. Those identities are not important in this section, but we will nevertheless use them in the following chapters. We can now formulate our first lemma:
 
\begin{Lema} \label{Lema 1.5}
\begin{equation}
\nabla_\nu H_{ij}^{\mu\nu} = b h_k{}^\mu \left( T^k{}_{ij} - \delta^k_i T_j + \delta^k_j T_i \right)
\end{equation}
\end{Lema}

\begin{Dokaz} \normalfont
\begin{eqnarray} 
\nabla_\nu H_{ij}^{\mu\nu} &=& \p_\nu H_{ij}^{\mu \nu} 
                            -\omega^s{}_{i\nu} H_{sj}^{\mu\nu}
                            -\omega^s{}_{j\nu} H_{is}^{\mu\nu} \NB
                           &=&
                             \p_\nu H_{ij}^{\mu \nu} + (b \omega^k{}_{ji} h_k{}^\mu - \omega^k{}_{ij}b h_k{}^\mu) - bh_i{}^\mu \omega^s{}_{js}+ \omega^s{}_{is}bh_j{}^\mu \label{Lema 1.5 PomRel 1}
\end{eqnarray}

\begin{eqnarray}
\p_\nu H_{ij}^{\mu \nu}
        &=&
            b \p_\nu \left[ h_i{}^\mu h_j{}^\nu - h_i{}^\nu h_j{}^\mu \right] +(\p_\nu b) \left[ h_i{}^\mu h_j{}^\nu - h_i{}^\nu h_j{}^\mu \right]
            \label{Lema 1.5 PomRel 2}
\end{eqnarray}
Using the relations~\eqref{Grupa III a} and~\eqref{Grupa III c}, as well as the chain rule, it is easy to see that: $\p_\nu b = \fr{\p b}{\p b^k{}_\rho}\p_\nu b^k{}_\rho = b h_k{}^\rho \p_\nu b^k{}_\rho$ and similarly $\p_\nu \left[ h_i{}^\mu h_j{}^\nu \right] = -h_j{}^\nu h_i{}^\rho h_k{}^\mu \p_\nu b^k{}_\rho - h_i{}^\mu h_j{}^\rho h_k{}^\nu \p_\nu b^k{}_\rho$. If we now use these relations in~\eqref{Lema 1.5 PomRel 2} it follows that:

\begin{equation*}
   \p_\nu H_{ij}^{\mu \nu} = b h_k{}^\mu \left( c^k{}_{ij} - \delta^k_i c_j + \delta^k_j c_i \right) 
\end{equation*}
where  $c^i{}_{\mu\nu}= \p_\mu b^i{}_\nu - \p_\nu b^i{}_\mu,\quad c_k = c^m{}_{mk}$. Let us now use that in~\eqref{Lema 1.5 PomRel 1}:
\begin{eqnarray*}
   \nabla_\nu H_{ij}^{\mu \nu}
        &=& \left( b h_k{}^\mu c^k{}_{ij} +  b \omega^k{}_{ji} h_k{}^\mu - \omega^k{}_{ij}b h_k{}^\mu\right) 
        -\left( b h_k{}^\mu \delta^k_i c_j  + bh_i{}^\mu \omega^s{}_{js}\right)\\
        &&+\left( b h_k{}^\mu \delta^k_j c_i + \omega^s{}_{is} bh_j{}^\mu \right)\\
        &=& bh_k{}^\mu T^k{}_{ij} - bh_k{}^\mu \delta^k_i T_j + bh_k{}^\mu \delta^k_j T_i
\end{eqnarray*}
\flushright
\qedsymbol

\end{Dokaz}
$H_{ij}^{\mu\nu}$ can be particularly useful because it turns out that $\fr{1}{2}H_{ij}^{\mu\nu} R^{ij}{}_{\mu\nu} = b R$.\newline
In order to formulate the next lemma, it is useful to have in mind the third item in lemma~\ref{Prva Lema}, which states that $\omega_{ij\mu} = \Delta_{ij\mu} + K_{ij\mu}$. We can see that $\Delta$ is actually Levi-Civita contribution to the spin connection, while $K$ is torsion contribution.

\begin{Lema} \normalfont
Let $R^{ij}{}_{\mu \nu} (\omega)$ be the curvature tensor with respect to the spin connection, $R^{ij}{}_{\mu \nu} (\Delta)$ the curvature tensor with respect to Levi-Civita connection, while $\Tilde{\nabla}$ is the covariant derivative with respect to Levi-Civita connection. Then:
\begin{equation} \label{Lema 1.6}
   R^{ij}{}_{\mu \nu} (\omega) =  R^{ij}{}_{\mu \nu} (\Delta) + \left[ \Tilde{\nabla}_\mu K^{ij}{}_\nu + K^i{}_{s\mu} K^{sj}{}_\nu - (\mu\nu) \right]
\end{equation}
\end{Lema}

\begin{Dokaz}
\begin{eqnarray*}
R^{ij}{}_{\mu\nu} (\omega) 
        &=&
            \p_\mu \omega^{ij}{}_\nu - \p_\nu \omega^{ij}{}_\mu 
            +\omega^i{}_{s\mu} \omega^{sj}{}_\nu
            -\omega^i{}_{s\nu} \omega^{sj}{}_\mu \\
        &=& \p_\mu (\Delta^{ij}{}_\nu + K^{ij}{}_\nu)
            -\p_\nu(\Delta^{ij}{}_\mu + K^{ij}{}_\mu) \\
        && + (\Delta^i{}_{s\mu} + K^i{}_{s\mu})(\Delta^{sj}{}_\nu + K^{sj}{}_\nu)
            -(\Delta^i{}_{s\nu}+K^i{}_{s\nu})(\Delta^{sj}{}_\mu + K^{sj}{}_\mu)\\
        &=& R^{ij}{}_{\mu\nu}(\Delta) + \left[ \left( \p_\mu K^{ij}{}_\nu + \Delta^i{}_{s\mu} K^{sj}{}_\nu - \Delta^{sj}{}_\mu K^i{}_{s\nu}\right) + K^i{}_{s\mu} K^{sj}{}_\nu - (\mu\nu) \right] \\
        &=& R^{ij}{}_{\mu\nu}(\Delta) +\left[ \Tilde{\nabla}_\mu K^{ij}{}_\nu + K^i{}_{s\mu} K^{sj}{}_\nu - (\mu\nu) \right]
\end{eqnarray*}
\flushright
\qedsymbol
\end{Dokaz}
We are now ready to prove~\eqref{relacija TEOTR}. Having in mind that $\fr{1}{2} H_{ij}^{\mu\nu} R^{ij}{}_{\mu\nu}=bR$ and\footnote{Lemma~\ref{Lema 1.5} states that  $\Tilde{\nabla}_\mu H_{ij}^{\mu\nu}$ is proportional to the torsion tensor. However, $\Tilde{\nabla}$ is defined with respect to Levi-Civita connection, so there is no torsion.} $\Tilde{\nabla}_\mu H_{ij}^{\mu\nu}=0$, we can multiply each side of equation~\eqref{Lema 1.6} by $H_{ij}^{\mu\nu}$:

\begin{eqnarray}
bR(\omega) 
        &=&
            bR(\Delta) + \Tilde{\nabla}_\mu \left( K^{ij}{}_\nu H^{\mu\nu}_{ij}\right)+ \fr{1}{2}H_{ij}^{\mu\nu} \left[ K^i{}_{s\mu}K^{sj}{}_\nu - K^i{}_{s\nu}K^{sj}{}_\mu \right]\NB
        &=& bR(\Delta) + \p_\mu (2b h_i{}^\mu h_j{}^\nu K^{ij}{}_\nu)+
            bh_i{}^\mu h_j{}^\nu \left( K^i{}_{s\mu}K^{sj}{}_\nu - K^i{}_{s\nu}K^{sj}{}_\mu \right) \NB
        &=& bR(\Delta) + 2\p_\mu (bK^\mu) + b\left(K^i{}_{si}K^{sj}{}_j - K^i{}_{sj}K^{sj}{}_i\right) \label{Lema 1.6 Pom Rel}
\end{eqnarray}
Using  $K^i{}_{jk} = -\fr{1}{2} \left( T^i{}_{jk} - T_k{}^i{}_j + T_{jk}{}^i  \right)$, we can transform the last term in ~\eqref{Lema 1.6 Pom Rel} :
\begin{eqnarray*}
bK^i{}_{si} K^{sj}{}_j &=& -b T^s T_s \\
-bK^i{}_{sj}K^{sj}{}_i &=& \fr{b}{2} T_{ijk}T^{jik} + \fr{b}{4} T_{ijk}T^{ijk}
\end{eqnarray*}
Hence~\eqref{Lema 1.6 Pom Rel} can be written as:

\begin{equation*}
  bR(\omega) =
            bR(\Delta)  + 2\p_\mu (bK^\mu)-b T^s T_s+ \fr{b}{2} T_{ijk}T^{jik} + \fr{b}{4} T_{ijk}T^{ijk}
\end{equation*}
which is exactly what we wanted to prove. As we mentioned earlier (in \S~\ref{Teleparalelna}), having in mind that this relation holds, it is now valid to conclude that the equivalence between GR and TEGR also holds.

\section{Electrodynamics in the External Gravitational Field}
\label{Elektrodinamika u spolj. grav. polju}

In this section, we will construct Hamiltonian in the ADM basis for the electrodynamics. In \S~\ref{Elektrodinamika} we will give further examination of electrodynamics by constructing generators of gauge symmetries. This example illustrates practically the same procedure that we will use in the case of teleparallel gravity, but the calculations are much simpler.\par\null\par

Hamiltonian is defined as a Legendre transformation of a Lagrangian. Since the Hamiltonian is supposed to be a function of coordinates and momenta, it is necessary to eliminate generalized velocities. From~\eqref{ADM grupa I a} we see that in the coordinate basis, velocities can be found in $F_{0\alpha}$ term, whereas in AMD basis $F_{\perp \B{m}}$ is the term with velocities. That's why our task is to eliminate those terms.\newline
EM Lagrangian is defined as $\mathcal{L}^{EM}=-\frac{1}{4}b F_{\mu\nu} F^{\mu\nu}$. Hence:
\begin{equation} \label{KanonskiImpulsED}
    \pi^\alpha=\frac{\partial \mathcal{L}^{EM}}{\partial \partial_0 A_\alpha}=bF^{\alpha 0},
    \qquad \pi^0=0
\end{equation}
Equation $\pi^0 =0$ is the so-called \textit{primary constraint}. Canonical Hamiltonian is thus:

\begin{eqnarray} \label{KanonskiHamiltonijanED}
 \mathcal{H}_c&=&\pi^\alpha \partial_0 A_\alpha-\mathcal{L}^{EM}  
\end{eqnarray}
In order to express equation \eqref{KanonskiHamiltonijanED} in the ADM basis, we will rewrite the first term as:
\begin{eqnarray*}
\pi^\alpha \partial_0 A_\alpha = \pi^\alpha F_{0\alpha} + \pi^\alpha \p_\alpha A_0 = \p_\alpha (\pi^\alpha A_0)-A_0 \p_\alpha \pi^\alpha + \pi^\alpha F_{0\alpha}
\end{eqnarray*}
Using~\eqref{ADM grupa I a} and~\eqref{ADM grupa I c} we can make some additional transformations:

\begin{eqnarray*}
    \pi^\alpha \partial_0 A_\alpha &=&\p_\alpha (\pi^\alpha A_0)-A_0 \p_\alpha \pi^\alpha + \pi^\alpha \left(
    N F_{\perp\alpha}+N^\beta F_{\beta\alpha}
    \right) \\
    &=& \partial_\alpha(\pi^\alpha A_0)-A_0\partial_\alpha \pi^\alpha+N\pi^{\bar{m}} F_{\perp\bar{m}}+N^\beta \pi^\alpha F_{\beta\alpha},
\end{eqnarray*}
where we used the fact that  $\pi^\alpha F_{\perp \alpha}=\pi^{\bar{m}} F_{\perp\bar{m}}$. The second term in~\eqref{KanonskiHamiltonijanED} should also be in ADM form, so we will write the Lagrangian as  $\mc{L}^{EM} = -\frac{1}{4}b F_{mn}F^{mn}$ and then use the following decomposition:

\begin{equation} \label{Ralaganjepolja}
    F_{mn}=F_{\bar{m} \bar{n}}+F_{\bar{m} \perp} n_n-F_{\bar{n} \perp}n_m
\end{equation}
It turns out that it is useful to introduce the so-called \textit{parallel momenta} as:
\begin{equation} \label{Paralelniimpulsi}
    \hat{\pi}^{\bar{l}}:=b^{l}_{\; \alpha} \pi^\alpha = J\frac{\partial \mathcal{L}^{em}}{\partial F_{\perp \bar{l}}}=J F^{\bar{l} \perp},
\end{equation}
where $J=\frac{b}{N}$, $ b:=\mathrm{det}b$. Lagrangian can now be rewritten using:
\begin{equation} \label{F2}
    F_{mn}F^{mn}=F_{\bar{m} \bar{n}} F^{\bar{m} \bar{n}} +\frac{2}{J^2} \pi^{\bar{m}} \pi_{\bar{m}}
\end{equation}
It is now straightforward to express the canonical Hamiltonian in the so-called \textit{Dirac-ADM}\cite{DirakADM,ADMoriginal} form.
\begin{equation} \label{ADMHamiltonijanED}
    \mathcal{H}_c=N\mathcal{H}_\perp + N^\alpha \mathcal{H}_\alpha-A_0 \partial_\alpha \pi^\alpha+\partial_\alpha D^\alpha,
\end{equation}
where:
\begin{subequations}
\begin{eqnarray}
D^\alpha &=& A_0 \pi^\alpha \\
\mathcal{H}_\alpha &=& \pi^\beta F_{\alpha \beta} \\
\mathcal{H}_\perp &=& -\frac{1}{2J} \pi^{\bar{m}} \pi_{\bar{m}}+\frac{J}{4}F_{\bar{m}\bar{n}}F^{\bar{m}\bar{n}}
\end{eqnarray}
\end{subequations}
The ADM Hamiltonian in the teleparallel case will have the same form.
\begin{Napomena} $\quad$ \normalfont
\begin{itemize}
    \item $F_{\bar{m} \bar{n}}=h_{\bar{m}}{}^{\mu} h_{\bar{n}}{}^{\nu} F_{\mu \nu}$ cannot depend on velocities because $h_{\bar{m}}{}^{0}=0$.
    \item The "perpendicular sign" $\perp$ represents contraction with $n_m$, so it is irrelevant whether we write it as an upper or lower index $F_{\perp m}=F^{\perp}_{\quad m}$.
    \item Throughout this section, we raised and lowered all Lorentz indices (both with and without bar above them) with $\eta^{km}$. Let us show that it is valid to do so:

\begin{eqnarray*}
    \eta^{km}A_{\bar{k}}&=&\eta^{km}\delta^l_{\bar{k}}A_l=\eta^{km}(\delta^l_k-n^l n_k)A_l \\
    &=&(\delta^m _k -n^m n_k)\eta^{kl}A_l=\delta^{\bar{m}}_k \eta^{kl}A_l=A^{\bar{m}}
\end{eqnarray*}
\end{itemize}
\end{Napomena}
It is important to note that \eqref{ADMHamiltonijanED} is the Hamiltonian density, whereas the Hamiltonian is $H = \int d^3 x \; \mc{H}_c$. However, that does not mean that the term with total divergence is insignificant because there may exist some nontrivial boundary conditions. Luckily, this term will be irrelevant in most cases (but not always!). As we will see in the following chapters, one of the most time-consuming parts will be a calculation of Poisson brackets. It is easy to show that  $\{F,\int d^3 x' \; \p^{'}_\alpha {D'}^\alpha  \} = 0$, where  $D^\alpha$ can depend on the canonical variables $b^k{}_\mu, \pi_l{}^\nu$, but not on their derivatives. On the other hand, $F$ can depend on both the canonical variables and their derivatives. For example, if $F$ and $D^\alpha$ do not depend on derivatives of canonical variables, then there exists $ f^\alpha\left(b^k{}_\mu, \pi_l{}^\nu\right)$  such that:

\begin{eqnarray*}
\{F, \int dx' \; \p^{'}_\alpha {D'}^\alpha \} &\sim& \int dx' \; \p^{'}_\alpha \left[ f^\alpha\left(b^k{}_\mu, \pi_l{}^\nu\right) \delta(x-x') \right] \\
&=& f^\alpha \left(b^k{}_\mu, \pi_l{}^\nu\right) \int dx' \; \p^{'}_\alpha \delta(x-x') = 0
\end{eqnarray*}
It is slightly more complicated, but still pretty simple to prove the same result in the case when $F$ can also depend on derivatives of canonical variables. It is because of this result  $\{F,\int d^3 x' \; \p^{'}_\alpha {D'}^\alpha  \} = 0$ that we say that total divergence is irrelevant in most cases. That is why we will sometimes omit the total divergence term in Hamiltonian if we know that it does not contribute to the end result. On the other hand, if the $\{F,\int d^3 x' \; \p^{'}_\alpha {D'}^\alpha  \}$ has some noncanonical variables, like parameter in gauge transformation $\xi(x)$, then the Poisson bracket can be nonzero. The only such place, where the total divergence term is important is in chapter \ref{glava konstrukcija generatora}. There we calculate Poisson brackets of the canonical variables $b^k{}_\mu, \pi_l{}^\nu$ with a generator of gauge symmetries, which depends on gauge parameter (noncanonical variable).\par\null\par 

From a theoretical point of view, canonical Hamiltonian \eqref{ADMHamiltonijanED} should give us a complete description of the physical system. For example, equations of motion can be derived straightforwardly from the Hamiltonian. However, the canonical analysis of constrained systems is a bit different from the "standard" canonical analysis from classical mechanics. That is why we will give a short review of systems with constraints in the next chapter.

\clearpage
\chapter{Canonical Analysis of Constrained Systems}
\label{Hamiltonov formalizam sa vezama}
In~\S \;\ref{Elektrodinamika u spolj. grav. polju} we saw that in the case of electrodynamics there is a phase-space restriction   $\pi^0 = 0$. These kinds of restrictions are called \textit{constraints}. \par\null\par 

In order to work in the canonical formalism, we need to construct the Hamiltonian $H= \sum p_i {\Dot{q}}_i - L$ as a function of the generalized coordinates and momenta. It is thus necessary to eliminate all generalized velocities from Lagrange's formalism by solving the system of equations $p_i := \frac{\p L}{\p \Dot{q}_i}$. However, this system of equations does not have a solution in the case of the systems with constraints, so it seems that the canonical formalism is not applicable. But, it turns out that there is, in fact, a method to solve this seemingly unsolvable problem straightforwardly. The main goal of this chapter will thus be to give a short review of the canonical analysis for the constrained systems. This formalism is particularly useful for gauge theories because they are an example of a constrained system.\par\null\par

Why do we even want to formulate our problem in the canonical formalism?

\begin{itemize}
    \item The canonical formalism gives a well defined, nonperturbative procedure for calculating the number of degrees of freedom \cite{Blagojevic}. 
    \item In the canonical formalism, there is a so-called Castellani's algorithm\cite{Kastelani} which gives a straightforward way to find the generators of gauge symmetries.
    \item There is a well-known way to quantize theory from the canonical formalism\cite{Dirak}. 
    \item There is also a well-defined way to calculate conserved charges\cite{Teitelboim}. In Lagrange's case, conserved charges are usually obtained using the Noether theorem, but there are cases in which that formula is not well defined.   
    \item A large number of numerical methods are especially useful in the canonical formalism\cite{NumerickeMetode1,NumerickeMetode2,NumerickeMetode3}.
\end{itemize}

Historically speaking, Rosenfeld was the first man to consider the canonical formalism of constrained systems~\cite{Rozenfeld} by applying it to electrodynamics. However, it was after the works of Bergmann (with his collaborators)~\cite{Bergman1,Bergman2,Bergman3,Bergman4} and  Dirac~\cite{DirakRad1,DirakRad2} that this formalism started to attract larger attention. Dirac also wrote a book~\cite{Dirak}, which represents one of the finest reviews of this method.
\section{Primary Constraints}

\label{Primarne}
We will only consider systems with finitely many degrees of freedom. It is not difficult to generalize this whole section to the case of infinitely many degrees of freedom.\par\null\par

As we said earlier, in order to work in the canonical formalism it is necessary to eliminate generalized velocities, which is why we need to solve the system of equations:

\begin{equation} \label{genImpulsi}
    p_n = \frac{\p L}{\p \Dot{q}_n}(q_i,\Dot{q}_i,t), \quad i=1, 2...N
\end{equation}
If we now apply implicit function theorem~\cite[page 455]{Fihtengoljc}, we can see that our system of equations has a solution if the determinant of the Hessian matrix $W_{ij}$ is nonzero:
\begin{equation}
  W = det (W_{ij}) := det  \bigg\rvert \fr{\p^2 L}{\p \Dot{q}_i  \p \Dot{q}_j} \bigg\rvert \neq 0
\end{equation}
In the case $W \neq 0$ there is always a way to eliminate generalized velocities and thus represent all physical variables as functions of the generalized coordinates and momenta. This is a well-known case thoroughly studied in every classical mechanics course. On the other hand, the case $W = 0$ is seemingly more complicated and, as such, it will be the focus of our attention in this chapter. In this case, the Hessian matrix has rank $R < N$. That means that we can solve~\eqref{genImpulsi} in terms of exactly  $R$ generalized momenta.

\begin{equation} \label{genBrzine}
    \Dot{q}_a = f_a(q, \Dot{q}_{R+1}...\Dot{q}_N, p_\alpha,t),\quad a=1,2...R,
\end{equation}
where we assumed, without the loss of generality, that $a = 1,2 ... R$. Let $\{ \alpha \}$ denote the set of values that the symbol  $\alpha$  can take. That set contains $R$ elements, but those are not necessarily  $1,2 ... R$.\par\null\par

Let $ i\not\in \{\alpha\}$:

\begin{equation} \label{ostaliImpulsi}
    p_i = \fr{\p L}{\p \Dot{q}_i}(q,\Dot{q}_1...\Dot{q}_N,p_\alpha,t) \equiv h_i(q,\Dot{q}_{R+1}...\Dot{q}_N,p_\alpha,t),
\end{equation}
where we used~\eqref{genBrzine}. It is important to notice that $h_i(q,\Dot{q}_{R+1}...\Dot{q}_N,p_\alpha,t)$ cannot depend on $\Dot{q}_{R+1}...\Dot{q}_N$. Otherwise, we could solve~\eqref{ostaliImpulsi} for another generalized velocity, and that is impossible because the rank of Hessian matrix is $R$. Hence,   $h_i(q,\Dot{q}_{R+1}...\Dot{q}_N,p_\alpha,t) = h_i(q,p_\alpha,t)$, so the expression~\eqref{ostaliImpulsi} becomes:

\begin{equation}
    \phi_i(p,q) \equiv p_i - h_i(q,p_\alpha,t) = 0
\end{equation}

Relations $\phi_i(p,q)$ are called \textit{primary contraints}. To simplify the labeling, we will suppose that all $M= N-R$ constraints are independent, although it is not necessary to make such an assumption.

\begin{Primer} \normalfont
Let:
\begin{equation*}
    L = \fr{1}{2} \Dot{q}_1^2 + C\Dot{q}_2 \cdot q_1 - \fr{m}{2} \left( q_1^2 + q_2^2+q_3^2 \right)
\end{equation*}
We can calculate the generalized momenta from the definition:
\begin{equation*}
    p_1 \equiv \fr{\p L}{\p \Dot{q}_1} = \Dot{q}_1;\quad
    p_2 \equiv \fr{\p L}{\p \Dot{q}_2} = C q_1; \quad 
    p_3 \equiv \fr{\p L}{\p \Dot{q}_3} = 0
\end{equation*}
The Hessian matrix is  $W_{ij} = \big\rvert \big\rvert \fr{\p^2 L}{\p \Dot{q}_i  \p \Dot{q}_j} \big\rvert\big\rvert = \big\rvert \big\rvert\fr{\p p_i}{\p \Dot{q}_j} \big\rvert\big\rvert = \mathrm{diag}[1,0,0] $, so it is obvious that it's rank is 1. Two primary constraints in this system are $\phi_1(p,q)=p_2 - Cq_1 =0$ and $\phi_2(p,q) = p_3 =0$.
\flushright
\qedsymbol
\end{Primer}

\subsection*{Strong and Weak Equalities}

The equations of primary constraints $\phi_m (p,q)=0$ define hypersurface $\Gamma$. This surface is the background on which the evolution of a physical system occurs. However, it turns out that in order to fully describe the system it is necessary to take into consideration not only $\Gamma$ but also a small neighborhood around it $\Gamma_\varepsilon$ \footnote{$\Gamma_\varepsilon$ includes $\Gamma$}. 
\begin{figure}[h]
\centering
\includegraphics[width=9cm]{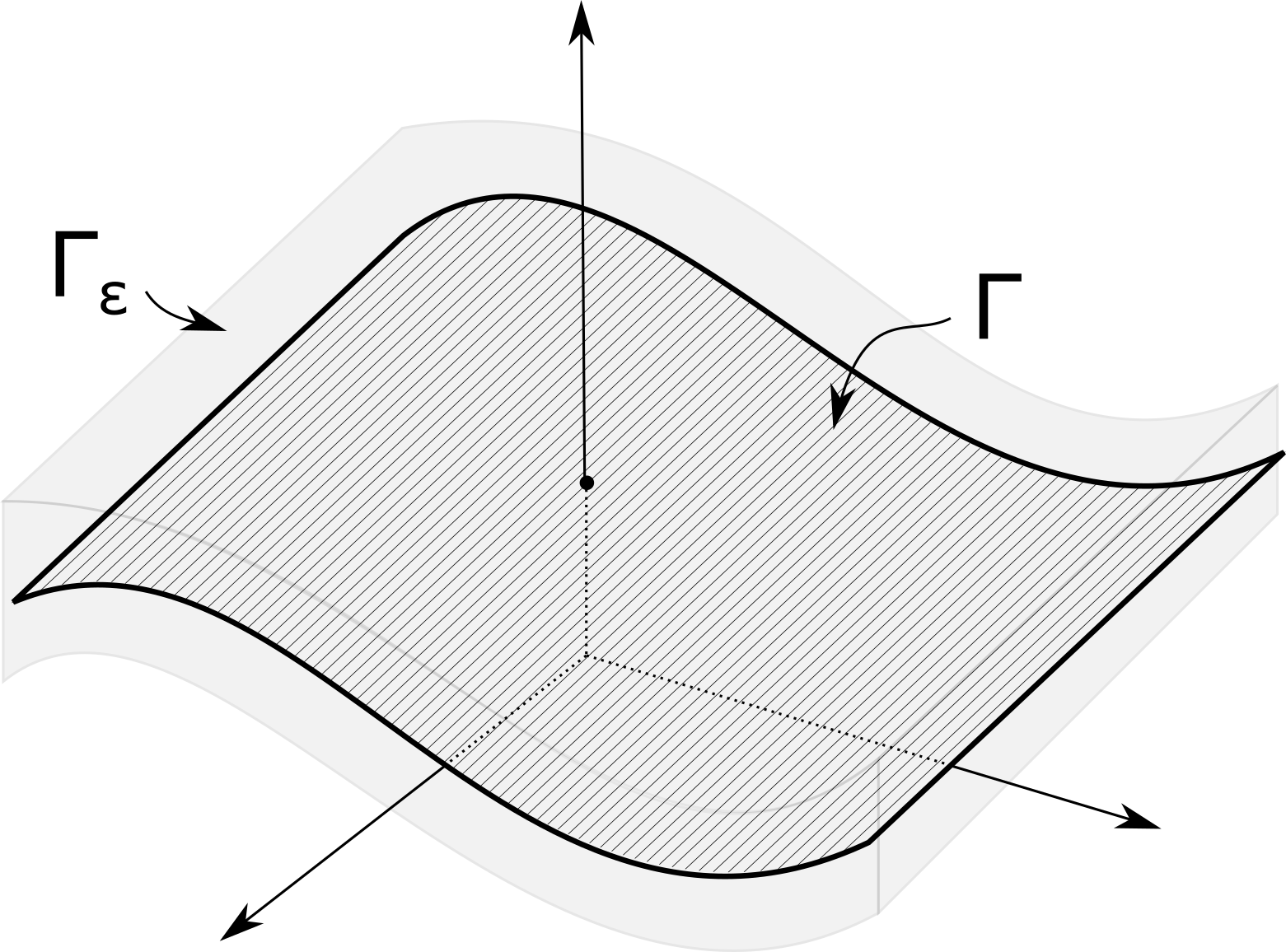}
\caption{Hypersurface $\Gamma$ and its small neighborhood  $\Gamma_\varepsilon$}
\label{hip}
\end{figure}
It is thus very important to specify if equality is valid on $\Gamma$ or $\Gamma_\varepsilon$. That is the reason why we define the notions of strong and weak equalities:
\vspace{1cm}
{
\begin{Definicija} $\quad$\normalfont
\begin{itemize}
    \item $f$ and $g$ are \textit{weakly equal} $(f \approx g)$ if the equality holds on $\Gamma$: $f \big\rvert_\Gamma=g \big\rvert_\Gamma$.
    \item  $f$ and $g$ are \textit{strongly equal} $(f =g)$ if the equality holds on $\Gamma_\varepsilon$:$\;f \big\rvert_{\Gamma_\varepsilon}=g \big\rvert_{\Gamma_\varepsilon}$.\newline
    In other words ${f \approx g  \;\text{and}\; df\approx dg }$.
\end{itemize}
\end{Definicija}
}
\noindent The phase space outside of $\Gamma_\varepsilon$ is practically irrelevant. Let us now state two important theorems:

\begin{Teorema} \label{Teorema slabo jednako} \normalfont
If $F \approx 0$, then we can always find functions $f^m$ such that $F = f^m \phi_m$.
\end{Teorema}

\begin{Dokaz}
\normalfont
Since the constraints are independent, we can choose them as coordinates in the submanifold $\Gamma_\varepsilon$. There are only $M$ of these coordinates, so we need to define additional $2N-M$ coordinates, which we will denote as $y$. Thus, we defined the coordinates $x=\left(\phi, y \right)$. We know that the hypersurface $\Gamma$ is defined by $\phi = 0$, so we can conclude that  $F(\phi=0,y) = 0$. Using that equation we obtain:

\begin{equation*}
    F(\phi,y) = \int_0^1 \fr{d}{dt} F(t \phi,y) dt = \phi_m \underbrace{ \int_0^1 \p_m F(t\phi, y)}_{f^m} dt
\end{equation*}
\flushright
\qedsymbol
\end{Dokaz}

\begin{Teorema} \label{teorema varijacija} \normalfont
   If $\alpha^i \delta q_i + \beta^i \delta p_i = 0$ for arbitrary variations tangent to $\Gamma$, then 
    \begin{equation}
        \alpha^i = u^m \fr{\p \phi_m}{\p q_i};\quad 
        \beta^i = u^m \fr{\p \phi_m}{\p p_i}
    \end{equation}
\end{Teorema}

\begin{Dokaz}
\normalfont
By assumption, vector $\Vec{A} = \alpha^i \Vec{e}_{q_i} + \beta^{i} \Vec{e}_{p_i}$ is directed along the tangent to $\Gamma$. That's why vectors $\Vec{e}_{q_i}$ and $\Vec{e}_{p_i}$ are not independent, so it would be wrong to conclude that $\alpha^i = \beta^i =0$. But, we can interpret our problem differently: we can let vectors $\Vec{e}_{q_i}$ and $\Vec{e}_{p_i}$ to be independent, but then we need to subtract component orthogonal to $\Gamma$:

\begin{equation*}
    0=\alpha^i \Vec{e}_{q_i} + \beta^{i} \Vec{e}_{p_i} - u^m \nabla \phi_m = \left( \alpha^i - u^m \fr{\p \phi_m}{\p q_i} \right) \Vec{e}_{q_i} + \left( \beta^{i} - u^m \fr{\p \phi_m}{\p p_i} \right)\Vec{e}_{p_i}
\end{equation*}

Since $\Vec{e}_{q_i}$ and $\Vec{e}_{p_i}$ are now independent,
\begin{equation*}
        \alpha^i = u^m \fr{\p \phi_m}{\p q_i};\quad 
        \beta^i = u^m \fr{\p \phi_m}{\p p_i}
    \end{equation*}
\flushright
\qedsymbol    
\end{Dokaz}

\subsection*{Construction of the Canonical Hamiltonian}

Canonical Hamiltonian is defined as:
\begin{equation*}
    H_c= p_n \Dot{q}_n - L
\end{equation*}
On hypersurface $\Gamma$ canonical Hamiltonian depends only on $p,q$, which can be seen from:
\begin{eqnarray}
    \delta H_c &=& p_n \delta \Dot{q}_n + \Dot{q}_n \delta p_n - \fr{\p L}{\p q_n} \delta q_n - \fr{\p L}{\p \Dot{q}_n} \delta \Dot{q}_n \NB \label{varijacijaH}
   &=& \Dot{q}_n \delta p_n - \fr{\p L}{\p q_n} \delta q_n \label{varijacijaH2}
\end{eqnarray}
where we used~\eqref{genImpulsi}. Since the primary constraints are derived from~\eqref{genImpulsi} we see that the same argument doesn't hold outside of $\Gamma$. That means that the variations $\delta q_n, \delta p_n$ are directed along the tangent of $\Gamma$.
\noindent
From \eqref{varijacijaH}:
\begin{equation}
    \left(\fr{\p H_c}{\p p_n} - \Dot{q}_n \right) \delta p_n +
    \left( \fr{\p H_c}{\p q_n} + \fr{\p L}{\p {q}_n}\right)\delta {q}_n =0 
\end{equation}
Using theorem~\ref{teorema varijacija} we get: 
\begin{subequations}
\begin{eqnarray}
\Dot{q}_n &\approx& \fr{\p H_c}{\p p_n} + u_m \fr{\p \phi_m}{\p p_n} \\
-\fr{\p L}{\p q_n} &\approx& \fr{\p H_c}{\p q_n} + u_m \fr{\p \phi_m}{\p q_n} \label{druga hamiltonova jna}
\end{eqnarray}
\end{subequations}
Lagrange multipliers $u_m$ are used to replace generalized velocities that we can't express in terms of momenta. The number of momenta and coordinates is $2N-M$, while the number of Lagrange multipliers is $u_m$. That is why we define $q,p,u$ to be dynamical variables of our theory. They are called \textit{Hamilton's variables}. \par\null\par

Using~\eqref{druga hamiltonova jna}, as well as Euler-Lagrange equations: 
\begin{equation}
    \Dot{p}_n \approx -\fr{\p H}{\p q_n} - u_m \fr{\p \phi_m}{\p q_n}
\end{equation}
It is important to notice that if  $g(p,q)$ is a function of the canonical variables, then its evolution $\Dot{g} = \fr{\p g}{\p q_n}\Dot{q}_n + \fr{\p g}{\p p_n}\Dot{p}_n$ can be written in terms of the Poisson bracket:
\begin{equation} \label{evolucija g}
    \Dot{g} = \{ g, H_c \} + u_m \{ g, \phi_m \}
\end{equation}
We note that we must not use any constraints before working out the Poisson bracket. In order to make our formulas more concise, sometimes it is useful to write $\{ g,  u_m \phi_m \}\approx u_m\{ g,   \phi_m \}$ because the term $\phi_m\{ g,u_m\}$, despite not being well-defined, is zero (it is multiplied by the constraint outside of the Poisson bracket). As a result of that, we can write the evolution equation in a more familiar form:
\begin{equation}
    \Dot{g} \approx \{ g,H_c+  u_m \phi_m \} \equiv \{ g, H_T \},
\end{equation}
where we defined \textit{total Hamiltonian} as  $H_T=H_c+  u_m \phi_m $.

\section{Consistency Conditions and Secondary Constraints}
\label{sekundarne}

Primary constraints have to be zero throughout all time, but the equation~\eqref{evolucija g} predicts that they should evolve. That is why we need to impose the  \textit{consistency conditions}:
\begin{equation}
    \{ \phi_m, H_c \} + u_n \{\phi_m, \phi_n \} \approx 0 
\end{equation}
Solving them can lead to:
\begin{itemize}
    \item inconsistencies (for example $1=0$). If this happens, Lagrangian itself has been chosen poorly. Such Lagrangians imply inconsistent Euler-Lagrange equations (for example $L=q$). We will assume that our Lagrangian does not lead to inconsistencies.
    \item identities (for example $0=0$). In this case, the consistency conditions are automatically satisfied and thus, they are not implying any further restrictions to the system.
    \item equations that are independent of Lagrange multipliers $u_m$. Such equations depend only on $p$ and $q$, so they are of the form  $\chi(p,q)=0$. They are completely analogues to the primary constraints, which is the reason we call them \textit{secondary constraints}.
    \item equations that depend on Lagrange multipliers $u_m$. Such equations impose certain restrictions on $u_m$.
\end{itemize}
The procedure is now repeated iteratively - we impose the consistency conditions on secondary constraints \footnote{All additional constraints are also called secondary constraints.}... The procedure ends when no additional constraints are being found. The primary and secondary constraints should be treated on the same footing. It is thus useful to introduce a new notation for all constraints $\varphi_k,\quad k=1...M+S$, where $S$ is the number of secondary constraints. All of these constraints now define a new hypersurface in the phase space which represents a submanifold of $\Gamma$. It is thus necessary to redefine the notions of a weak and strong equation with respect to this new hypersurface\footnote{The old hypersurface is now irrelevant, so from now on, symbol $\Gamma$ will always be used to denote this new hypersurface}. \par\null\par
 
\begin{Napomena}
\normalfont
Even though we treat primary and secondary constraints equally, we will still define total Hamiltonian using only primary constraints. It can be shown that this formalism is completely equivalent to Lagrange's formalism~\cite{Mukunda}. On the other hand, if we used both primary and secondary constraints in the definition of total Hamiltonian, then the equations of motion would be different from Euler-Lagrange equations, but the difference would be unphysical. 
\end{Napomena}
 
 Let's now examine equations which impose restrictions on $u_m$ (we ignore the ones that lead to secondary constraints and identities):
 
 \begin{equation} \label{uslovi konzistentnosti}
     \{ \varphi_k, H_c \} + u_m \{ \varphi_k, \phi_m \} = 0,
 \end{equation}
These equations have to have particular solution  $u_m = U_m (p,q)$. Otherwise, equations would be inconsistent which is in contradiction with our assumptions. General solution is of the form $u_m = U_m + v_a V_{am}, \quad a=1...A$, where $V_{am}$ are independent solutions of homogeneous equations:
 \begin{equation} \label{kernel}
     V_{am} \{ \varphi_k, \phi_m \} = 0 
 \end{equation}
We have started with $M$ independent variables $u_m$ and have now successfully replaced them with a fewer number of independent variables $v_a$. In terms of these new variables, the total Hamiltonian can be written as:
 \begin{equation} \label{redefinisanje H}
    H_T = H' + v_a \phi_a, \quad H' = H_c +U_m \phi_m,\quad \phi_a = V_{am}\phi_m
 \end{equation}

\section{A New Classification of the Constraints}\label{novaklasifikacija}

The mere existence of constraints hints that the number of \textit{real physical} degrees of freedom is probably smaller than $2N$ in the phase space. That means that some (but not all!) of the constraints are responsible for the existence of nonphysical degrees of freedom, that can be eliminated by fixing the gauge. In order to determine which constraints imply the existence of gauge symmetry, it is useful to introduce the new classification of constraints\footnote{The old classification is useless because primary and secondary constraints are always treated equally, except in the definition of the total Hamiltonian.}.\par\null\par 

It is important to notice that in order to solve~\eqref{kernel} it is sufficient to find the basis of the kernel of the matrix $C_{km} = \{ \varphi_k, \phi_m \}$. It is now obvious that $\phi_a = V_{am} \varphi_m$  has the vanishing Poisson bracket with each constraint:
\begin{equation}
    \{ \varphi_k, \phi_a \} = 0
\end{equation}
Further, since $V_{am}$ represents a basis\footnote{Index $a$ runs over basis vectors, whereas $m$ runs over components of the vector}, that means that vectors $\phi_a$ constitute a complete set of primary constraints that have the vanishing Poisson bracket with each constraint. We can use that as a motivation for the following definition: 
\begin{Definicija} \normalfont
A dynamical variable $A(p,q)$ is said to be \textit{fist-class} if its Poisson bracket with each constraint is weakly equal to zero:
\begin{equation}
    \{ A, \varphi_k \} \approx 0
\end{equation}
Otherwise, $A(p,q)$ is \textit{second-class}.
\end{Definicija}
\noindent

\begin{Teorema} \label{PZ velicina prve klase} \normalfont
   The Poisson bracket of two first-class quantities is also first-class.
\end{Teorema}
\begin{Dokaz}
\normalfont
Let $A$ and $B$ be two first-class quantities. Then, using Jacobi's identity, we easily obtain:
\begin{eqnarray*}
\{ \{ A,B \}, \varphi_k\}
        &=& 
            \{ \{ A,\varphi_k \} , B\} - \{ \{ B,\varphi_k \},A \} \\*
        &=& \{ c_{km} \varphi_m, B\} - \{ d_{km}\varphi_m,A \} \\*
        &\approx& c_{km} \{ \varphi_m,B \} - d_{km}\{ \varphi_m,A \} \approx 0
\end{eqnarray*}
\flushright
\qedsymbol
\end{Dokaz}    
\subsection*{Properties}
\begin{itemize}
    \item From theorem~\ref{Teorema slabo jednako} we see that if $A$ is first-class, then $\{ A, \varphi_k \} = c_{km}\varphi_m$.
    \item $H'$ and $\phi_a$ from~\eqref{redefinisanje H}  are first-class, so that the Poisson  brackets between the total Hamiltonian and first-class constraints are vanishing. In other words, the consistency condition for first-class constraints is trivial $0=0$. 
    \end{itemize}

From now on, we will use a new notation, in which $\gamma_a$ stands for first-class constraints (whether they are primary or not), while $\chi_\alpha$ stands for second-class constraint.

\begin{enumerate}
    \item $\{ \chi_\alpha, \chi_\beta \}$ is nonsingular. Otherwise, we could find a vector $p^\beta$, such that $0 \approx \{ \chi_\alpha, \chi_\beta \} p^\beta \approx \{ \chi_\alpha, \chi_\beta p^\beta \}$. Then, the Poisson bracket of the constraint  $p^\beta \chi_\beta$ with every other constraint would be vanishing. That means that we could construct an additional first-class constraint, which is impossible. Inverse matrix of $\{ \chi_\alpha, \chi_\beta \}$ will be denoted as $\{ \chi_\alpha, \chi_\beta \}^{-1}$.
    \item The number of second-class constraints is even. That can be seen from the fact that $\{ \chi_\alpha, \chi_\beta \}$ is nonsingular, antisymmetric matrix because for every antisymmetric matrix $M, M^T = -M$:
    \begin{equation*}
        \mathrm{det} M = det(-M^T) = det(-M) = (-1)^{dim(M)} detM
    \end{equation*}
\end{enumerate}

Let us now examine the consistency conditions for second-class constraints\footnote{We already noted that the consistency conditions for first-class constraints are automatically fulfilled $0=0$}. 
\begin{equation} \label{pomocna jednacina 123}
    \{ \chi_\alpha, H_c \} + \{ \chi_\alpha, \chi_\beta \} u^\beta \approx 0,
\end{equation}
It is important to emphasize that multipliers $u_\beta$ are defined only when the corresponding constraint $\chi_\beta$ is primary. In other words, $u_\beta=0$ if $\chi_\beta$ is secondary constraint. Now, from \eqref{pomocna jednacina 123} we get:
\begin{equation} \label{uslovi konzistentnosti za sekundarne veze}
    \{ \chi_\beta, \chi_\alpha \}^{-1}\{ \chi_\alpha, H_c \} =   \begin{cases*}
     -u^{\beta} & \text{if $\chi_\beta$ is a primary constraint}  \\
      0      & \text{if $\chi_\beta$ is a secondary constraint} 
    \end{cases*}
\end{equation}
We come to a very important conclusion:\par\null\par

\textit{The number of arbitrary multipliers is equal to the number of primary first-class (PFC) constraints. Thus, they are very closely related to gauge symmetries. On the other hand, primary second-class constraints have predetermined multipliers. In other words, it is like their gauge is fixed by the consistency condition.}\par\null\par

We can now rewrite the equation for total Hamiltonian, using \eqref{uslovi konzistentnosti za sekundarne veze}:
\begin{equation}
    \Dot{g} \approx \{ g, H_c \} + \{ g,\phi_a \} v_a  - \{ g, \chi_\alpha \} \{ \chi_\alpha, \chi_\beta \}^{-1} \{\chi_\beta, H_c\}
\end{equation}
where $\phi_a$ are primary first-class constraints. Since the multipliers, that correspond to primary second-class constraints are predetermined, they cannot represent dynamical variables and are thus physically irrelevant. We can eliminate them from the theory, in a very convenient way, by defining the so-called \textit{Dirac bracket}:
\begin{Definicija} \normalfont
Dirac bracket between quantities $A$ and $B$ is defined as:
\begin{equation}
    \{ A,B \} ^* = \{ A,B \} - \{ A, \chi_\alpha \}  \{ \chi_\alpha, \chi_\beta \}^{-1} \{\chi_\beta,B \}
\end{equation}
\end{Definicija} 
\noindent
It is not necessary to define the Dirac bracket, but it is a useful notion that we wish to implement. Dirac bracket enables us to write the equations more neatly because it absorbs cumbersome terms that we usually have. For example, the evolution of any canonical variable $g$ can now be written as:
\begin{equation}
    \Dot{g} \approx \{ g, H_T\}^*
\end{equation}
which is easy to see if we notice that the Dirac bracket between any second-class constraint and arbitrary variable vanishes.

\begin{Napomena} $\quad$
 \normalfont
\begin{enumerate}
   \item We have already seen that first-class constraints have arbitrary multipliers. We can use this fact to determine the first or second class nature of constraints, without calculating Poisson brackets.
    \item The number of physical degrees of freedom is:
    \begin{equation} \label{Broj stepeni slobode}
        \text{the number of physical degrees of freedom} = 2N - 2N_1 - N_2,
    \end{equation}
    where $N_1$ and $N_2$ are the numbers of first-class and second-class constraints respectively. We can easily see that this formula holds - from the initial $2N$ degrees of freedom we need to subtract all of the constraints  ($N_1+N_2$) and then we also need to subtract the number of arbitrary multipliers ($N_1$) because we are free to fix a gauge (for any of them).
\end{enumerate}
\end{Napomena}

\section{Castellani's Algorithm}
\label{Kastelani}

In \S~\ref{novaklasifikacija} we saw that primary first-class constraints are very closely related to gauge symmetries, but that does not mean they themselves are generators. Luckily, Leonardo Castellani\cite{Kastelani} found a straightforward way to find all generators of gauge symmetries. In this section, we will give a short review of his work. \par\null\par

\begin{figure}[h]
\centering
\includegraphics[width=12cm]{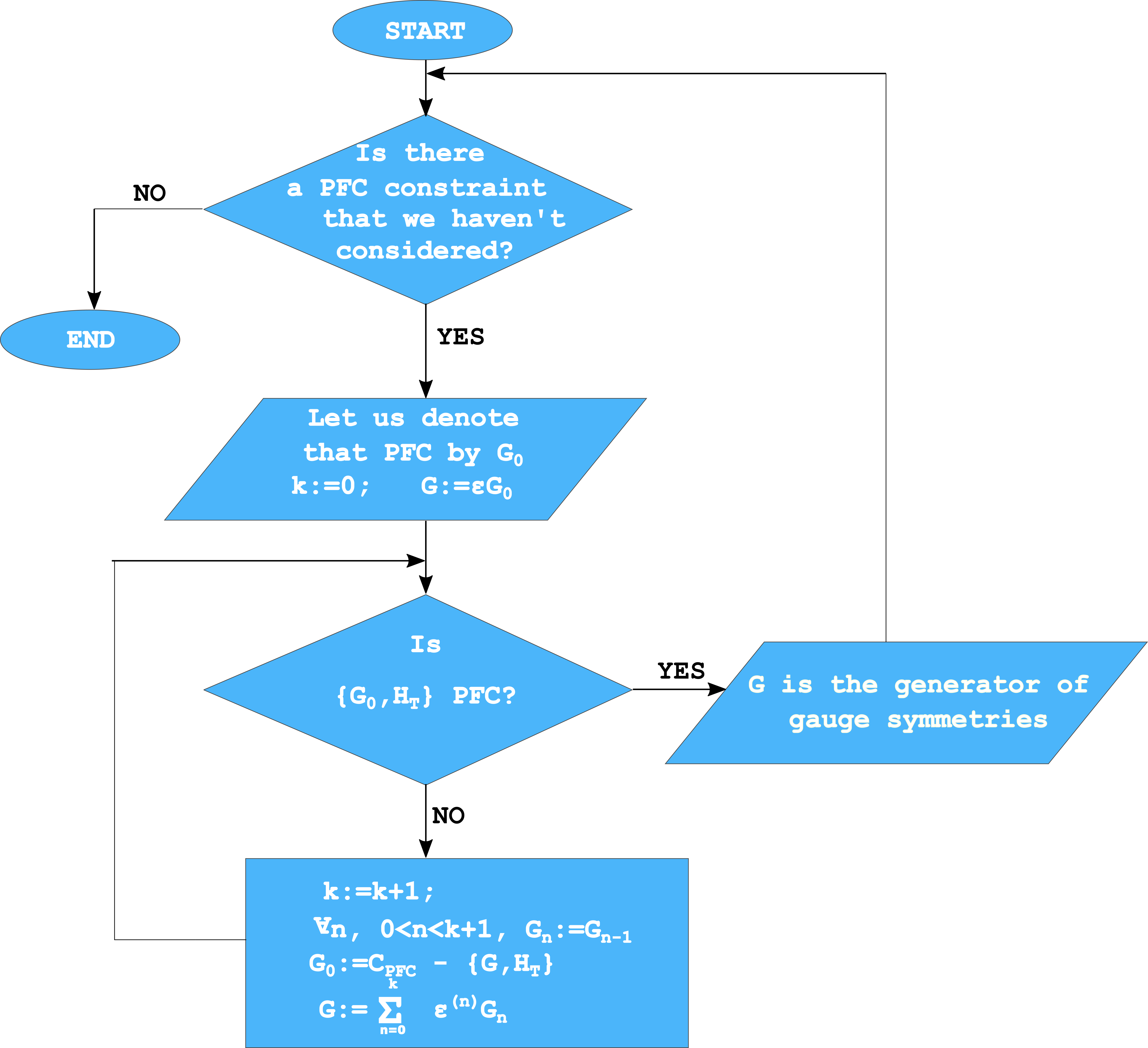}
\caption{Castellani's algorithm}
\label{dijagram}
\end{figure}

We know that a generator preserves constraints:
\begin{equation} \label{prva bitna}
    \{ \phi_k , G \} \approx 0
\end{equation}
That means that it is a first-class quantity. To determine it exactly, we will assume that the generator takes the form $G= \varepsilon G_0 + \Dot{\varepsilon}G_1 + ... + \varepsilon^{(k)}G_k$, where $\varepsilon$ is an arbitrary function of time\footnote{In field theory, $\varepsilon$ will be an arbitrary function of the space-time coordinates, but it will not depend on dynamical fields}. If $G$ really generates gauge symmetries, that means that if  $q(t), p(t)$ is a solution of Hamilton's equations $\Dot{q} \approx \{q, H_T \}\;$(similarly for momenta as well), then $q(t) + \delta q(t), p(t) +\delta p(t)$ is a solution as well, where  $\delta q= \{ q,G \}$,  $\delta p= \{ p,G \}$. It is important to note that multipliers  $v_a$   in  \eqref{redefinisanje H} also transform (into $v_a + \delta v_a$). That is why the transformation of the total Hamiltonian has two terms - the first term accounts for the transformation of fields   $\delta H_T = \{ H_T, G \}$, while the second term accounts for the transformation of multipliers $\phi_a \delta v_a$.

\begin{eqnarray} 
\fr{d}{dt}(q+\delta q) 
        &\approx& \{ q+\delta q, H_T+ \delta H_T + \phi_a \delta v_a \} \\
\delta \Dot{q} 
        &\approx& \{ \delta q ,H_T \} + \{q,\delta H_T \} + \{q, \phi_a \delta v_a \} \NB
        &=& \{ \{ q,G\},H_T \} + \{ q,\{ H_T,G\} \} + \{ q,\phi_a\delta v_a \} \NB
        &=& \left[ \varepsilon \{ \{ q,G_0 \}, H_T \} + ...+ \varepsilon^{(k)}\{\{q,G_k \},H_T \} \right] + \NB
        &&  \left[  \varepsilon\{q,\{H_T,G_0\}\}+ ... +\varepsilon^{(k)} \{q,\{H_T,G_k \} \} \right] + \NB
        && \{q,\phi_a \delta v_a \} \label{Kastelani leva strana}
\end{eqnarray}
On the other hand $\delta \Dot{q} $ can also be written as:
\begin{eqnarray}
\delta \Dot{q} 
        &=& \fr{d}{dt} \delta q = \fr{d}{dt} \{q,G \} \NB
        &=& \fr{d}{dt} \{q, \varepsilon G_0 + ... + \varepsilon^{(k)}G_k \}\NB
        &=& \left[\varepsilon \{ \{ q,G_0 \}, H_T\} +...+\varepsilon^{(k)} \{ \{q,G_k\},H_T \} \right] + \NB
         &&   \left[ \Dot{\varepsilon} \{q,G_0 \} + ... + \varepsilon^{(k+1)}\{q,G_k\} \right] \label{Kastelani desna strana}
\end{eqnarray}
If we now compare~\eqref{Kastelani leva strana} and \eqref{Kastelani desna strana} we get  $\{q,A \} \approx 0$, where $A$ is determined by:
\begin{multline} \label{def A}
A=\varepsilon^{(k+1)} G_k + \varepsilon^{(k)} \left[ G_{k-1} + \{ G_k,H_T \} \} \right] + ... + \Dot{\varepsilon} \left[ G_0 + \{G_1, H_T \} \right] + \\
+\varepsilon\{G_0,H_T\} - \phi_a \delta v_a
\end{multline}
Similarly, using Hamilton's equation for momenta we obtain  $\{p,A \} \approx 0$. We conclude that if $F(p,q)$ is only a function of the canonical variables $p$ and $q$, then: $\{F,A\} = \fr{\p F}{\p q} \{q,A \} + \fr{\p F}{\p p} \{ p, A \} \approx 0$. In other words $F(p,q)$ has a vanishing Poisson bracket with $A$:
\begin{equation} \label{druga bitna}
    \{F,A\} \approx 0
\end{equation}
That means that $A=0$ (up to a constant or some power of a constraint). The last term in $A$ is a primary first-class constraint. We conclude that arbitrary functions in \eqref{def A} are multiplied by primary first-class constraints. In other words:
\begin{equation} \label{Kastelanijev algoritam}
\begin{array}{r@{}l}
    G_k &{}= C_{\text{PFC}}\\
    G_{k-1} + \{ G_k, H_T\} &{}= C_\text{PFC}\\
    &\vdots \\
    G_{0} + \{G_1, H_T \} &{}= C_\text{PFC}\\
    \{ G_0, H_T \} &{}=C_\text{PFC}
\end{array}
\end{equation}
At last, we can finally formulate Castellani's algorithm (picture~\ref{dijagram})

The procedure is as follows: we start with arbitrary primary first-class constraint $G_k$. Then, we calculate $G_{k-1}$ using~\eqref{Kastelanijev algoritam}. Using the same formula we can continue calculating $G_{k-2}$... The procedure ends when the Poisson bracket between the last calculated term  $G_0$ and total Hamiltonian is vanishing.

\begin{Napomena}$\quad$ \normalfont
\begin{enumerate}
    \item Formula~\eqref{Kastelanijev algoritam} tells us that every $G_i$ is determined only up to a primary first-class constraint. That property can be used used to our advantage - by adding convenient terms to $G_i$, Castellani's procedure can be shortened significantly.
    \item In a specific problem, we do not know in advance how many steps will be needed for the procedure to end. In other words, we do not know in advance what $k$ is. But, once the procedure ends, we can just count how many steps did it take (that's what $k$ is). That is why algorithm~\ref{dijagram} always labels newly calculated terms as $G_0$. If it turns out that its Poisson bracket with the total Hamiltonian is not $C_{PFC}$ then we just relabel  $G_n:= G_{n-1}$. In this paper, $k$ will be either 1 or 0.
\end{enumerate}
\end{Napomena}

\section{Electrodynamics}
\label{Elektrodinamika}

Let us consider the application of the canonical formalism and Castellani's algorithm using the example of electrodynamics in the external gravitational field. In \S$\;$\ref{Elektrodinamika u spolj. grav. polju} we already constructed canonical Hamiltonian~\eqref{ADMHamiltonijanED}. Besides, in~\eqref{KanonskiImpulsED}  we found the primary constraint  $\pi^0 = 0$, hence the total Hamiltonian is:
\begin{equation}
    H_T =\int d^3 x \left[  N \mc{H}_{\perp} + N^\alpha \mc{H}_\alpha - A_0 \p_\alpha \pi^\alpha  \right] + \int d^3 x u(x) \pi^0 (x),
\end{equation}{}
where $\mathcal{H}_\alpha = \pi^\beta F_{\alpha \beta}, \quad
\mathcal{H}_\perp = -\frac{1}{2J} \pi^{\bar{m}} \pi_{\bar{m}}+\frac{J}{4}F_{\bar{m}\bar{n}}F^{\bar{m}\bar{n}}$. \newline
One could now use total Hamiltonian to find the equations of motion. However, we will not do that because our main task is to classify the constraints and construct the generator of gauge symmetries. \par\null\par

One can find secondary constraints from the consistency conditions. That is very easy to do from the ADM form of the Hamiltonian:
\begin{equation}
    0 \approx \{ \pi^0, H_T \} = \p_\alpha \pi^\alpha
\end{equation}{}
This is the secondary constraint. It is easy to show that the consistency condition for the secondary constraint is automatically fulfilled, thus $\pi^0$ and $\p_\alpha \pi^\alpha$ are the only constraints in our theory. Both of them are first-class:
\begin{equation}
    \{ \pi^0, \p_\alpha \pi^\alpha \} = 0
\end{equation}{}
It remains to construct the generator of gauge symmetries. Since $\pi^0$ is the only primary first-class constraint\footnote{It will turn out that $k=1$} $G_1 = \pi^0$. The next step in Castellani's algorithm is to calculate $G_0$:
\begin{equation}
    G_0 = -\{ G_1 ,H_T \} = - \p_\alpha \pi^\alpha
\end{equation}{}
It can be easily shown that $G_0$ has a vanishing Poisson bracket with the total Hamiltonian (up to a primary first-class constraint), so Castellani's algorithm ends. The generator of gauge symmetries is:
\begin{equation}
    G = \int d^3 x \left( \Dot{\varepsilon}G_1 + \varepsilon G_0 \right) = \int d^3 x \left( \Dot{\varepsilon} \pi^0 - \varepsilon \p_\alpha \pi^\alpha \right) 
\end{equation}{}
We can now determine the gauge transformations  of the canonical variables $A^\mu, \pi^\mu$:
\begin{subequations}
\begin{eqnarray}
\delta_0 A^\mu &=& \{A^\mu, G \} = \p^\mu \varepsilon \\
\delta_0 \pi^\mu &=& \{ \pi^\mu, G \} = 0
\end{eqnarray}
\end{subequations}

One more interesting thing we can do is to calculate the number of degrees of freedom using the equation \eqref{Broj stepeni slobode}. From the $2N=8$ degrees of freedom we need to subtract\footnote{There are no secondary constraints, while the number of primary constraints is $N_1=2$} $2N_1 = 4$, so we obtain 4 degrees of freedom in the phase space. The number of degrees of freedom in configurational space is 4/2=2.\par\null\par

This was a very useful example because it illustrates the same type of the canonical analysis that we will use in the case of TG and TEGR:
\begin{itemize}
    \item Primary constraints are obtained by solving the system of equations $p_i  = \fr{\p L}{\p \Dot{q}_i}$. We can then construct the total Hamiltonian.
    \item From the consistency conditions $\Dot{\phi}_m = \{ \phi_m, H_T \} = 0$ we either get secondary constraints or restrictions on Lagrange multipliers.
   \item We impose consistency conditions on secondary constraints iteratively, until no more secondary conditions arise.
    \item By calculating the Poisson brackets between the constraints, we can classify first and second-class constraints.
    \item We can (but don't have to) construct the Dirac bracket in order to eliminate determined multipliers that correspond to primary second-class constraints.
    \item Using previously obtained results we calculate the number of degrees of freedom and construct generators of gauge symmetries.
    \item Finally, using the generator, we obtain gauge transformations of canonical variables.
\end{itemize}

Several different things can be done in the canonical formalism ( for example calculation of conserved charges), but in this paper, we will limit ourselves only to the procedure described in the list above.

\chapter{Canonical Analysis of TG}
\label{genericki slucaj teleparalelne teorije}

In this chapter, we will use the canonical formalism to analyze TG\cite{BlagojevicNikolic}, which is defined by the Lagrangian\footnote{Actually the Lagrangian is $\Tilde{\mc{L}}= b \mc{L}_T$}~\eqref{Lagranzijan teleparalelne gr}. The Lagrangian~\eqref{Lagranzijan teleparalelne gr} depends on $A, B, C$,  so it is not unexpected that there exist constraints that are present only for some specific values of $A, B, C$. These are known as \textit{if-constraints}. On the other hand, there are also \textit{sure constraints}, which are present no matter what the values of $A, B, C$ are. We define TG to be a theory with the Lagrangian~\eqref{Lagranzijan teleparalelne gr}, where $A, B, C$ are chosen such that we only have sure constraints. It turns out that those are not too restrictive conditions. In the following section, we will examine TEGR which in addition to sure constraints also has if constraints. \par\null\par

As a part of the canonical analysis, our task will be to find the canonical Hamiltonian and Poisson brackets between the constraints. The constraint algebra will be used to classify the constraints and calculate the number of degrees of freedom. However, the generator of gauge symmetries will be constructed in chapter \ref{glava konstrukcija generatora}.

\section{Construction of the Canonical Hamiltonian}
In \S~\ref{Teleparalelna} we found that the TG Lagrangian is:
\begin{equation} \label{Lagranzijan Teleparalelne1}
\mc{L}_T = A T_{ijk} T^{ijk} + BT_{ijk}T^{jik} + CT^k T_k 
\end{equation}
\subsection*{Calculating the Momenta}
The definition of momenta are given by:
\begin{equation} 
     \pi_i{}^{\mu}:=b\frac{\partial \mathcal{L}_{T}}{\partial \partial_0 b^i{}_{\mu}}=b\frac{\partial \mathcal{L}_{T}}{\partial T^i{}_{0\mu}}
\end{equation}
Actually, it turns out that it is useful to define slightly more general quantities, called \textit{covariant} momenta\footnote{We adopted the convention in which $\pi_i{}^\mu$ are called momenta, whereas $H_m{}^{\mu\nu}$ are called covariant momenta.}: $H_m{}^{\mu\nu} := \frac{\partial \Tilde{ \mathcal{L}}}{\partial T^m{} _{\mu \nu}}=\fr{\p \Tilde{\mc{L}}}{\p T^{mnp}} h^{n\mu} h^{p\nu} = H_{mnp}h^{n\mu} h^{p\nu} $. It is now straightforward to calculate:

\begin{eqnarray}
H_{mnp} 
        &=& 
            2ab \left[ AT_{ijk} \frac{\partial T^{ijk}}{\partial T^{mnp}}+ BT_{ijk} \frac{\partial T^{jik}}{\partial T^{mnp}}+ CT_k \frac{\partial T^k}{\partial T^{mnp}} \right] \NB
        &=&
            2ab \left[  AT_{ijk}  \delta^i_m (\delta^j_n \delta^k_p - \delta^j_p \delta^k_n)
            + BT_{ijk} \delta^j_m (\delta^i_n \delta^k_p -\delta^i_p \delta^k_n)\right] + \NB
        &&    
           2ab CT_k \left( \eta_{mn}\delta^k_p-\eta_{mp}\delta^k_n \right)
             \NB
        &=& 4ab(AT_{mnp}+B_{[nmp]}+C\eta_{m[n}T_{p]}) \label{kovarijantni impulsi}
\end{eqnarray}
Notice that the Lagrangian \eqref{Lagranzijan Teleparalelne1} can now be written as:
\begin{equation} \label{Lagranzijan Teleparalelne}
    \mc{L}_T = \fr{1}{4b} H_{ijk}T^{ijk}
\end{equation}
and momenta are easily obtained from \eqref{kovarijantni impulsi}:
\begin{equation}
    \pi_i{}^{\mu} = H_i{}^{mn}h_m{}^{0}h_n{}^{\mu}= \fr{1}{N}H_i{}^{\perp \mu} 
\end{equation}
Sure primary constraints are  $\pi_i{}^0 = 0$ because $H_i{}^{\perp 0} \sim H_i{}^{\perp\perp} =0$. They take the same form as in the case of electrodynamics. We now introduce new quantities called \textit{parallel momenta}\footnote{Parallel momenta are labeled with a hat to distinguish them from the quantity\newline ${\pi_i{}^\B{k} = \pi_i{}^k - n^k \pi_i{}^\perp}$, but they still have the property that  $\Hat{\pi}_i{}^\B{k} n_k = 0$.\newline That's why there is a bar over~$k$.}
\begin{equation}
    \Hat{\pi}_i{}^\B{k}:= \pi_i{}^{\alpha} b^k{}_{\alpha} = \fr{1}{N} H_i{}^{\perp n}h_n{}^{\alpha} b^k{}_\alpha = \fr{1}{N}  H_i{}^{\perp \B{k}}
\end{equation}
They account for the existence of sure constraints. Parallel momenta are expressed in terms of the covariant momenta $H_{i\perp \B{k}}$, which depend on torsion linearly. That means that they can be written as a sum of the term that depends on velocities $H_{i\perp \B{k}}(1)$ and a term that does not depend on velocities $H_{i\perp \B{k}}(0)$ :
\begin{equation}
    H_{i\perp \B{k}}(T) = H_{i\perp \B{k}}(0) + H_{i\perp \B{k}}(1)
\end{equation}
We can use that fact as a motivation to introduce new quantities, the \textit{generalized momenta}\footnote{Again, we note that we introduce this new terminology for convenience even though it is not widely used.}  $P_{i\B{k}}$, as parallel momenta without the term $H(0)$:
\begin{equation} \label{Def generalisanih impulsa}
    P_{i\B{k}} := \fr{1}{b}\left( N \hat{\pi}_{i\B{k}} - H_{i\perp \B{k}}(0)\right) =  \fr{1}{b} H_{i\perp \B{k}}(1)
\end{equation}
We could have defined generalized momenta without the factor $\fr{1}{b}$, but it turns out that this definition makes other equations look simpler.\par\null\par 

\noindent
Using \eqref{kovarijantni impulsi} we can express $H(0), H(1)$ and $ P_{i\B{k}}$ explicitly:
\begin{eqnarray}
\fr{1}{4b}H_{i\perp\Bar{k}} (T)
        &=&
            {aAT_{i\perp\Bar{k}}}+\frac{aB}{2}T_{\perp i\Bar{k}}-{\frac{aB}{2}T_{\Bar{k}i\perp}}+\frac{aC}{2}n^j\eta_{ij}T_{\Bar{k}}-{\frac{aC}{2}n^j\eta_{i\Bar{k}}T_j } \NB
        &=& {aAT_{i\perp\Bar{k}}} + \frac{aB}{2} \left( T_{\perp \BB{i}{k}} + n_i T_{\perp\perp \B{k}} \right) - \frac{aB}{2}T_{\BB{k}{i}\perp}+ \NB
        &&+ \frac{aC}{2} \left( n_i T^{\Bar{m}}{}_{\Bar{m} \Bar{k}}+{n_i T_{\perp \perp \Bar{k}}} \right) - \frac{aC}{2} \eta_\BB{i}{k}T^\B{m}{}_{\B{m}\perp}
\end{eqnarray}
Terms with a perpendicular second or third index depend on velocities, which can be seen from~\eqref{ADM grupa I a}. On the other hand, if the second and third indices have bars over them, then the corresponding term cannot depend on velocities because of $T^i{}_\BB{m}{n} = h_\B{m}{}^\alpha h_\B{n}{}^\beta T^i{}_{\alpha \beta}$. We conclude that:

\begin{subequations}
\begin{eqnarray} \label{betanula}
\fr{1}{4b}H_{i\perp \Bar{k}}(0) &=& \frac{aB}{2}T_{\perp\Bar{i}\Bar{k}}+\frac{aC}{2}n_i T^{\Bar{m}}{}_{\Bar{m}\Bar{k}} \\ \label{betajedan}
\fr{1}{4b}H_{i\perp \Bar{k}}(1) &=& aAT_{i \perp \Bar{k}}+\frac{aB}{2} T_{\Bar{k}\perp \Bar{i}}+\frac{aC}{2}\eta_{\Bar{i}\Bar{k}}T^{\Bar{m}}{}_{\perp \Bar{m}}+\frac{a}{2}(B+C)n_i T_{\perp\perp \Bar{k}} \\ \label{GeneralisaniImpuls}
P_{i \Bar{k}} &=& 4a\left[AT_{i\perp \Bar{k}}+\frac{B}{2}T_{\Bar{k}\perp \Bar{i}}+\frac{C}{2}\eta_{\Bar{i}\Bar{k}}T^{\Bar{m}}{}_{\perp \Bar{m}}+\frac{1}{2}(B+C)n_iT_{\perp\perp\Bar{k}}\right] \label{Gen brzine}
\end{eqnarray}
\end{subequations}
We note that generalized momenta provide only a technical convenience that will be helpful when considering irreducible decomposition. It was not necessary to introduce them. Also, the fact that $H(0)$ does not depend on velocities, while $H(1)$ does, is not a property that will be used in the future. In that sense, we could have just defined them as \eqref{betanula} and \eqref{betajedan}, but we chose not to, because it would look very unnatural.

\subsection*{Euler-Lagrange Equations}
In field theory, the Euler-Lagrange equations are given by:
\begin{equation}
-\p_\nu \left( \fr{\p\left( b \mc{L_T}\right)}{\p \left( \p_\nu b^m{}_\mu \right)} \right) + \fr{\p \left( b \mc{L_T}\right)}{\p b^m{}_\mu} = 0
\end{equation}
Using the definition of covariant momenta, it is easy to calculate the first term:
\begin{equation*}
    -\p_\nu \left( \fr{\p\left( b \mc{L_T}\right)}{\p \left( \p_\nu b^m{}_\mu \right)} \right)  =- \p_\nu H_m{}^{\nu\mu}
\end{equation*}
The second term can be calculated using \eqref{Grupa III c} and the chain rule: $\fr{\p \left( b \mc{L_T}\right)}{\p b^m{}_\mu} = h_m{}^\mu b \mc{L_T} + b \fr{\p \mc{L}_T}{\p b^m{}_\mu}$. In order to calculate $\fr{\p \mc{L}_T}{\p b^m{}_\mu}$ it is useful to rewrite the Lagrangian as:
\begin{equation*}
    \mc{L}_T =aA T_{i\rho\sigma} T^{i}{}_{\kappa\sigma} g^{\kappa\rho} g^{\tau\sigma} +aBT_{i\rho\sigma}T^j{}_{\kappa\tau} g^{\tau\sigma} h^{i\kappa}h_j{}^\rho + aC T^p{}_{\rho\sigma} T^n{}_{\kappa\tau} h_p{}^\rho h_n{}^\kappa g^{\tau\sigma}
\end{equation*}
and then use  \eqref{Grupa III a}:
\begin{eqnarray*}
 b \fr{\p \mc{L}_T}{\p b^m{}_\mu} &=&-4aAb T_i{}^{\mu\tau} T^i{}_{m\tau} - 2aBb T_{ijm}T^{ji\mu} - 2aBb T_{\mu ji}T^{jmi} \\
 && - 2aCb T^\mu{}_{mk}T^k - 2aCb T_m T^\mu \\
 &=& - H^{ij\mu}T_{ijm}
\end{eqnarray*}
Finally, we conclude that the E-L equations are:
\begin{equation} \label{jednacine kretanja TTG}
- \p_\nu H_m{}^{\nu\mu} - H^{ij\mu}T_{ijm} + h_m{}^\mu b \mc{L}_T = 0   
\end{equation}

\subsection*{Construction of the Hamiltonian}
As in the case of electrodynamics, we use primary constraints $\pi_i{}^0 =0$ to construct the canonical Hamiltonian. 
\begin{eqnarray} \label{RacunKH}
    \mathcal{H}_c
        &=&
            \pi_i{}^{\alpha}\partial_0 b^i{}_{\alpha}-b\mathcal{L}_{T} 
        = \pi_i{}^{ \alpha}T^i{}_{0\alpha}+\pi_i{}^{ \alpha}\partial_\alpha b^i{}_{0} -b\mathcal{L}_{T}\NB
        &=& \pi_i{}^\alpha \left( NT^i{}_{\perp\alpha}+N^\beta T^i{}_{\beta\alpha} \right) + \p_\alpha \left( \pi_i{}^\alpha b^i{}_0 \right) - b^i{}_0 \p_\alpha \pi_i{}^\alpha -b\mathcal{L}_{T}
\end{eqnarray}
In the last step, we used~\eqref{ADM grupa I a}. This is not the end result, so we need to transform it further. In the first term, we use $   \pi_i{}^{\alpha}T^i{}_{ \perp \alpha}=\hat{\pi}_i{}^{\Bar{m}}T^i{}_{ \perp \Bar{m}}$ (which is valid because of \eqref{ADM grupa I c}), whereas in the second-to-last term we use the identity  $b^i{}_0 = N n^i + N^\beta b^i{}_\beta $ (which can be obtained from~\eqref{uvodjenje N i N alfa}):

\begin{eqnarray}
    \mc{H}_c 
            &=&
                N \hat{\pi}_i{}^\B{m} T^i{}_{\perp \B{m}} + N^\beta \pi_i{}^\alpha T^i{}_{\beta\alpha} + \p_\alpha \left( \pi_i{}^\alpha b^i{}_0 \right)- \NB
            &&    \left(N n^i + N^\beta b^i{}_\beta \right) \p_\alpha         \pi_i{}^\alpha - b\mc{L}_T \NB
            &=& 
                N\mathcal{H}_{\perp}+N^\beta \mathcal{H}_\beta +\partial_\alpha D^\alpha
\end{eqnarray}
where:
\begin{subequations}
\begin{eqnarray} \label{H ortogonalno}
\mathcal{H}_\perp &:=& \hat{\pi}_i{}^{\Bar{m}}T^i{}_{\perp\Bar{m}}-n^i\partial_\alpha \pi_i{}^{\alpha}-J\mathcal{L}_{T}\\ \label{ H beta}
\mathcal{H}_\beta &:=& \pi_i{}^{\alpha}T^i{}_{\beta\alpha}-b^i{}_{\beta}\partial_\alpha \pi^\alpha{}_{ i}\\
D^\alpha &:=& \pi_i{}^{\alpha} b^i{}_{ 0}
\end{eqnarray}
\end{subequations}
The total Hamiltonian is given by:
\begin{equation} \label{TOTALNI hamiltonijan genericki slucaj}
    \mc{H}_T = \mc{H}_c + \pi_k{}^0 u^k
\end{equation}

As in the case of electrodynamics, we successfully represented canonical Hamiltonian in the Dirac-ADM form. However, we still need to eliminate the velocities because orthogonal Hamiltonian ~\eqref{H ortogonalno} still has velocity dependence in the first and last term. We thus make some additional transformations on those terms using $T^{ijk}=T^{i \Bar{j} \Bar{k}}+2T^{i [\Bar{j} \perp}n^{k]}$:

\begin{eqnarray}
\hat{\pi}_i{}^{\Bar{m}}T^i{}_{\perp\Bar{m}}-J\mathcal{L}_{T}
        &=&
            \hat{\pi}_i{}^{\Bar{m}}T^i{}_{\perp\Bar{m}}-\fr{J}{4b}H_{ijk}T^{ijk} \NB
        &=&
           \hat{\pi}_i{}^{\Bar{m}}T^i{}_{\perp\Bar{m}}-\fr{1}{4N}H_{ijk}(T^{i \Bar{j} \Bar{k}}+2T^{i [\Bar{j} \perp}n^{k]})\NB \label{RacumHort}
        &=&
            \hat{\pi}_i{}^{\Bar{m}}T^i{}_{\perp\Bar{m}}-\fr{1}{2N}H_{i \Bar{j}\perp}T^{i \Bar{j}\perp}-\fr{1}{4N}H_{ijk}T^{i\Bar{j}\Bar{k}} \NB
        &=&
            \frac{1}{2}\hat{\pi}^{i \Bar{m}}T_{i\perp \Bar{m}}-\fr{1}{4N}H_{i\Bar{j}\Bar{k}}T^{i\Bar{j}\Bar{k}} \label{pom rel 2}
\end{eqnarray}
Let us analyze the last term:
\begin{equation}
    \fr{1}{4b}H_{i\Bar{j}\Bar{k}}T^{i \Bar{j} \Bar{k}}=aAT_{i\Bar{j}\Bar{k}}T^{i \Bar{j} \Bar{k}}+aB\underbrace{T_{\Bar{j}i\Bar{k}}T^{i \Bar{j} \Bar{k}}}_{II}+aC\underbrace{\eta_{i \Bar{j}}T_{\Bar{k}}T^{i\Bar{j}\Bar{k}}}_{III}
\end{equation}
Using $T_{\Bar{j}i \Bar{k}}=T_{\Bar{j}\Bar{i}\Bar{k}}+n_i T_{\Bar{j}\perp \Bar{k}}$ we get $II=T_{\Bar{j}\Bar{i}\Bar{k}}T^{\Bar{i} \Bar{j} \Bar{k}} +T_{\Bar{j}\perp \Bar{k}}T^{\perp \Bar{j}\Bar{k}}$.\newline
Using $T^{\Bar{m}}{}_{\Bar{m}\Bar{k}}=T^m{}_{m\Bar{k}}-T_{\perp\perp\Bar{k}}$ we get $III=T_{\perp\perp\Bar{k}}T_{\Bar{j}}{}^{\Bar{j}\Bar{k}}+T^{\Bar{m}}{}_{\Bar{m}\Bar{k}}T_{\Bar{j}}{}^{\Bar{j}\Bar{k}}$.
Combining those results:
\begin{equation*} \label{DefLT}
   \fr{1}{4ab} H_{i\Bar{j}\Bar{k}}T^{i \Bar{j} \Bar{k}}=\underbrace{AT_{i\Bar{j}\Bar{k}}T^{i\Bar{j}\Bar{k}}+BT_{\Bar{j}\Bar{i}\Bar{k}}T^{\Bar{i}\Bar{j}\Bar{k}}+CT^{\Bar{m}}{}_{\Bar{m}\Bar{k}}T_{\Bar{j}}{}^{\Bar{j}\Bar{k}}}_{\frac{1}{a}\mathcal{L}_{T}(\overline{T})}+\underbrace{BT_{\Bar{j}\perp\Bar{k}}T^{\perp \Bar{j}\Bar{k}}+CT_{\perp\perp\Bar{k}}T_{\Bar{j}}{}^{\Bar{j}\Bar{k}}}_{\fr{1}{4ab}H_{ijk}(1)T^{i\Bar{j}\Bar{k}}}
\end{equation*}
The first term is independent of velocities, whereas the second one depends them, so it is useful to rewrite it as:
\begin{equation} \label{pom rel 1}
    \fr{1}{4b}H_{ijk}(1)T^{i\Bar{j}\Bar{k}}=aBT_{\perp\Bar{i}\Bar{k}}T^{\Bar{i}\perp\Bar{k}}+aCT^{\perp\perp\Bar{k}}T^{\Bar{m}}{}_{\Bar{m}\Bar{k}}=a(BT_{\perp\Bar{i}\Bar{k}}+Cn_i T^{\Bar{m}}{}_{\Bar{m}\Bar{k}})T^{i\perp\Bar{k}}
\end{equation}
Using~\eqref{pom rel 1} and \eqref{pom rel 2}, the orthogonal part of Hamiltonian~\eqref{H ortogonalno} can be written as: 
\begin{eqnarray}
\mathcal{H}_\perp
        &=&
            -n^i \partial_\alpha \pi_i{}^{\alpha} -J\mathcal{L}_{T}(\overline{T}) \NB
        && +\frac{1}{2}T_{i\perp \Bar{m}}\left[\hat{\pi}^{i\Bar{m}}-aJ(2BT^{\perp\Bar{i}\Bar{m}}+2Cn^i T_{\Bar{n}}{}^{\Bar{n}\Bar{m}})\right] \NB
    &=& \frac{J}{2} P^{i \Bar{m}}T_{i\perp\Bar{m}}-n^i \partial_\alpha \pi_i{}^{\alpha}-J\mathcal{L}_{T}(\overline{T}) \label{H ort rez}
\end{eqnarray}
Now, only the first term depends on the velocities. In order to eliminate them, we need to solve equation~\eqref{Gen brzine}. That can be done using the irreducible decomposition of $P^{i\B{m}}$.
\subsection*{Irreducible Decomposition}
Let us express $P_{i\B{k}}$ as a sum of a term in the hypersurface $x^0 = const.$ and a term that is perpendicular to it: $P_{i\B{k}} = P_\BB{i}{k} + n_i P_{\perp \B{k}}$. Then, we use the well-known decomposition of $P_\BB{i}{k}$ in a three-dimensional space:
\begin{equation}
    P_\BB{i}{k} = P_\BB{i}{k}^A + P_\BB{i}{k}^T + \fr{1}{3} \eta_\BB{i}{k}P
\end{equation}
where $P^A_\BB{i}{k}, P^T_\BB{i}{k}, P$ are antisymmetric, symmetric traceless and trace parts of the $P_\BB{i}{k}$, respectively:
\begin{equation}
    P^A_\BB{i}{k} = P_{[\BB{i}{k}]}, \quad P= P^\B{m} {}_\B{m},\quad P^T_\BB{i}{k} = P_{(\BB{i}{k})}-\fr{1}{3} \eta_\BB{i}{k} P
\end{equation}
This represents the irreducible decomposition of the $P_{i\B{k}}$. Irreducible parts are easily calculated using  $n_\B{i} = 0, T_{\perp \B{m}}{}^\B{m}=0$ as well as relations \eqref{betanula} - \eqref{Gen brzine}: 
\begin{subequations}\label{ireducibilna dekompozicija jna}
\begin{eqnarray}
    P_{\perp \Bar{k}} &=& \frac{1}{J}\hat{\pi}_{\perp \Bar{k}}-2CT^{\Bar{m}}{}_{\Bar{m}\Bar{k}}=2a\left(2A+B+C\right)T_{\perp\perp\Bar{k}} \label{ired 1}\\
    P^A_{\Bar{i}\Bar{k}} &=& \frac{1}{J}\hat{\pi}^A_{\Bar{i}\Bar{k}}-2BT_{\perp \Bar{i}\Bar{k}}=2a\left(2A-B\right)T^A_{\Bar{i}\perp\Bar{k}} \label{ired 2}\\
    P^T_{\Bar{i}\Bar{k}} &=& \frac{1}{J}\hat{\pi}^T_{\Bar{i}\Bar{k}}=2a\left(2A+B\right)T^T_{\Bar{i}\perp\Bar{k}}\label{ired 3}\\
    P^{\Bar{m}}{}_{\Bar{m}}&=&\frac{1}{J}\hat{\pi}^{\Bar{m}}{}_{\Bar{m}}=2a\left(2A+B+3C\right)T^{\Bar{m}}{}_{\perp\Bar{m}} \label{ired 4}
\end{eqnarray}
\end{subequations}
None of the expressions on the right-hand side can be zero. Otherwise,  there would exist additional constraints, which is impossible because we defined TG to be a theory with only sure constraints.\par\null\par 

If we notice that the irreducible components are orthogonal, it immediately follows that the first term in~\eqref{H ort rez} can be written as:
\begin{eqnarray}
P^{i\Bar{m}}T_{i\perp \Bar{m}}
        &=&
            P^{\perp \Bar{m}}T_{\perp\perp \Bar{m}}+P_A^{\Bar{i}\Bar{m}}T^A_{\Bar{i}\perp\Bar{m}}+P_T^{\Bar{i}\Bar{m}}T^T_{\Bar{i}\perp \Bar{m}}+\frac{1}{3}P^{\Bar{m}}{}_{\Bar{m}}T^{\Bar{n}}{}_{\perp \Bar{n}} \NB
        &=& 
            \fr{1}{2a(2A+B+C)} P^{\perp \B{m}}P_{\perp \B{m}}+\fr{1}{2a(2A-B)}P_A^{\BB{i}{m}}P^A_{\BB{i}{m}}+ \NB
        &&  
            \fr{1}{2a(2A+B)} P_T^{\BB{i}{m}}P^T_{\BB{i}{m}} + \fr{1}{6a(2A+B+3C)}P^\B{m}{}_\B{m} P^{\B{n}}{}_{\B{n}} \label{prvi red ired dekom}
\end{eqnarray}
Finally, we obtain the orthogonal part of the hamiltonian~\eqref{H ort rez}: 
\begin{eqnarray}
\mathcal{H}_\perp
        &=&    -n^i \partial_\alpha \pi_i{}^{\alpha}-J\mathcal{L}_{T}(\overline{T}) +
            \fr{J}{2}\left[ \fr{1}{2a(2A+B)} P_T^{\BB{i}{m}}P^T_{\BB{i}{m}} +\fr{1}{2a(2A-B)}P_A^{\BB{i}{m}}P^A_{\BB{i}{m}}\right]+ \NB
        &&
           \fr{J}{2}\left[ \fr{1}{6a(2A+B+3C)}P^\B{m}{}_\B{m} P^{\B{n}}{}_{\B{n}}  + \fr{1}{2a(2A+B+C)} P^{\perp \B{m}}P_{\perp \B{m}}\right] \label{H ortogonalno konacno}
\end{eqnarray}
It is now straightforward to express it in terms of the momenta using \eqref{ireducibilna dekompozicija jna}. However, we will not do that because the expression will just become more cumbersome and complicated. In the following section, we will see that~\eqref{H ortogonalno} is the most useful form of $\mathcal{H}_\perp$ for calculating the Poisson brackets, where the velocities are interpreted as functions of momenta.

\subsection*{Consistency Conditions}

In~\S\ref{sekundarne} we saw that it is now necessary to impose the consistency conditions on the primary constraints $\pi_i{}^0 = 0$. Since the primary constraints $\pi_i{}^0 = 0$ have vanishing Poisson brackets between themselves the consistency conditions reduce to $\{ \pi_{}^0,\mc{H}_c \} = 0$. The only variable that gives a nonvanishing Poisson bracket with $\pi_i{}^0 = 0$ is $b^k{}_0$, so it is important to examine the dependence of ADM variables on $b^k{}_0$:

\begin{Teorema} \label{TeoremaZavisnosti1} \normalfont
 $\quad$
 \begin{itemize}
     \item $n^k, h_{\Bar{k}}{}^{ \mu}, \mathcal{H}_\alpha,\mathcal{H}_\perp,J=\frac{b}{N}$ do not depend on $b^k{}_{ 0}$
     \item $N,N^\alpha,b$ linearly depend on $b^k{}_{0}$.
 \end{itemize}
 \end{Teorema}
 \noindent
 Proof of this theorem can be found in appendix \ref{DodatakB}. Now, using this, as well as lemma~\ref{ N i N alfa} we can easily impose the following consistency conditions:
 
 \begin{eqnarray} \label{komutatorPrimarne}
 \{ \pi_i{}^{0}, \mathcal{H}_c \} &=& \{\pi_i{}^{0}, N  \mathcal{H}_\perp+N^\beta \mathcal{H}_\beta \} \NB
 &=& \{\pi_i{}^{ 0},N \} \mathcal{H}_\perp 
 +\{ \pi_i{}^{0},N^\beta\} \mathcal{H}_\beta \NB
 &=& \{ \pi_i{}^{ 0},b^k{}_{0 } \}n_k \mathcal{H}_\perp + \{ \pi_i{}^{ 0}, h_{\Bar{k}}{}^{\beta}b^k{}_{ 0} \}\mathcal{H}_\beta \NB \label{uslov pv}
 &=&- n_i \mathcal{H}_\perp -h_{\Bar{i}}{}^{\beta}\mathcal{H}_\beta \approx 0
 \end{eqnarray}
By multiplying each side of equation \eqref{uslov pv} by $n^i$, one obtains
 \begin{equation}
     \mathcal{H}_\perp \approx 0
 \end{equation}
 On the other hand, multiplying the same equation with $b^\B{i}{}_\alpha$ gives:
 \begin{eqnarray}
 0 \approx h_{\Bar{l}}{}^{ \beta} b^l{}_{\alpha} \mathcal{H}_\beta 
    =
        \left(h_l{}^\beta -n_l h_\perp{}^\beta \right) b^l{}_{\alpha}\mathcal{H}_\beta  = \delta^\beta_\alpha \mc{H}_\beta = \mc{H}_\alpha
 \end{eqnarray}
 Thus, we conclude that the secondary constraints are given by:
 \begin{equation}
     \mathcal{H}_\perp \approx 0;\quad \mathcal{H}_\alpha \approx 0
 \end{equation}
 
This proves that the Hamiltonian is, in fact, a constraint up to a total divergence term. Thus, by calculating the Poisson brackets\cite{NikolicPuason} between the constraints, we automatically examine the consistency conditions for the secondary constraints. Theorem~\ref{TeoremaZavisnosti1} tells us that the Poisson brackets between $\pi_i{}^0$ and any other constraint are vanishing, so the next few sections will be devoted to computing the Poisson brackets $\{ \mathcal{H}_\alpha (x), \mathcal{H}_\beta (x') \}$, $\{\mc{H}_\alpha (x), \mc{H}_\perp(x')\}$, as well as $\{\mc{H}_\perp(x), \mc{H}_\perp(x') \}$.

\section{Poisson Bracket \texorpdfstring{$\{ \mathcal{H}_\alpha (x), \mathcal{H}_\beta (x') \}$}{Ha(x),Hb(x')}}

It turns out that in our specific examples, rather than using the definition of the Poisson bracket, it is much more useful to use the following properties:

\begin{enumerate}
    \item $\{ b^i{}_\mu (x), {\pi_j{}^\nu} (x') \} = \delta^i_j \delta^\nu_\mu \delta(x-x')$ 
    \item Chain rules states that if $f$ and $g$ depend on $\xi_a$ and $\chi_b$ respectively, then  $\{f, g \} = \fr{\p f}{\p \xi_a} \{\xi_a, \chi_b \} \fr{\p g}{\p \chi_b}$
\end{enumerate}
For simplicity, in the following chapters we will adopt abbreviated notation:
\begin{equation*}
\delta (x-x')\equiv \delta,\quad \pi_i{}^\mu (x') \equiv {\pi'}_i{}^\mu    
\end{equation*}
We are now ready to start calculating. From:
\begin{subequations}
\begin{eqnarray}
    \mathcal{H}_\alpha &=&\pi_i {}^{\gamma}\partial_\alpha b^i{}_{ \gamma}-\partial_\gamma \left( \pi_i{}^{\gamma}b^i{}_{ \alpha}  \right) \\
    \mathcal{H'}_\beta &=&\left({\pi}_j {}^{\epsilon} \partial^{}_\beta {b}^j{}_{\epsilon}\right)' -\partial^{'}_\epsilon \left( {\pi}_j{}^{\epsilon}{b}^j{}_{ \beta} \right)'
\end{eqnarray}
\end{subequations}
we get:

\begin{eqnarray}
\{ \mc{H}_\alpha, \mc{H'}_\beta \}
        &=&
            \{ \pi_i {}^{\gamma}\partial_\alpha b^i{}_{ \gamma},
            \left({\pi}_j {}^{\epsilon} \partial^{}_\beta {b}^j{}_{\epsilon}\right)'\} -
            \partial^{'}_\epsilon \{\pi_i {}^{\gamma}\partial_\alpha b^i{}_{ \gamma}, \left( {\pi}_j{}^{\epsilon}{b}^j{}_{ \beta} \right)' \} \NB
        &&
            - \p_\gamma \{  \pi_i{}^{\gamma}b^i{}_{ \alpha}, \left({\pi}_j {}^{\epsilon} \partial^{}_\beta {b}^j{}_{\epsilon}\right)'\}
            +\p_\gamma \partial^{'}_\epsilon \{ \pi_i{}^{\gamma}b^i{}_{ \alpha},\left( {\pi}_j{}^{\epsilon}{b}^j{}_{ \beta} \right)' \}\NB
        &=& 
            \pi_i{}^\gamma \left( \p_\beta b^j{}_\epsilon \right)' \{ \p_\alpha b^i{}_\gamma, {\pi'}_j{}^\epsilon \} + \left(\p_\alpha b^i{}_\gamma\right){\pi'}_j{}^\epsilon \{\pi_i{}^\gamma, \left( \p_\beta b^j{}_\epsilon \right)' \} \NB
        && -\p^{'}_\epsilon 
            \left[ 
                \pi_i{}^\gamma {b '}^j{}_\beta \{ \p_\alpha b^i{}_\gamma, {\pi '}_j{}^\epsilon  \} + \left( \p_\alpha b^i{}_\gamma  \right) {\pi'}_j{}^\epsilon \{\pi_i{}^\gamma, {b '}^j{}_\beta \} 
            \right] \NB
        &&
            -\p_\gamma\left[
            \pi_i{}^\gamma \left( \p_\beta b^j{}_\epsilon \right)' \{ b^i{}_\alpha, {\pi'}_j{}^\epsilon \} + b^i{}_\alpha {\pi'}_j{}^\epsilon \{ \pi_i{}^\gamma, \left(\p_\beta b^j{}_\epsilon \right)' \}
            \right] \NB
        && 
           + \p_\gamma \p^{'}_\epsilon \left[
            \pi_i{}^\gamma {b'}^j{}_\beta \{ b^i{}_\alpha, {\pi'}_j{}^\epsilon \} + b^i{}_\alpha {\pi'}_j{}^\epsilon \{ \pi_i{}^\gamma, {b'}^j{}_\beta \}
            \right] \NB
        &=& 
            \pi_j{}^\gamma \left( \p_\beta b^j{}_\gamma \right)' \p_\alpha \delta- \p^{'}_\gamma \left[\pi_j{}^\gamma {b'}^j{}_\beta \p_\alpha \delta - (\p_\alpha b^j{}_\beta)\pi_j{}^\gamma \delta \right] \NB
        &&
            +\p_\gamma \p^{'}_\alpha \left[ \pi_j{}^\gamma b^j{}_\beta \delta \right] - \left(\alpha \leftrightarrow \beta, x \leftrightarrow x' \right)\label{Prva Poason 1}
\end{eqnarray}
The symbol $\left(\alpha \leftrightarrow \beta, x \leftrightarrow x' \right)$ denotes the term that is obtained from the rest of the expression by interchanging $\alpha$ and $\beta$, as well as $x$ and $x'$.\par\null\par

It is now important to determine if it is possible to rewrite~\eqref{Prva Poason 1} as a linear combination of primary first-class constraints. That is how we check if the constraints are first-class or not. 
Terms of the form $f(x) g(x) \p_\mu \delta$ are much easier to deal with than terms of the form $f(x) g(x') \p^{'}_\mu \delta$, which are common in \eqref{Prva Poason 1}. The following theorem can thus be very useful:

\begin{Teorema}{} \label{Teorema sa deltom}
\begin{equation}
    f(x)g(x') \p^{'}_\mu \delta =-f(x) \p_\mu \left[g(x) \delta \right] = -\delta(x-x') f(x) \p_\mu g(x) - f(x)g(x) \p_\mu \delta 
\end{equation}{}
\end{Teorema}

\begin{Dokaz}{}
\begin{eqnarray*}
f(x)g(x') \p^{'}_\mu \delta 
        &=&
            -f(x)g(x') \p_\mu \delta = -f(x) \p_\mu \left[g(x') \delta \right] =-f(x) \p_\mu \left[g(x) \delta \right]
\end{eqnarray*}{}
\normalfont
\flushright
\qedsymbol
\end{Dokaz}
Using this theorem, the first term in~\eqref{Prva Poason 1} becomes:
\begin{eqnarray}
\pi_j{}^\gamma \left( \p_\beta b^j{}_\gamma \right)' \p_\alpha \delta = \pi_j{}^\gamma \left( \p_\beta b^j{}_\gamma \right) \p_\alpha \delta +
\cancelto{0}{\pi_j{}^\gamma \left(\p_\alpha \p_\beta b^j{}_\gamma \right) \delta} \label{Prva Poason 2}
\end{eqnarray}{}
The last term in the previous equation is not really equal to zero, but it does not give a contribution to the end result because of the  $\left(\alpha \leftrightarrow \beta, x \leftrightarrow x' \right)$ term. Let us now examine the remaining terms in~\eqref{Prva Poason 1}: 

\begin{eqnarray}
&&- \p^{'}_\gamma \left[\pi_j{}^\gamma {b'}^j{}_\beta \p_\alpha \delta - (\p_\alpha b^j{}_\beta)\pi_j{}^\gamma \delta \right] 
            +\p_\gamma \p^{'}_\alpha \left[ \pi_j{}^\gamma b^j{}_\beta \delta \right] \NB
&=& - \p^{'}_\gamma \left[ \pi_j{}^\gamma {b}^j{}_\beta \p_\alpha \delta + \pi_j{}^\gamma (\p_\alpha b^j{}_\beta) \delta  - (\p_\alpha b^j{}_\beta)\pi_j{}^\gamma \delta \right] - \p_\gamma\left[ \pi_j{}^\gamma b^j{}_\beta \p^{}_\alpha \delta \right] \NB
&=& - \p_\gamma \left( \pi_j{}^\gamma b^j{}_\beta \right) \p_\alpha \delta\label{Prva Poason 3}
\end{eqnarray}
Using ~\eqref{Prva Poason 2} and \eqref{Prva Poason 3} in \eqref{Prva Poason 1} we get: 
\begin{eqnarray}
\{ \mc{H}_\alpha, \mc{H'}_\beta \}
        &=& \left[ \pi_j{}^\gamma\p_\beta b^j{}_\gamma   - \p_\gamma \left( \pi_j{}^\gamma b^j{}_\beta \right)  \right]\p_\alpha \delta - \left(\alpha \leftrightarrow \beta, x \leftrightarrow x' \right)
\end{eqnarray}
Finally, the end result is:
\begin{equation}
    \{ \mc{H}_\alpha, \mc{H'}_\beta \} =
    \left(\mc{H}_\beta \p_\alpha + \mc{H'}_\alpha \p_\beta  \right) \delta(x-x')
\end{equation}

\section{Poisson Bracket \texorpdfstring{$\{ \mathcal{H}_\alpha , \mathcal{H'}_\perp\}$}{Ha,H'}}

This Poisson bracket is much more complicated than the last one. If we tried to calculate it straightforwardly, there would be too much work to do because of~\eqref{H ortogonalno konacno}. That is why we will use a different approach. At the beginning, we will introduce new quantities which will help us to organize our calculations. Some of them will be helpful only in the following sections, but we believe that it is simpler to define all of them just at one place.

\subsection*{New Quantities and Relations Among Them}

\begin{Definicija}
$\quad$
\normalfont
\begin{itemize}
    \item $T^\mu{}_{\nu} :=\frac{\partial \mathcal{L}}{\partial T^k{}_{\mu \gamma }}T^k{}_{\nu \gamma }-\delta^\mu _\nu \mathcal{L_T} \qquad$ Notice that we sum over $1, 2, 3$. If we summed over $0, 1 , 2, 3$, then $T^\mu{}_{\nu}$ would be a covariant generalization of the canonical energy-momentum tensor.
    \item $\Delta^\mu {}_{\nu }:=-b^k{}_{\nu }\partial_\gamma H_k{}^{\mu \gamma }\quad$ Again, we emphasize that we sum over $1, 2, 3$. It is in this way that we account for the existence of the primary constraints $\pi_i{}^0$.
\end{itemize}
\end{Definicija}
\noindent
Let us rewrite $T^\mu{}_{\nu}$ in a more convenient form:
\begin{eqnarray}
 T^{\mu} {}_{\nu } &=& \frac{1}{b} H_k{}^{ \mu \gamma }T^k{}_{\nu \gamma }-\delta^\mu _\nu  \mathcal{L_T}= \frac{1}{b} H_k{}^{ \mu \gamma } \partial_\nu b^k{}_{\gamma }-\frac{1}{b} H_k{}^{ \mu \gamma } \partial_\gamma b^k{}_{\nu }-\delta^\mu _\nu  \mathcal{L_T}\NB
 &=& \frac{1}{b} H_k{}^{ \mu \gamma } \partial_\nu b^k{}_{\gamma }-\frac{1}{b} \partial_\gamma \left( H_k{}^{\mu \gamma} b^k{}_{\nu} \right)-\frac{1}{b}\Delta^\mu{}_{\nu}-\delta^\mu_{\nu}\mathcal{L_T}
\end{eqnarray}
Introducing another new variable $\tau$, the last expression can be written as:
\begin{equation} \label{TAU}
    bT^\mu{}_{\nu}+\Delta^\mu{}_{\nu}=H_k{}^{\mu\gamma} \partial_\nu b^k{}_{\gamma}-\partial_\gamma\left( H_k{}^{\mu \gamma} b^k{}_{\nu} \right)-\delta^\mu_\nu b\mathcal{L_T}=:\tau^\mu{}_{\nu}
\end{equation}
We use $\tau$ only to state and prove the following theorem:
\begin{Teorema} \label{Osnovna teorema}
\normalfont
$\quad$
\begin{enumerate}
    \item $\tau^0{}_{ \alpha }=\mathcal{H}_\alpha =JT_{\perp \alpha }+\Delta^0{}_{\alpha }$
    \item $\tau^0{}_{\perp}=\mathcal{H}_\perp=J T_{\perp \perp}+\Delta^0{}_{\perp}$
    \item $T^k{}_{\perp \Bar{l}}=J\frac{\partial T_{\perp \perp}}{\partial\hat{\pi}_k{}^{ \Bar{l}}}$
    \item $\frac{\partial \mathcal{L}}{\partial T^i{}_{ \Bar{k}\Bar{m}}}=-\frac{\partial T_{\perp \perp}}{\partial T^i{}_{ \Bar{k}\Bar{m}} } \quad$ Legendre identities
\end{enumerate}
where $\Delta^0{}_{\perp}=-n^i \partial_\alpha \pi_i{}^{ \alpha}$, which can be easily seen from the definition of $\Delta^\mu {}_{\nu }$.
\end{Teorema}
\noindent
It will turn out that this theorem will be particularly useful in this and the next section. Its proof can be found in appendix~\ref{DodatakC}.\par\null\par
 
The second item of this theorem enables us to naturally divide our task into three smaller ones:
 \begin{equation}
 \{ \mathcal{H}_\alpha , \mathcal{H'}_\perp\} = \{ \mathcal{H}_\alpha , {\Delta}^{'0}{}_\perp \} + \{ \mathcal{H}_\alpha , J' \} {T'}_{\perp \perp} + \{ \mathcal{H}_\alpha , {T'}_{\perp \perp} \} J' \label{Druga Puas podela posla}
 \end{equation}
We will now calculate each of them separately.

\subsection*{Calculating  ${ \{ \mathcal{H}_\alpha , {\Delta}^{'0}{}_\perp \}}$}
 Using the definition of  $\mathcal{H}_\alpha$ and ${\Delta}^{'0}{}_\perp$ we can see that:
 
 \begin{eqnarray}
  \{ \mathcal{H}_\alpha , {\Delta}^{'0}{}_\perp \}
        &=&
            -\pi_l{}^\gamma {n }^{'i}\{T^l{}_{\alpha\gamma},\p_\beta^{'} {\pi'}_i{}^\beta \} 
            -T^l{}_{\alpha\gamma} \left( {\p}_\beta  {\pi}_i{}^\beta \right)' \{ \pi_l{}^\gamma, {n'}^i \} \NB
        &&  +b^l{}_\alpha \left( {\p}_\beta  {\pi}_i{}^\beta \right)' \{ \p_\gamma \pi_l{}^\gamma, {n'}^i \}
            + \left(  \p_\gamma \pi_l{}^\gamma \right) {n'}^i \{ b^l{}_\alpha,\left( {\p}_\beta  {\pi}_i{}^\beta \right)' \} \label{Druga Puas 1}
 \end{eqnarray}
The first term is vanishing:
\begin{equation}
\{T^l{}_{\alpha\gamma},\p_\beta^{'} {\pi'}_i{}^\beta \} =
\p_\beta^{'} \{ \p_\alpha b^l{}_\gamma - \p_\gamma b^l{}_\alpha,{\pi'}_i{}^\beta \}=\p_\beta^{'} \left[\delta^\beta_\gamma  \p_\alpha \delta - \delta^\beta_\alpha \p_\gamma \delta  \right] \delta^l_i = 0
\end{equation}
so \eqref{Druga Puas 1} reduces to:
{
\begin{eqnarray}
  \{ \mathcal{H}_\alpha , {\Delta}^{'0}{}_\perp \}
        &=& T^l{}_{\alpha\gamma}  \left( {\p}_\beta  {\pi}_i{}^\beta \right) \frac{\p n^i}{\p b^l{}_\gamma} \delta
            -b^l{}_\alpha \left( {\p}_\beta  {\pi}_i{}^\beta \right)' \left(\fr{\p n^i }{\p b^l{}_\gamma} \right)' \p_\gamma \delta \NB*
        &&
            + \left(  \p_\gamma \pi_i{}^\gamma \right)  {n'}^i \p^{'}_\alpha \delta \label{Druga Puas 2}
 \end{eqnarray}
 }
Now, we apply theorem~\ref{Teorema sa deltom} in the second and third term and thus obtain four additional terms. However, one of them is vanishing, which can be seen from\footnote{This relation is derived using~\eqref{Grupa III b}:}  $\fr{\p n^i}{\p b^l{}_\gamma} b^l{}_\alpha = -n_l h^{\B{i}\gamma}b^l{}_\alpha \sim \delta^0_\alpha =0$.
 
 \begin{eqnarray}
  \{ \mathcal{H}_\alpha , {\Delta}^{'0}{}_\perp \}
        &=&    T^l{}_{\alpha\gamma}  \left( {\p}_\beta  {\pi}_i{}^\beta \right) \frac{\p n^i}{\p b^l{}_\gamma} \delta
            +\left( \p_\gamma b^l{}_\alpha \right)\left( \p_\beta \pi_i{}^\beta \right) \fr{\p n^i}{\p b^l{}_\gamma} \delta \NB
        &&  - \left(  \p_\gamma \pi_i{}^\gamma \right) \left(\p_\alpha  n^i  \right) \delta - \left(  \p_\gamma \pi_i{}^\gamma \right) n^i \p_\alpha \delta \label{Druga Puas 3}
 \end{eqnarray}
 Using $\p_\alpha n^i = \fr{\p n^i}{\p b^l{}_\gamma} \p_\alpha b^l{}_\gamma$, it is now clear that the first three terms in~\eqref{Druga Puas 3} cancel each other out.
 \begin{equation}
      \{ \mathcal{H}_\alpha , {\Delta}^{'0}{}_\perp \}
      =- \left(  \p_\gamma \pi_i{}^\gamma \right) n^i \p_\alpha \delta
      = \Delta^0{}_\perp \p_\alpha \delta \label{Druga Puas jedan}
 \end{equation}

\subsection*{Calculating$ \{ \mathcal{H}_\alpha , J' \}$}
Starting from the definitions of $ \mathcal{H}_\alpha $ and $ J'$, we get:
\begin{eqnarray}
  \{ \mc{H}_\alpha, J' \}&=&  T^i{}_{\alpha\beta} \{ \pi_i{}^\beta, J' \} -  b^i{}_\alpha \{ \p_\beta \pi_i{}^\beta, J' \} \NB
  &=& -T^i{}_{\alpha\beta} \fr{\p J}{\p b^i{}_\beta} \delta + b^i{}_\alpha \p_\beta \left( \fr{\p J}{\p b^i{}_\beta} \delta \right) \NB
  &=& \p_\beta \left( b^i{}_\alpha \fr{\p J}{\p b^i{}_\beta} \delta    \right) -\left( \p_\alpha b^i{}_\beta \right) \fr{\p J}{\p b^i{}_\beta} \delta \NB
  &=&  \p_\beta \left( b^i{}_\alpha \fr{\p J}{\p b^i{}_\beta} \delta    \right) - \left( \p_\alpha J \right) \delta \label{Druga Puas 4}
\end{eqnarray}
We transform the first term using~\eqref{Grupa III c}:
\begin{equation}
 \{ \mc{H}_\alpha, J' \} = 
 \p_\alpha \left( J \delta \right) -\left(\p_\alpha J\right) \delta
 = J \p_\alpha \delta \label{Druga Puas dva}
\end{equation}

\subsection*{Calculating $ \{ \mathcal{H}_\alpha , {T'}_{\perp \perp} \} $}
This is by far the most complicated Poisson bracket we have encountered so far because $T_{\perp \perp}$ itself is complicated. Luckily, there is a way to calculate this Poisson bracket without using an explicit expression for $T_{\perp \perp}$. From~\eqref{H ortogonalno konacno} we can see that  $T_{\perp \perp}$ can be interpreted as a function of $n^i$, $T^k{}_\BB{l}{m}$ and $\fr{1}{J}\hat{\pi}_k{}^\B{m}$. Therefore, we will examine Poisson brackets between each of those variables with $\mathcal{H}_\alpha$ separately.\par\null\par

Formal substitution  $J \rightarrow f$ in equation~\eqref{Druga Puas 4} leads to a completely correct equation if $f$ depends only on tetrads\footnote{$f$ cannot depend neither on $\p b^k{}_\mu$ nor on $\pi_k{}^\mu$}. This can easily be seen from the fact that we didn't use any specific property of  $J$ when deriving~\eqref{Druga Puas 4}. Hence:
\begin{eqnarray}
 \{ \mc{H}_\alpha, n^k \} &=& \p_\beta \left( b^i{}_\alpha \fr{\p n^k}{\p b^i{}_\beta} \delta    \right) - \left( \p_\alpha n^k \right) \delta =   - \left( \p_\alpha n^k \right) \delta 
\end{eqnarray}
where we used that $b^i{}_\alpha \fr{\p n^k}{\p b^i{}_\beta} = -n_i h^{\B{k}\beta} b^i{}_\alpha \sim \delta^0_\alpha =0$. \par\null\par

Unfortunately,  $ T^k{}_\BB{l}{m}$ depends on $\p b^k{}_\mu$, so we can't use the same formula, but we can always start from scratch using only the definitions of our variables:
\begin{eqnarray}
 \{ \mc{H}_\alpha, {T'}^k{}_\BB{l}{m} \}
        &=& 
            T^r{}_{\alpha \gamma} \{ \pi_r{}^\gamma,  {T'}^k{}_\BB{l}{m} \}
        -b^r{}_\alpha  \p_\gamma \{ \pi_r{}^\gamma ,{T'}^k{}_\BB{l}{m} \}\NB
        &=& 
            - \left( \p_\alpha b^r{}_\gamma \right) \fr{\delta \left[ {h'}_\B{l}{}^\beta {h'}_\B{m}{}^\epsilon {T'}^k{}_{\beta \epsilon}\right]}{\delta b^r{}_\gamma} + \p_\gamma \left[ b^r{}_\alpha  \fr{\delta {T'}^k{}_\BB{l}{m}}{\delta b^r{}_\gamma}\right] \NB
        &=& 
            -\p_\alpha \left[{h}_\B{l}{}^\beta {h}_\B{m}{}^\epsilon  \right] {T}^k{}_{\beta \epsilon} \,\delta - {h'}_\B{l}{}^\beta {h'}_\B{m}{}^\epsilon \left( \p_\alpha b^r{}_\gamma \right) \fr{\delta{T'}^k{}_{\beta \epsilon}}{\delta b^r{}_\gamma} \NB
        && + \p_\gamma \left[ b^r{}_\alpha  \fr{\delta {T'}^k{}_\BB{l}{m}}{\delta b^r{}_\gamma}\right] \NB
        &=&
             -\p_\alpha \left[{h}_\B{l}{}^\beta {h}_\B{m}{}^\epsilon  \right] {T}^k{}_{\beta \epsilon} \,\delta +\p_\gamma \left[ b^r{}_\alpha  \fr{\delta {T'}^k{}_\BB{l}{m}}{\delta b^r{}_\gamma}\right] \NB
        &&
            - {h'}_\B{l}{}^\beta {h'}_\B{m}{}^\epsilon \left[ \left(\p_\alpha b^k{}_\beta \right) \p^{}_\epsilon \delta - \left(\p_\alpha b^k{}_\epsilon \right) \p_\beta \delta \right]
\end{eqnarray}
In the last line we use theorem~\ref{Teorema sa deltom} in the direction $x \rightarrow x'$:
\begin{eqnarray}
 \{ \mc{H}_\alpha, {T'}^k{}_\BB{l}{m} \}
        &=& 
             -\p_\alpha \left[{h}_\B{l}{}^\beta {h}_\B{m}{}^\epsilon  \right] {T}^k{}_{\beta \epsilon} \,\delta +\p_\gamma \left[ b^r{}_\alpha  \fr{\delta {T'}^k{}_\BB{l}{m}}{\delta b^r{}_\gamma}\right] \NB
        &&
            - {h'}_\B{l}{}^\beta {h'}_\B{m}{}^\gamma \left[
            \left( \p^{'}_\alpha {b'}^k{}_\gamma  \right) \p^{'}_\beta \delta - 
            \left( \p^{'}_\alpha {b'}^k{}_\beta  \right) \p^{'}_\gamma \delta
            \right] \NB
        && - {h}_\B{l}{}^\beta {h}_\B{m}{}^\epsilon  \left( \p_\alpha T^k{}_{\beta\epsilon} \right) \delta \NB
        &=&
            -\left(\p_\alpha T^k{}_\BB{l}{m} \right) \delta +\p_\gamma \left[ b^r{}_\alpha  \fr{\delta {T'}^k{}_\BB{l}{m}}{\delta b^r{}_\gamma}\right] \NB
        &&
            - {h'}_\B{l}{}^\beta {h'}_\B{m}{}^\gamma \left[
            \left( \p^{'}_\alpha {b'}^k{}_\gamma  \right) \p^{'}_\beta \delta - 
            \left( \p^{'}_\alpha {b'}^k{}_\beta  \right) \p^{'}_\gamma \delta
            \right]  \label{Druga Puas 5}
\end{eqnarray}
The second term and the last line cancel each other out. That can be proved using \eqref{Grupa III e}, as well as ${T'}^k{}_\BB{l}{m} = {h'}_\B{l}{}^\beta {h'}_\B{m}{}^\epsilon {T'}^k{}_{\beta \epsilon}$:
\begin{eqnarray}
  \p_\gamma \left[ b^r{}_\alpha  \fr{\delta {T'}^k{}_\BB{l}{m}}{\delta b^r{}_\gamma}\right]
        &=&
            \p_\gamma \left[ - \delta^\beta_\alpha h_\B{l}{}^\gamma h_\B{m}{}^\epsilon T^k{}_{\beta \epsilon} \delta - h_\B{l}{}^\beta \delta^\epsilon_\alpha h_\B{m}{}^\gamma T^k{}_{\beta\epsilon} \delta  \right] \NB
        &&
            + \p_\gamma \left[ b^k{}_\alpha {h'}_\B{l}{}^\beta {h'}_\B{m}{}^\epsilon \left(\delta^\gamma_\epsilon \p^{'}_\beta \delta  -\delta^\gamma_\beta \p^{'}_\epsilon \delta  \right) \right] \NB
        &=& -{h'}_\B{l}{}^\gamma {h'}_\B{m}{}^\epsilon {T'}^k{}_{\alpha\epsilon} \p_\gamma \delta - {h'}_\B{l}{}^\beta {h'}_\B{m}{}^\gamma {T'}^k{}_{\beta\alpha} \p_\gamma \delta \NB
        &&
           + \left( \p_\gamma b^k{}_\alpha \right) {h'}_\B{l}{}^\beta {h'}_\B{m}{}^\gamma \p^{'}_\beta \delta 
            -\left( \p_\gamma b^k{}_\alpha \right) {h'}_\B{l}{}^\gamma {h'}_\B{m}{}^\beta\p^{'}_\beta \delta \NB
        &=&
             {h'}_\B{l}{}^\beta {h'}_\B{m}{}^\gamma \left[
            \left( \p^{'}_\alpha {b'}^k{}_\gamma  \right) \p^{'}_\beta \delta - 
            \left( \p^{'}_\alpha {b'}^k{}_\beta  \right) \p^{'}_\gamma \delta
            \right] 
\end{eqnarray}
That means that equation~\eqref{Druga Puas 5} is reduced to:
\begin{equation}
    \{ \mc{H}_\alpha, {T'}^k{}_\BB{l}{m} \} = -\left(\p_\alpha T^k{}_\BB{l}{m} \right) \delta 
\end{equation}
Our last task in this section is to calculate the Poisson bracket $\{ \mc{H}_\alpha,  \fr{1}{J'}{\hat{\pi}'}_k{}^\B{m} \}$. Since $\fr{1}{J}\hat{\pi}_k{}^\B{m}$ depends on both tetrads and momenta\footnote{But there is no dependence on $\p b^k{}_\mu$}, the Poisson brackets will have four terms. The first two only account for the tetrad dependence, so we can use the formula analogues to ~\eqref{Druga Puas 4}, whereas the other two terms have to be calculated using the definitions:
\begin{eqnarray}
\{ \mc{H}_\alpha,  \fr{1}{J'}{\hat{\pi}'}_k{}^\B{m} \}
        &=&
            T^i{}_{\alpha\gamma} \{ \pi_i{}^\gamma, \fr{1}{J'}{\hat{\pi}'}_k{}^\B{m} \} - b^i{}_\alpha \p_\epsilon \{ \pi_i{}^\epsilon, \fr{1}{J'}{\hat{\pi}'}_k{}^\B{m} \} \NB
        && +\pi_i{}^\gamma \{ T^i{}_{\alpha\gamma},  \fr{1}{J'}{\hat{\pi}'}_k{}^\B{m} \} - \left(\p_\epsilon \pi_i{}^\epsilon \right) \{ b^i{}_\alpha,  \fr{1}{J'}{\hat{\pi}'}_k{}^\B{m} \} \NB
        &=&
            -\left(\p_\alpha b^i{}_\mu \right) \fr{\p \left( \fr{1}{J} \hat{\pi}_k{}^\B{m} \right)}{\p b^i{}_\mu} \delta + \p_\beta \left( b^i{}_\alpha \fr{\p \left( \fr{1}{J} \hat{\pi}_k{}^\B{m}\right)}{\p b^i{}_\beta}  \delta \right) \NB
        &&
            +\p_\alpha \left[ \pi_i{}^\gamma \fr{\p \left(\fr{1}{J}\hat{\pi}_k{}^\B{m} \right)}{\p \pi_i{}^\gamma} \delta \right] - \left(\p_\alpha \pi_i{}^\mu\right) \fr{\left(\fr{1}{J} \hat{\pi}_k{}^\B{m} \right)}{\p \pi_i{}^\mu} \delta \NB
        &&  -\pi_i{}^\gamma \p_\gamma \left[ \fr{\p \left( \fr{1}{J}\hat{\pi}_k{}^\B{m} \right)}{\p \pi_i{}^\alpha}\delta \right] - \left( \p_\epsilon \pi_i{}^\epsilon \right) \fr{\p \left(\fr{1}{J}\hat{\pi}_k{}^\B{m} \right)}{\p \pi_i{}^\alpha} \delta\NB
        &=&
            - \p_\alpha\left(\fr{1}{J}\hat{\pi}_k{}^\B{m} \right)\delta+   {b'}^i{}_\alpha{\pi'}_k{}^\delta \fr{\p}{\p {b'}^i{}_\beta} \left( \fr{1}{J'}  {b'}^\B{m}{}_\delta \right)  \p_\beta \delta \NB
        && + \fr{1}{J'}{\pi'}_k{}^\gamma {b'}^\B{m}{}_\gamma \p_\alpha \delta - \fr{1}{J'} {\pi'}_k{}^\gamma {b'}^\B{m}{}_\alpha \p_\gamma \delta \NB
        &=&  - \p_\alpha\left(\fr{1}{J}\hat{\pi}_k{}^\B{m} \right)\delta 
\end{eqnarray}
where we used the fact that:
\begin{equation}
b^i{}_\alpha\fr{\p}{\p b^i{}_\beta} \left( \fr{1}{J} b^\B{m}{}_\delta \right) =
b^i{}_\alpha \left( -\fr{1}{J} h_\B{i}{}^\beta b^\B{m}{}_\delta + \fr{1}{J} \delta^\beta_\delta \delta^\B{m}_i \right),\quad b^i{}_\alpha h_\B{i}{}^\beta = \delta^\beta_\alpha
\end{equation}
The obtained results can now be written as $\{ \mc{H}_\alpha, \xi_a \} = - \left( \p_\alpha \xi_a\right) \delta$, where $\xi_a = \{n^i, T^k{}_\BB{l}{m} , \fr{1}{J}\hat{\pi}_k{}^\B{m} \}$. Using the fact that $T_{\perp \perp}$ is a function of $\xi_a$, as we noted earlier, it follows that:
\begin{equation}
    \{ \mc{H}_\alpha, {T'}_{\perp \perp }\} = -\left(\p_\alpha T_{\perp \perp} \right) \delta \label{Druga Puas tri}
\end{equation}
Using \eqref{Druga Puas jedan}, \eqref{Druga Puas dva} and \eqref{Druga Puas tri} in \eqref{Druga Puas podela posla}, we can finally calculate the most important Poisson bracket of this section:

\begin{eqnarray}
 \{ \mathcal{H}_\alpha , \mathcal{H'}_\perp\} &=& \{ \mathcal{H}_\alpha , {\Delta}^{'0}{}_\perp \} + \{ \mathcal{H}_\alpha , J' \} {T'}_{\perp \perp} + \{ \mathcal{H}_\alpha , {T'}_{\perp \perp} \} J' \NB*
 &=&
        \Delta^0{}_\perp \p_\alpha \delta + J {T'}_{\perp \perp}\p_\alpha \delta -J' \left(\p_\alpha T_{\perp \perp} \right) \delta \NB*
&=&
         \Delta^0{}_\perp \p_\alpha \delta +
         J T_{\perp \perp} \p_\alpha \delta \NB*
&=&
        \mc{H}_\perp \p_\alpha \delta
\end{eqnarray}

\section{Poisson Bracket \texorpdfstring{$\{ \mathcal{H}_\perp , \mathcal{H'}_\perp\}$}{H,H'}}

Notice that $\{ \mathcal{H}_\perp (x),\mathcal{H}_\perp (x') \}$ cannot have the terms proportional to the Dirac delta function $\delta(x-x')$ because they do not have the correct symmetry under $x \leftrightarrow x'$, hence they cancel each other out. Thus, $\{ \mathcal{H}_\perp (x),\mathcal{H}_\perp (x') \}$ has to be proportional to $\p_\alpha \delta$, so those will be the only terms that will be kept in the following calculations\footnote{In other words, in this section, we temporarily redefine the equality sign $A=B$ to be valid only up to terms proportional to Dirac delta function $A=B+C\delta(x-x')$. That is good enough for us because terms of the form $C\delta(x-x')$ will eventually cancel each other out, so they will not give a contribution to the end result.}. With that in mind, theorem~\ref{Teorema sa deltom} can now be written as $g(x)f(x')\partial_\alpha \delta=\delta(x-x')g(x)\partial_\alpha f+f(x)g(x)\partial_\alpha \delta=f(x)g(x)\partial_\alpha \delta$. It tells us that we can interchange $x$ and $x'$ without consequences, so there is no need to use $x'$, but nonetheless, we will keep $x'$ in some places, to make the results look simpler in the end. \par\null\par

In the last section, using the theorem~\eqref{Osnovna teorema}, we successfully managed to solve the Poisson bracket, without using the explicit expression for $T_{\perp \perp}$. We will try to do the same thing here:

\begin{eqnarray}
    \{ \mathcal{H}_\perp,\mathcal{H}'_\perp \}&=&\{JT_{\perp\perp}+\Delta^0{}_{\perp},J'T_{\perp\perp}'+{\Delta^0{}_{\perp}}' \} \NB
    &=&\{JT_{\perp\perp},J'T_{\perp\perp}' \}+\{JT_{\perp\perp},{\Delta^0{}_{\perp}}' \} \NB
    &&
    +\{\Delta^0{}_{\perp},J'T_{\perp\perp}'\}+\{\Delta^0{}_{\perp}, {\Delta^0{}_{\perp}}' \} \label{Treca Puas 1}
\end{eqnarray}
This can be further simplified if we notice that:
\begin{enumerate}
    \item $\{ \Delta^0{}_{\perp},J' \}=-n^i\{\partial_\alpha \pi_i{}^{\alpha},J'\}=n^i\p_\alpha \left( \frac{\p J}{\p b^i{}_\alpha}\delta \right) = n^i J h_\B{i}{}^\alpha \p_\alpha\delta = 0$
    \item $T_{\perp \perp}$ does not depend on $\p\pi_k{}^\mu$, while $J$ depends only on $b^k{}_{ \mu}$, so we can conclude that $\{ T_{\perp \perp},J' \}=0$.
\end{enumerate}
Using these two items, the equation~\eqref{Treca Puas 1} is reduced to: 
\begin{equation} \label{TRAZENI REZULTAT}
     \{ \mathcal{H}_\perp,\mathcal{H}'_\perp \}= {\{\Delta^0{}_{\perp}, {\Delta^0{}_{\perp}}' \}}+
     JJ'{\{T_{\perp\perp},T_{\perp\perp}' \}}+
      J'{\{\Delta^0{}_{\perp},T_{\perp\perp}'\}}+
      J{\{T_{\perp\perp},{\Delta^0{}_{\perp}}' \}}
\end{equation}
The fourth term can be obtained from the third term by multiplying it with $-1$ and substituting $x \leftrightarrow x'$, so we only need to calculate the first three terms:
\subsection*{Calculating $ {\{\Delta^0{}_{\perp}, {\Delta^0{}_{\perp}}' \}}$}

\begin{eqnarray}
        {\{\Delta^0{}_{\perp}, {\Delta^0{}_{\perp}}' \}}&=&\{ -n^i\partial_\alpha \pi_i{}^{\alpha },-{n'}^j{\partial_\beta}'{{\pi'}_j{}^{\beta}} \} \NB
        &=&{n^i {\partial_\beta}'{\pi'}_j{}^{\beta} \{\partial_\alpha  \pi_i{}^{\alpha }, {n'}^j \}}+{ \partial_\alpha  \pi_i{}^{\alpha }{n'}^j \{n^i,{\partial'}_\beta {\pi'}_j{}^{\beta} \}} \label{Treca Puas 2}
\end{eqnarray}
The second term can be obtained from the first one by substituting $x \leftrightarrow x'$, so it is sufficient to calculate only one of them:
\begin{eqnarray}
{n^i {\partial_\beta}'{\pi'}_j{}^{\beta} \{\partial_\alpha  \pi_i{}^{\alpha }, {n'}^j \}}
        &=&
            n^i \partial_\beta \pi_j{}^{\beta }n_i b^j{}_{\gamma} \;^3g^{\gamma \alpha} \partial_\alpha \delta  \NB
            &=& \;^3g^{\gamma \alpha} b^j{}_{\gamma} \partial_\beta \pi_j{}^{\beta }\partial_\alpha \delta =-\;^3g^{\gamma \alpha}\Delta^0{}_{\gamma} \partial_\alpha \delta 
\end{eqnarray}
Hence:
\begin{equation} \label{REZULTAT I}
     {\{\Delta^0{}_{\perp}, {\Delta^0{}_{\perp}}' \}}=-\left( \;^3{g'}^{\gamma \alpha} {\Delta^0{}_{\gamma}}'+\;^3g^{\gamma \alpha } \Delta^0{}_{\gamma } \right) \partial_\alpha \delta 
\end{equation}
where we used that ${\partial '}_\alpha \delta =-\partial_\alpha \delta $.
\subsection*{Calculating$ {\{T_{\perp\perp},T_{\perp\perp}' \}}$}
It is obvious that $\{T_{\perp\perp},{n'}_i \}=0$, since $T_{\perp\perp}$ does not depend neither on $\pi_k{}^\mu$ nor on $\p_\alpha\pi_k{}^\mu$. As a consequence, we get quite a useful property: $\{T_{\perp\perp},\delta_\B{i}^{\B{j}} \}=0$. That means that $T_{\perp\perp}$ can be considered a function of  $\xi_A=\{ \frac{1}{J} \hat{\pi}_k{}^{\Bar{m}},T^k{}_{\Bar{l},\Bar{m}} \}$ when using the chain rule:

\begin{equation} \label{Lancano pravilo}
    \{ T_{\perp\perp},{T'}_{\perp\perp} \}=\sum_{A,B} \frac{\partial T_{\perp\perp}}{\partial \xi_A} \{ \xi^A,{\xi'}^B \} \frac{\partial {T'}_{\perp\perp}}{\partial {\xi '}_B}
\end{equation}
There are only two Poisson brackets in \eqref{Lancano pravilo}, proportional to the derivative of the Dirac delta function. The first one is:
\begin{eqnarray}
 \{T^k{}_{\Bar{l}\Bar{m}},\frac{1}{J}\hat{\pi}_p{}^{\Bar{r}} \} &=& \frac{1}{J}h_{\Bar{l}}{}^{\alpha }h_{\Bar{m}}{}^{\beta }b^{{r}}{}_{\gamma }\{T^k{}_{\alpha \beta} ,\pi_p{}^{\gamma } \}=\frac{1}{J}\delta^k_p \left( \delta^{\Bar{r}}_{\Bar{m}} h_{\Bar{l}}{}^{\alpha }-\delta^{\Bar{r}}_{\Bar{l}}h_{\Bar{m}}{}^{\alpha } \right) \partial_\alpha \delta 
\end{eqnarray}
while the second one is obtained by substituting  $x \leftrightarrow x'$ in \eqref{Lancano pravilo}.  Hence:
\begin{eqnarray}
  \{ T_{\perp\perp},{T'}_{\perp\perp} \}=\underbrace{
  \frac{1}{2} \frac{\partial T_{\perp \perp}}{\partial T^k{}_{\Bar{l}\Bar{m}}} \frac{\partial T_{\perp\perp}}{\partial (\frac{1}{J}\hat{\pi}_p{}^{\Bar{r}})} \{T^k{}_{ \Bar{l} \Bar{m}}, \frac{1}{J}\hat{\pi}_p{}^{\Bar{r}} \}}_{I(x,x')}
  -I(x', x) \label{Treca Puas 3}
\end{eqnarray}
Note that we have inserted one half in the previous expression because $T^k{}_\BB{l}{m}$ is antisymmetric with respect to  $ l $ and $ m $, so summing over them results in repetition. Using \eqref{Osnovna teorema} we can calculate $I(x',x)$:

\begin{subequations}
\begin{eqnarray}
  I(x,x') &=&\frac{1}{2} \frac{\partial T_{\perp \perp}}{\partial T^k{}_{\Bar{l}\Bar{m}}} \frac{\partial T_{\perp\perp}}{\partial (\frac{1}{J}\hat{\pi}_p{}^{\Bar{r}})}
  \frac{1}{J}\delta^k_p \left( \delta^{\Bar{r}}_{\Bar{m}} h_{\Bar{l}}{}^{\alpha }-\delta^{\Bar{r}}_{\Bar{l}}h_{\Bar{m}}{}^{\alpha } \right) \partial_\alpha \delta \\
  &=& -\frac{1}{2}\frac{1}{J} \frac{\partial \mathcal{L}}{\partial T^k{}_{ \Bar{l}\Bar{m}}} T^k{}_{\perp \Bar{r}} \left( \delta^{\Bar{r}}_{\Bar{m}} h_{\Bar{l}}{}^{\alpha }-\delta^{\Bar{r}}_{\Bar{l}}h_{\Bar{m}}{}^{\alpha } \right) \partial_\alpha \delta \\ \label{II1}
  &=& -\frac{1}{J} T^{\Bar{l}}{}_{\perp} h_{\Bar{l}}{}^{\alpha }\partial_\alpha \delta 
\end{eqnarray}
\end{subequations}
Using~\eqref{ADM grupa II e}, the previous expression can be written as: 
\begin{equation} \label{I pom rel za T perp}
    I(x,x')= -\frac{1}{J}T_{\beta \perp} \;^3g^{\beta \alpha }\partial_\alpha \delta 
\end{equation}
Replacing \eqref{I pom rel za T perp} in~\eqref{Treca Puas 3}: 
\begin{equation} \label{REZULTAT II}
    JJ'\{T_{\perp \perp},{T'}_{\perp\perp} \}=-\left( JT_{\beta \perp} \;^3g^{\beta \alpha  }+J'{T'}_{\beta \perp} {\;^3g'}^{\beta \alpha  } \right) \partial_\alpha \delta 
\end{equation}

\subsection*{Calculating ${\{\Delta^0{}_{\perp},T_{\perp\perp}'\}}$}

We will once again use the chain rule  $\{\Delta^0{}_{\perp},{T'}_{\perp\perp} \}=\sum_{A}\{\Delta^0{}_{\perp},{\xi'}_A\}\frac{\partial {T'}_{\perp\perp}}{\partial {\xi'}_A}$, where $\xi_a=\{b^k{}_{\beta }$ , $\pi_k{}^{\mu  } \}$. Note that $b^k{}_0$ is not the element of $\xi_a$, because $\Delta^0{}_{\perp}$ does not depend on $\pi_k{}^0$, so the corresponding Poisson bracket is vanishing. Calculations can be further simplified using:
\begin{enumerate}
    \item $\{ \Delta^0{}_{\perp},{\pi'}_k{}^{\mu} \}=0$, because $\Delta^0{}_{\perp}$ \mbox{ does not depend on ${\p b^k{}_\mu}$.}
    \item \begin{eqnarray}
      \{ \Delta^0{}_{\perp},{T'}^k_{\alpha \beta } \} &=& -n^i\{\partial_\gamma \pi_i{}^{\gamma },\partial_\alpha b^k{}_{\beta }\}+n^i\{\partial_\gamma \pi_i{}^{\gamma },\partial_\beta b^k{}_{\alpha } \}\NB &=& -n^k\partial_\alpha \partial_\beta \delta +n^k\partial_\alpha \partial_\beta \delta =0
    \end{eqnarray}
\end{enumerate}
These two relations tell us that for the use of the chain rule, the dependence of $ T_{\perp \perp} $ on the momenta is irrelevant, as well as the implicit dependence on the tetrad inside the torsion (with spatial indices) $T^k_{\alpha \beta }$. Hence:
\begin{equation} \label{medjurezultat u III}
    \{\Delta^0{}_{\perp},T_{\perp\perp} \}=\{\Delta^0{}_{\perp},b^k{}_{\alpha } \} \frac{\partial T_{\perp\perp}}{\partial b^k{}_{\alpha }}=n^k\frac{\partial T_{\perp \perp}}{\partial b^k{}_{ \alpha }} \partial_\alpha \delta 
\end{equation}
where we used that:
\begin{equation}
    \{ \Delta^0{}_{\perp},b^k{}_{\beta }\}=-n^i\{\partial_\alpha \pi_i{}^{\alpha },b^k{}_{\beta } \}=n^k\partial_\beta \delta 
\end{equation}
\subsubsection{Calculating $n^k\frac{\partial T_{\perp \perp}}{\partial b^k{}_{ \alpha }}$}

Since:
\begin{equation}
 T_{\perp \perp}=\frac{1}{J} \pi_k{}^{\beta } T^k{}_{\perp \beta }-\mathcal{L}  \label{T ort ort oblik} 
\end{equation}
it is clear that the hardest part is to calculate $n^k \frac{\partial \mathcal{L}}{\partial b^k{}_{\alpha }}$, so that will be our first task:

\begin{eqnarray}
   n^k \frac{\partial \mathcal{L}}{b^k{}_{\alpha }} &=& n^k \frac{1}{2} \frac{\partial \mathcal{L}}{\partial T^i{}_{mn}} \frac{\partial T^i{}_{mn}}{\partial b^k{}_{\alpha }}\NB
   &=&n^k
   \frac{1}{2} \frac{\partial \mathcal{L}}{\partial T^i{}_{mn}}\frac{\partial }{\partial b^k{}_{\alpha }} \left( T^i{}_{\Bar{m} \Bar{n}}+T^i{}_{\Bar{m}
   \perp} n_n -T^i{}_{\Bar{n}\perp} n_m \right)  \NB
   &=& n^k\frac{1}{2b} H_i{}^{mn} \frac{\partial}{\partial b^k{}_{ \alpha }}
   \left( h_{\Bar{m}}{}^{\beta } h_{\Bar{n}}{}^{ \gamma } T^i{}_{ \beta \gamma } +h_{\Bar{m}}{}^{\beta } T^i{}_{\beta \perp} n_n -h_{\Bar{n}}{}^{\beta }T^i{}_{\beta \perp} n_m \right)
\end{eqnarray}
Using~\eqref{Grupa III d} and \eqref{Grupa III f}:
\begin{eqnarray}
   n^k \frac{\partial \mathcal{L}}{b^k{}_{\alpha }} 
        &=&
            \frac{1}{2b} H_i{}^{mn}  \left( \;^3g^{\alpha \beta }n_m h_{\Bar{n}}{}^{\gamma } T^i{}_{\beta \gamma } +\;^3g^{\alpha \gamma }n_n h_{\Bar{m}}{}^{\beta } T^i{}_{\beta \gamma }
            -h_{\Bar{m}}{}^{\beta }b_{n\rho } \;^3g^{\rho \alpha }T^i{}_{\beta \perp} \right)+\NB
        &&\frac{1}{2b} H_i{}^{mn}   \left(h_{\Bar{m}}{}^{\beta } n_n n^k \frac{\partial T^i{}_{ \beta \perp}}{\partial b^k{}_{\alpha }} + h_{\Bar{n}}{}^{\beta } b_{m\rho }\;^3g^{\rho \alpha }T^i{}_{\beta \perp}- h_{\Bar{n}}{}^{\beta }n_m n^k \frac{\partial T^i{}_{\beta \perp}}{\partial b^k{}_{\alpha }} \right)\NB
     &=& \underbrace{\frac{1}{b} H_i{}^{\perp n} \;^3g^{\alpha \beta } h_{\Bar{n}}{}^{\gamma } T^i{}_{\beta \gamma }}_{A_1} \underbrace{-\frac{1}{b}H_i{}^{ mn} h_{\Bar{m}}{}^{\beta } b_{n\rho } \;^3g^{\rho \alpha } T^i{}_{\beta \perp}}_{A_2}+\underbrace{\frac{1}{b}H_i{}^{mn} h_{\Bar{m}}{}^{\beta }n_n n^k \frac{\partial T^i{}_{\beta \perp}}{\partial b^k{}_{\alpha }}}_{A_3} \nonumber
\end{eqnarray}
We divided the previous line into three parts to make the calculations clearer:
\begin{subequations}
\begin{equation}
    A_1=\frac{1}{J} \pi_i{}^{\gamma } T^i{}_{\beta \gamma } \;^3g^{\alpha \beta }=\frac{b}{J} \left( T^0{}_{\beta } +\delta^0{}_{\beta } \mathcal{L} \right)  \;^3g^{\alpha \beta }=T_{\perp \beta } \;^3g^{\alpha \beta }
\end{equation}

\begin{equation}
    A_2=-\frac{1}{b} H_i{}^{\beta n}T^i{}_{ \beta \perp}b_{n\rho } \;^3g^{\rho \alpha }=-T^n{}_{\perp}b_{n\rho } \;^3g^{\rho \alpha }=-T_{\gamma \perp} \;^3g^{\gamma \alpha }
\end{equation}
\begin{equation}
    A_3=\frac{1}{b} H_i{}^{m\perp} h_{\Bar{m}}{}^{\beta } n^k \frac{\partial T^i{}_{\beta \perp}}{\partial b^k{}_{\alpha }}
    =-\frac{1}{J}\pi_i{}^{m} h_{\Bar{m}}{}^{\beta }n^k \frac{\partial T^i{}_{\beta\perp}}{\partial b^k{}_{\alpha }}
    =-\frac{1}{J}\pi_i{}^{\beta} n^k \frac{\partial T^i{}_{\beta\perp}}{\partial b^k{}_{\alpha }}
\end{equation}
\end{subequations}
Acting with $n^k \frac{\p}{\p b^k{}_\alpha}$ on each side of equation \eqref{T ort ort oblik}, one can see\footnote{We note that $J$ acts like a constant because of $n^k \fr{\p J}{\p b^k{}_\alpha}= n^k J h_\B{k}{}^\alpha = 0$} that  $\frac{1}{J} \pi_i{}^\beta n^k  \frac{\partial T^i{}_{\beta\perp}}{\partial b^k{}_{\alpha }}$ and $A_3$ cancel each other out. Thus, the result contains only  $A_1$ and $A_2$, but with an extra minus sign, because of \eqref{T ort ort oblik}.

\begin{equation}
    n^k \frac{\partial T_{\perp \perp}}{\partial b^k_{\; \alpha }}=
    \left( T_{\beta \perp}-T_{\perp \beta} \right) \;^3g^{\beta \alpha }
\end{equation}
Replacing this in (\ref{medjurezultat u III}):
\begin{equation} \label{REZULTAT III}
    J\{\Delta^0_{\;\perp},T_{\perp\perp}'\}= J\left( T_{\beta \perp}-T_{\perp \beta} \right) \;^3g^{\beta \alpha } \partial_\alpha \delta 
\end{equation}
The last term in (\ref{TRAZENI REZULTAT}) is now obtained from \eqref{REZULTAT III} by multiplying it with $-1$ and substituting $x \leftrightarrow x'$
\begin{equation} \label{REZULTAT IV}
    J\{T_{\perp\perp},{\Delta^0_{\;\perp}}' \}=J' \left( {T'}_{\beta \perp}-{T'}_{\perp \beta} \right) \;^3{g'}^{\beta \alpha } \partial_\alpha \delta 
\end{equation}
Finally, replacing  (\ref{REZULTAT I}), (\ref{REZULTAT II}), (\ref{REZULTAT III}), (\ref{REZULTAT IV}) in (\ref{TRAZENI REZULTAT}):
\begin{eqnarray}
    \{ \mathcal{H}_\perp,\mathcal{H}'_\perp \}&=&-\left(
    JT_{\perp \beta}+\Delta^0_{\;\beta} \right) \;^3g^{\beta \alpha }\partial_\alpha \delta 
    -\left(
    J'{T'}_{\perp \beta}+{\Delta'}^0_{\;\beta} \right) \;^3{g'}^{\beta \alpha }\partial_\alpha \delta \NB
    &=& - \left( \mathcal{H}_\beta \;^3g^{\beta \alpha } +{\mathcal{H}'}_\beta \;^3{g'}^{\beta \alpha } \right) \partial_\alpha \delta 
\end{eqnarray}
where we used the theorem~\ref{Osnovna teorema} in the last line.
\subsection*{Constraint Algebra}
Summarizing the results from this chapter:
\begin{subequations} \label{algebra veza genericki}
\begin{empheq}[box=\widefbox]{align}
\{ \mc{H}_\alpha, {\mc{H'}}_\beta \}&= \left( {\mc{H'}}_\alpha \p_\beta + \mc{H}_\beta \p_\alpha \right) \delta(x-x') \\
\{ \mc{H}_\alpha, {\mc{H'}}_\perp \} &= \mc{H}_\perp \p_\alpha \delta \\
\{ \mathcal{H}_\perp,\mathcal{H}'_\perp \}&= - \left( \mathcal{H}_\beta \;^3g^{\beta \alpha } +{\mathcal{H}'}_\beta \;^3{g'}^{\beta \alpha } \right) \partial_\alpha \delta 
\end{empheq}
\end{subequations}
These relations represent the constraints algebra. Now it is obvious\footnote{Having in mind that the Hamiltonian itself is a constraint, up to a total divergence term, it does not give any kind of contribution in this case.} that the consistency conditions for the secondary constraints are identically fulfilled. In other words, there will be no new constraints nor restrictions to Lagrange multipliers.\par\null\par

\subsection*{Number of Degrees of Freedom}
From the constraint algebra, we conclude that all 8 constraints  $\pi_i{}^0, \mc{H}_\perp, \mc{H}_\alpha$ are first-class. Since the tetrad $b^k{}_\mu$ has 16 independent components, using  \eqref{Broj stepeni slobode} we see that the number of degrees of freedom is:
\begin{equation} \label{teleparalelna broj stepeni slobode}
    \text{number of degrees of freedom} = 2\cdot 16 - 2 \cdot 8 = 16
\end{equation}
This is a number of degrees of freedom in the phase space, whereas the number of degrees of freedom in the configuration space is equal to $16/2 = 8$. Further analysis of the canonical formalism, such as the construction of the gauge generator by Castellani's algorithm, will be given in chapter \ref{glava konstrukcija generatora}.

\chapter{Canonical Analysis of TEGR}
\label{specijalni slucaj teleparalelne teorije}
In this chapter, we will analyze TEGR, which is defined by the Lagrangian~\eqref{Lagranzijan teleparalelne gr}, but with a special choice of the coefficients~\eqref{spec koef}. We already proved that TEGR is equivalent to GR. In particular, the number of degrees of freedom in configurational space is 2. Given that all the constraints found in the previous chapter are still valid in this theory, as well as the constraint algebra~\eqref{algebra veza genericki}, equation  \eqref{teleparalelna broj stepeni slobode} indicates that it is necessary to have additional constraints in TEGR (relative to TG), in order to have the correct number of degrees of freedom.

\section{New Constraints}
From \eqref{ired 1} and \eqref{ired 2} we can see that these new constraints are: 

\begin{subequations}\label{extra veze}
\begin{eqnarray}
P_{\perp \B{k}}&=&\fr{1}{J}\hat{\pi}_{\perp \B{k}} + 2a T^{\B{m}}{}_\BB{m}{k} \approx 0 \label{extra veza 1} \\
P_{\BB{}{ik}}^A&=&\fr{1}{J}\hat{\pi}_\BB{i}{k}^A - aT_{\perp \BB{i}{k}} \approx 0 \label{extra veza 2}
\end{eqnarray}
\end{subequations}

They can be written in a more convenient form using:
\begin{Lema} \normalfont

Constraints \eqref{extra veza 1} and \eqref{extra veza 2} can be equivalently written as:
\begin{equation} \label{extra veze ekv oblik}
    \hat{\pi}_{i\B{k}} - \hat{\pi}_{k\B{i}} \approx 2aJ \left( T_{\perp \BB{i}{k}} - n_i T^\B{m}{}_\BB{m}{k} + n_k T^\B{m}{}_\BB{m}{i} \right)
\end{equation}
\end{Lema}
\begin{Dokaz}
\normalfont

This proof is divided into two parts. In the first part, we prove that equation \eqref{extra veze}  implies \eqref{extra veze ekv oblik}, whereas the second part is the other way around.
\subsubsection*{first part \eqref{extra veze} $\Rightarrow$ \eqref{extra veze ekv oblik}}
\begin{eqnarray*}
\hat{\pi}_{i \B{k}} -\hat{\pi}_{k \B{i}} &=& \hat{\pi}_\BB{i}{k} + n_i \hat{\pi}_{\perp \B{k}} - (i \leftrightarrow k) \approx 2\hat{\pi}_\BB{i}{k}^A - 2aJ\left( n_i T^\B{m}{}_\BB{m}{k} - n_k T^\B{m}{}_\BB{m}{i} \right) \NB
&\approx& 2aJT_{\perp \BB{i}{k}} -  2aJ\left( n_i T^\B{m}{}_\BB{m}{k} - n_k T^\B{m}{}_\BB{m}{i} \right) 
\end{eqnarray*}
\subsubsection*{second part \eqref{extra veze ekv oblik}$\Rightarrow$\eqref{extra veze}  }
\noindent
Relation  \eqref{extra veza 1} can be obtained by multiplying each side of \eqref{extra veze ekv oblik}  with $n^i$. On the other hand, \eqref{extra veza 2} can be derived by putting bars over  $i$ and  $k$ in \eqref{extra veze ekv oblik} and then using that  $n_\B{i} = 0$.

\flushright
\qedsymbol
\end{Dokaz}
\noindent
Using \eqref{Lema 1.5}, relation \eqref{extra veze ekv oblik} can be rewritten as:
\begin{equation}
   \phi_{ik}:=\hat{\pi}_{i\B{k}} - \hat{\pi}_{k\B{i}}- 2a\p_\alpha H^{0\alpha}_{ik}
   \approx 0 
\end{equation}
We can obtain even more convenient form if we use \eqref{B=2H kompaktno}:
\begin{equation} \label{dodatne veze phi}
   \phi_{ik}:= \hat{\pi}_{i\B{k}} - \hat{\pi}_{k\B{i}} + a\p_\alpha B^{0\alpha}_{ik}
   \approx 0 
\end{equation}
where $B^{\mu\nu}_{mn} = \varepsilon^{\mu\nu\rho\sigma}_{mnik} b^i{}_\rho b^k{}_\sigma$. Since TEGR has additional constraints (compared to TG), canonical Hamiltonian~\eqref{H ortogonalno konacno} should be rewritten in a slightly different way. From \eqref{extra veze}, we can see that the generalized momenta  $P_{\perp \B{k}}$ и $P_\BB{i}{k}^A$ are vanishing, whereas:
 \begin{eqnarray} \label{dodatneveze}
     P^T_{\Bar{i}\Bar{k}} &=& \frac{1}{J}\hat{\pi}^T_{\Bar{i}\Bar{k}}=2aT^T_{\Bar{i}\perp\Bar{k}};\quad  P^{\Bar{m}}_{\;\Bar{m}}=\frac{1}{J}\hat{\pi}^{\Bar{m}}{}_{\Bar{m}}=-4aT^{\Bar{m}}_{\;\perp\Bar{m}}
 \end{eqnarray}
 Replacing these relations\footnote{We note that relations $P_{\perp \B{k}}\approx 0$ and $P_\BB{i}{k}^A\approx 0$ should actually be substituted into the first line of \eqref{prvi red ired dekom} even though we can also obtain the correct result from \eqref{H ortogonalno konacno} by ignoring the fact that there is a division by zero.} into \eqref{H ortogonalno konacno} we obtain:
 \begin{equation}
     \mc{H}_\perp = -n^i \partial_\alpha \pi_i{}^{\alpha} -J\mathcal{L}_{T}(\overline{T})+\underbrace{\frac{1}{4aJ}\left(\hat{\pi}^{(\Bar{i}\Bar{m})}\hat{\pi}_{(\Bar{i}\Bar{m})}-\frac{1}{2}\hat{\pi}^{\Bar{m}}{}_{\Bar{m}} \hat{\pi}^{\Bar{n}}{}_{\Bar{n}}\right)}_{\fr{1}{2}P^2}
 \end{equation}
 \subsection*{Equations of Motion}
 
 Given that we have already shown the equivalence between TEGR and GR in \S~\ref{Ekvivalentnost}, it is clear that the equations of motion of the gravitational field in vacuum can be written as:

 \begin{equation} \label{jednacine kretanja OTR}
     R^{ik}(\Delta) - \fr{1}{2}\eta^{ik}R(\Delta) = 0
 \end{equation}
 We can also prove this differently. The equations of motion in TG case for the specific choice of coefficients $A=\fr{1}{4},\; B=\fr{1}{2},\; C=-1$ are the same as in TEGR case, so it is sufficient to prove that in the teleparallel case  $R^{ij}{}_{\mu\nu}(\omega) = 0$, as follows from the identity:
 \begin{equation} \label{identitet za ekvivalentnost}
     2ab\left[ R^{ik}(\Delta) - \fr{1}{2}\eta^{ik}R(\Delta) \right] = -\nabla_\mu H^{i\mu k} - H_{mn}{}^k T^{mni} + \fr{1}{2}H^{imn}T^k{}
     _{mn} + \eta^{ik} b \mc{L}_T
 \end{equation}
 In our specific case, the covariant derivative should be replaced with partial derivative\footnote{Having in mind that the connection is vanishing} ${\nabla_\mu \rightarrow \p_\mu}$,  so the right-hand side of \eqref{identitet za ekvivalentnost} becomes the left-hand side of \eqref{jednacine kretanja TTG}. 
 Identity \eqref{identitet za ekvivalentnost}  can be proved if we rewrite \eqref{Lema 1.6} as:
 \begin{equation}
     R^{ij}{}_{\mu\nu}(\omega) =  R^{ij}{}_{\mu\nu}(\Delta) + \left[ \nabla_\mu K^{ij}{}_\nu - K^i{}_{s\mu}K^{sj}{}_\nu-(\mu\leftrightarrow\nu)\right] 
 \end{equation}
 and then multiply each side by  $\fr{1}{2} H_{kj}^{\mu\nu}$.\par\null\par
 

In the presence of matter, the right-hand side of equation \eqref{jednacine kretanja OTR} has
an additional term, the matter energy-momentum tensor $\tau^{ik}$.
Since equation \eqref{jednacine kretanja OTR} is symmetric, $\tau^{ik}$ is expected to be also
symmetric. However, this is not always true. In particular, this is
not true for the Dirac field. If we don't want to impose an extra
(non-dynamical!) condition $\tau^{[ik]}=0$ on matter, the only
conclusion is that TEGR is not consistent in the presence of matter
with $\tau^{[ik]}\ne 0$\cite{Blagojevic}. Nevertheless, restricting our attention
to classical fluid matter, TEGR can be consistently interpreted as an
effective theory of the macroscopic gravitational phenomena equivalent
to GR.

\section{Poisson Bracket\texorpdfstring{$\{\phi_{ij},{\phi'}_{kl} \}$}{fij,f'kl}}

Using \eqref{dodatne veze phi} we can express this Poisson bracket as:
\begin{subequations}
\begin{eqnarray} \label{Sprva 1}
\{\phi_{ij},{\phi'}_{kl} \} &=&
\{ \hat{\pi}_{i\B{j}}, {\hat{\pi}'}_{k\B{l}} \} - \{ \hat{\pi}_{i\B{j}}, {\hat{\pi}'}_{l\B{k}} \}
-\{ \hat{\pi}_{j\B{i}}, {\hat{\pi}'}_{k\B{l}} \}+\{ \hat{\pi}_{j\B{i}}, {\hat{\pi}'}_{l\B{k}} \} \\
&&+a\{ \p_\alpha B^{0\alpha}_{ij},  {\hat{\pi}'}_{k\B{l}} \}
-a\{ \p_\alpha B^{0\alpha}_{ij},  {\hat{\pi}'}_{l\B{k}} \} \label{Sprva 2}\\
&&+a\{ \hat{\pi}_{i\B{j}}, {\p}^{'}_\alpha {B'}^{0\alpha}_{kl} \}
- a\{ \hat{\pi}_{j\B{i}}, {\p}^{'}_\alpha {B'}^{0\alpha}_{kl} \} \label{Sprva 3}\\
&\equiv& I + II \nonumber
\end{eqnarray}
\end{subequations}
where we defined new variables $I$ and $II$: 
\begin{equation}
I= A_{ijkl}- A_{ijlk} - A_{jikl} + A_{jilk}, \quad \text{}\; A_{ijkl}:= \{ \hat{\pi}_{i\B{j}}, {\hat{\pi}'}_{k\B{l}} \}
\end{equation}
\begin{eqnarray}
II&=&B_{ijkl}(x,x')-B_{ijlk}(x,x')-B_{klij}(x',x) + B_{klji}(x',x),\NB
&& \text{} \; \; \; B_{ijkl}(x,x') := a\{ \p_\alpha B^{0\alpha}_{ij},  {\hat{\pi}'}_{k\B{l}} \}
\end{eqnarray}
It is now clear that our original task is divided into calculating $A_{ijkl}$ and $B_{ijkl}(x,x')$.
\subsubsection*{Calculating $\mathbf{A_{ijkl}}$}
\begin{eqnarray}
A_{ijkl} &=& \{ \pi_i{}^\alpha b_{j\alpha}, {\pi'}_k{}^\beta {b'}_{l\beta} \} = \pi_i{}^\alpha  {b'}_{l\beta} \{ b_{j\alpha}, {\pi'}_k{}^\beta \} +  b_{j\alpha}  {\pi'}_k{}^\beta \{ \pi_i{}^\alpha ,  {b'}_{l\beta}  \} \NB
&=& \pi_i{}^\alpha b_{l\alpha} \eta_{jk} \delta - b_{j\alpha}\pi_k{}^\alpha \eta_{il} \delta = \hat{\pi}_{i\B{l}} \eta_{jk} \delta - \hat{\pi}_{k\B{j}} \eta_{il} \delta
\end{eqnarray}
Hence:
\begin{equation} \label{Sprva I}
    I= \left[\hat{\pi}_{i\B{l}} \eta_{jk} - \hat{\pi}_{k\B{j}}\eta_{il} - \hat{\pi}_{i\B{k}}\eta_{jl} + \hat{\pi}_{l\B{j}}\eta_{ik} - \hat{\pi}_{j\B{l}}\eta_{ik}+\hat{\pi}_{k\B{i}}\eta_{jl} + \hat{\pi}_{j\B{k}}\eta_{il}-\hat{\pi}_{l\B{i}}\eta_{jk} \right] \delta
\end{equation}
\subsubsection*{Calculating $\mathbf{B_{ijkl}(x,x')}$}
\begin{eqnarray}
B_{ijkl}(x,x') 
    &=&
        a\{ \p_\alpha B^{0\alpha}_{ij}, {\hat{\pi}'}_{k\B{l}} \} = a\varepsilon^{0\alpha\beta \gamma}_{ijmn} \p_\alpha\{ b^m{}_\beta b^n{}_\gamma, {\pi'}_k{}^\delta {b'}_{l\delta} \} \NB
    &=&
        2a\p_\alpha\left[\varepsilon^{0\alpha\beta \gamma}_{ijmn} b^n{}_\gamma b_{l\beta} \delta(x-x')  \right]
\end{eqnarray}
Using \eqref{G: relacija 1}:
\begin{equation}
 B_{ijkl}(x,x') = a\p_\alpha\left[ \left( B_{ij}^{0\alpha}\eta_{kl} + B_{ki}^{0\alpha}\eta_{jl} + B_{jk}^{0\alpha}\eta_{il} \right) \delta(x-x') \right]   
\end{equation}
It is now straightforward to calculate $II$:
\begin{equation} \label{Sprva II}
    II=a \p_\alpha \left[ B_{ki}^{0\alpha}\eta_{jl} + B^{0\alpha}_{jk}\eta_{il} - B^{0\alpha}_{li}\eta_{jk} - B^{0\alpha}_{jl}\eta_{ik}  \right] \delta(x-x')
\end{equation}
Adding \eqref{Sprva I} and \eqref{Sprva II} we finally obtain:
\begin{equation}
 \{\phi_{ij},{\phi'}_{kl} \} = 
 \left( \eta_{ik} \phi_{lj} + \eta_{jk}\phi_{il} \right) \delta - (k \leftrightarrow l)
\end{equation}

\section{Poisson Bracket \texorpdfstring{$\{\phi_{ij},\mc{H'}_\beta \}$}{fij,H'b}}
We divide our problem  into smaller and easier sections:
\begin{equation} \label{Sdruga 1}
\{ \phi_{ij}, \mc{H'}_\beta \} = \underbrace{\{\hat{\pi}_{i\B{j}}, \mc{H'}_\beta \}}_{I_{ij\beta}} -  \underbrace{\{ \hat{\pi}_{j\B{i}}, \mc{H'}_\beta \}}_{I_{ji\beta}} + \underbrace{ a\{\p_\gamma B^{0\gamma}_{ij}, \mc{H'}_\beta \}}_{II}
\end{equation}
Since the second term can be obtained from the first one by interchanging $i$ and $j$, we only need to calculate $I_{ij\beta}$ and $II$.

\subsection*{Calculating  $I_{ij\beta}$}
\begin{eqnarray}
I_{ij\beta}
    &=&
        \{ \pi_i{}^\gamma b_{j\gamma}, {\pi'}_k{}^\alpha {T'}^k{}_{\beta \alpha} - {b'}^k{}_\beta \p^{'}_\alpha {\pi'}_k{}^\alpha \} \NB
    &=&
        \pi_i{}^\gamma {T'}_{\beta \alpha} \{ b_{j\gamma}, {\pi'}_k{}^\alpha \} + b_{j\gamma}{\pi'}_k{}^\alpha \{ \pi_i{}^\gamma, {T'}^k{}_{\beta \alpha}\} \NB
    &&
        -\pi_i{}^\gamma {b'}^k{}_\beta \{ b_{j\gamma}, \p^{'}_\alpha {\pi'}_k{}^\alpha \}
        -b_{j\gamma}\left( \p^{'}_\alpha {\pi'}_k{}^\alpha \right) \{ \pi_i{}^\gamma, {b'}^k{}_\beta \} \NB
    &=&
        \pi_i{}^\alpha T_{j\beta\alpha} \delta - b_{j\alpha}{\pi'}_i{}^\alpha \p^{'}_\beta \delta + b_{j\beta} {\pi'}_i{}^\alpha \p^{'}_\alpha\delta - \pi_i{}^\alpha {b'}_{j\beta} \p^{'}_\alpha \delta + b_{j\beta} \p_\alpha \pi_i{}^\alpha \delta
\end{eqnarray}
Using theorem \ref{Teorema sa deltom}:
\begin{eqnarray}
I_{ij\beta}
    &=&
        \left( \pi_i{}^\alpha \p_\beta b_{j\alpha} + b_{j\alpha} \p_\beta \pi_i{}^\alpha \right) \delta + b_{j\alpha}\pi_i{}^\alpha \p_\beta \delta \NB
    &=&
        \p_\beta \left( \pi_i{}^\alpha b_{j\alpha} \delta \right) =  \p_\beta \left( \hat{\pi}_{i\B{j}} \delta \right)
\end{eqnarray}
Hence:
\begin{equation}
I_{ij\beta} - I_{ji\beta} = \p_\beta \left[ \left( \hat{\pi}_{i\B{j}} - \hat{\pi}_{j\B{i}} \right) \delta \right] \label{Sdruga I}
\end{equation}

\subsection*{Calculating $II$}
\begin{eqnarray}
II
    &=& a\{ \p_\gamma B_{ij}^{0\gamma}, {\pi'}_k{}^\alpha {T'}^k_{\beta \alpha} - {b'}^k{}_\beta {\p}^{'}_\alpha {\pi'}_k{}^\alpha \} \NB
    &=&
        a{T'}^k_{\beta\alpha} \p_\gamma \{ B_{ij}^{0\gamma},{\pi'}_k{}^\alpha\} - {b'}^k{}_\beta \p_\gamma \p_\alpha^{'} \{ B^{0\gamma}_{ij},{\pi'}_k{}^\alpha \} \NB
    &=&
        -a\p_\alpha^{'}\p_\gamma \left[ {b'}^k{}_\beta  \{ B^{0\gamma}_{ij}, {\pi'}_k{}^\alpha \} \right] + a \p_\gamma \left[ \left( \p_\beta^{'} {b'}^k{}_\alpha \right) \{ B_{ij}^{0\gamma}, {\pi'}_k{}^\alpha \} \right] \label{Sdruga 2}
\end{eqnarray}
Before moving on, it is useful to calculate:
\begin{eqnarray}
    \{ B_{ij}^{0\gamma}, {\pi'}_k{}^\alpha \} &=& \{\varepsilon^{0\gamma\delta\beta}_{ijnm} b^n{}_\delta b^m{}_\beta, {\pi'}_k{}^\alpha \} = 2 \varepsilon^{0\gamma\delta \alpha}_{ijnk} b^n{}_\delta \delta(x-x') \NB
    &=& \left[ B^{0\gamma}_{ij}h_k{}^\alpha + B^{\alpha 0 }_{ij} h_k{}^\gamma + B^{\gamma\alpha}_{ij}h_k{}^0 \right] \delta(x-x')
\end{eqnarray}
where we used $\eqref{G: lema 5}$. Now, the first term in~\eqref{Sdruga 2} is given by:
\begin{equation}
   -a\p_\alpha^{'}\p_\gamma \left[  b^k{}_\beta \{ B_{ij}^{0\gamma}, {\pi'}_k{}^\alpha \} \right]
    = -a\p_\alpha^{'}\p_\gamma \left[ \left( B^{0\gamma}_{ij}\delta^\alpha_\beta - B^{0\alpha}_{ij}\delta^\gamma_\beta \right) \delta(x-x') \right] \label{Sdruga 3}
\end{equation}
On the other hand, in order to calculate the second term in \eqref{Sdruga 2}, we first need to simplify the expression inside the square brackets:
\begin{eqnarray}
     \left( \p_\beta^{'} {b'}^k{}_\alpha \right) \{ B_{ij}^{0\gamma}, {\pi'}_k{}^\alpha \} 
     &=&
        2 \left(\p_\beta b^k{}_\alpha \right) \varepsilon^{0\gamma\alpha\delta}_{ijkn} b^n{}_\delta \delta(x-x')\NB
        &=& 2\p_\beta \left[ \varepsilon^{0\gamma\alpha\delta}_{ijkn} b^k{}_\alpha b^n{}_\delta \right]\delta -\underbrace{2 \varepsilon^{0\gamma\alpha\delta}_{ijkn}b^n{}_\delta \left( \p_\beta b^k{}_\alpha\right) \delta}_{\left( \p_\beta^{'} {b'}^k{}_\alpha \right) \{ B_{ij}^{0\gamma}, {\pi'}_k{}^\alpha \} }
\end{eqnarray}
Hence:
\begin{equation}
 \left( \p_\beta^{'} {b'}^k{}_\alpha \right) \{ B_{ij}^{0\gamma}, {\pi'}_k{}^\alpha \} 
 =\p_\beta \left[ \varepsilon^{0\gamma\alpha\delta}_{ijkn} b^k{}_\alpha b^n{}_\delta \right]\delta = \left(\p_\beta  B^{0\gamma}_{ij} \right) \delta \label{Sdruga 4}
\end{equation}
Substituting the results \eqref{Sdruga 3} and \eqref{Sdruga 4} in \eqref{Sdruga 2}:
\begin{eqnarray}
    II&=& -a\p_\alpha^{'}\p_\gamma \left[ \left( B^{0\gamma}_{ij}\delta^\alpha_\beta - B^{0\alpha}_{ij}\delta^\gamma_\beta \right) \delta(x-x') \right] +a
     \p_\gamma \left[\left(\p_\beta  B^{0\gamma}_{ij} \right)\delta\right] \NB
     &=&
        a\p_\beta \left[ \left( \p_\gamma B^{0\gamma}_{ij} \right) \delta(x-x') \right] \label{Sdruga II}
\end{eqnarray}
Finally, adding \eqref{Sdruga I} and \eqref{Sdruga II} gives
\begin{equation}
    \{ \phi_{ij}, \mc{H'}_\beta \} = \p_\beta \left[ \phi_{ij} \delta(x-x') \right]
    \label{druga puasonova zagrada spec slucaj}
\end{equation}

\section{Poisson Bracket \texorpdfstring{$\{\phi_{ij},\mc{H'}_\perp \}$}{fij,H'}}

In order to make the expressions in this section less cumbersome, it is useful to introduce:

\begin{subequations}
\begin{eqnarray}
B_{ij}        &:=& \p_\alpha B^{0\alpha }_{ij} \\
\mc{H}_{ij}   &:=& \KP_{i\B{j}} - \KP_{j\B{i}} 
\end{eqnarray}
\end{subequations}
Now, if-constraints can be written as $\phi_{ij}    = a B_{ij} + \mc{H}_{ij}$. We can now divide our problem into smaller and easier parts:

\begin{subequations}
\begin{eqnarray}
\{ \phi_{ij}, \mc{H}^{'}_\perp \}
        &=& \{ a B_{ij}, -(n^{k} \p_\beta \pi_k{}^\beta)' \} +
            \{ a B_{ij}, \fr{1}{2} {P^2}' \} - 
            \{ a B_{ij}, (J \B{\mc{L}}_T)' \}  \label{PS za B}\\
        && +\{ \mc{H}_{ij}, -(n^{k} \p_\beta \pi_k{}^\beta)' \} +
            \{ \mc{H}_{ij}, \fr{1}{2} {P^2}' \} - 
            \{ \mc{H}_{ij}, (J \B{\mc{L}}_T)' \} \label{PS za H}
\end{eqnarray}
\end{subequations}
\subsection*{Poisson Brackets Involving $\mathbf{B_{ij}}$}
In this section, we examine Poisson brackets in \eqref{PS za B}. Since $B_{ij}$ depends only on tetrads, it is useful to calculate:

\begin{eqnarray}
\{ B^{0\gamma}_{ij}, \pi_k{}^{\alpha'}\}
                          &=& 
                            2\varepsilon^{0\gamma \delta \alpha}_{ijnk}
                             b^n{}_\delta \delta (x-x') \NN \\
                        &=&\left(
                            B^{0\gamma}_{ij}h_k{}^\alpha +
                            B^{\alpha 0}_{ij}h_k{}^\gamma +
                            B^{\gamma \alpha}_{ij}h_k{}^0
                            \right)
                            \delta(x-x') \NN \\
                        &:=&
                        f^{0\gamma \alpha}_{ijk} \delta(x-x')
\end{eqnarray}
where we defined $f^{0\gamma \alpha}_{ijk}$ as a totally antisymmetric quantity with respect to both upper and lower indices. Now let us examine each term, one by one:
\boldmath
\subsection*{$\{a  B_{ij}, -(n^{k} \p_\beta \pi_k{}^\beta)' \}$}
\unboldmath

\begin{eqnarray}
\{ a B_{ij}, -(n^{k} \p_\beta \pi_k{}^\beta)' \}
&=&
-a {n^k}' \p_\alpha {\p_\beta}' \{ B^{0\alpha}_{ij}, {\pi_k{}^\beta}' \} \NN \\
&=&
- a{n^k}' \p_\alpha {\p_\beta}' 
\left(
     f^{0\alpha \beta}_{ijk} \delta(x-x')
\right) \NN \\
&=&a
 \p_\alpha 
\left[
     \p_\beta \left(
                     n^k \delta(x-x')
             \right)
                    f^{0\alpha  \beta}_{ijk}
\right] \NN \\
&=&
2 a\p_\alpha  
\left[
        \p_\beta \left(
                        n^k \delta(x-x')
                \right)
        \varepsilon^{0\alpha \beta \gamma}_{ijkm} b^m{}_\gamma 
\right] \NN \\
&=&
-a \p_\alpha 
\left[
         n^k \delta(x-x')
        \varepsilon^{0\alpha \beta \gamma}_{ijkm} T^m{}_{\beta  \gamma}
\right] \NN \\
&=& -a\p_\alpha 
\left[
        n^k \delta(x-x')
        \varepsilon^{\alpha 0 \beta \gamma  }_{ijmk} T^m{}_{  \beta \gamma  }
\right]
\end{eqnarray}

\boldmath
\subsection*{$\{ a B_{ij}, \fr{1}{2} {P^2}' \} $}
\unboldmath

\begin{eqnarray}
\{ a B_{ij}, \fr{1}{2} {P^2}' \}
            &=&
                \fr{1}{2J'} \{ B_{ij}, \hat{\pi'}_{\BBS{m}{n}} \} {\hat{\pi'}}^{\BBS{m}{n}} -
                \fr{1}{4J'} \{ B_{ij}, \hat{\pi}^\B{m}{}_\B{m}' \} \hat{\pi}^\B{p}{}_\B{p}' \NN \\
            &=&
                \fr{1}{2J'} \{ B_{ij}, {\pi}_{mn}' \} \hat{\pi}^{\BBS{m}{n}'} -
                \fr{1}{4J'} \eta^\BB{m}{n} \{ B_{ij}, {\pi}_{mn}' \}
                            \hat{\pi}^\B{p}{}_\B{p}' \label{3.2 I}
\end{eqnarray}
where we used $\{ B_{ij}, n^k \} = 0$ which implies $\{ B_{ij}, \delta_\B{k}^\B{m} \} = 0$. Now, from  (\ref{3.2 I}) one can see that our next step is to calculate $\{ B_{ij}, \pi_{mn}' \} $:

\begin{eqnarray}
\{ B_{ij}, \pi_{mn}' \} 
            &=&
                \{ B_{ij}, \pi_m{}^{\mu'} \} b_{n\mu}' \NN \\
            &=& \p_\alpha 
                \left[
                        f^{0\alpha \mu}_{ijm} b_{n\mu } \delta(x-x')
                \right] \NN \\
            &=& \p_\alpha 
                \left[
                \left(
                        B^{0\alpha}_{ij} \eta_{mn} +
                        B^{0\alpha}_{mi} \eta_{jn} +
                        B^{0\alpha}_{jm} \eta_{in} 
                \right)     \delta(x-x')
                \right]
\end{eqnarray}
We can now use this result in (\ref{3.2 I}). The first term becomes:
\begin{subequations}
\begin{equation}
\fr{1}{2J'} \{ B_{ij}, \pi_{mn}' \} \hat{\pi}^{\BBS{m}{n}'}
            =
                \p_\alpha 
                \left[
                \fr{1}{2J}
                \left(
                        B^{0\alpha}_{ij} \hat{\pi}^\B{n}{}_\B{n} +
                        B^{0\alpha}_{mi} \eta_\BB{j}{n} \hat{\pi}^\BBS{m}{n} +
                        B^{0\alpha}_{jm} \eta_\BB{i}{n} \hat{\pi}^\BBS{m}{n}
                \right)     \delta
                \right] \label{3.2 konacni deo I}
\end{equation}
whereas the second one reads:
\begin{equation}
-\fr{1}{4J'} \eta^\BB{m}{n} \{ B_{ij}, \pi_{mn}' \}\hat{\pi}^\B{p}{}_\B{p}'
            =
                -\p_\alpha 
                \left[
                \fr{1}{4J}
                \left(
                        3B^{0\alpha}_{ij} +
                        B^{0\alpha}_{mi} \delta^\B{m}_\B{j} +
                        B^{0\alpha}_{jm} \delta^\B{m}_\B{i}
                \right)
                \hat{\pi}^\B{n}{}_\B{n} \delta
                \right] \label{3.2 konacni deo II}
\end{equation}
Adding  (\ref{3.2 konacni deo I}) and (\ref{3.2 konacni deo II}) gives:
\begin{eqnarray}
\{ a B_{ij}, \fr{1}{2} {P^2}' \}
            &=&
                \p_\alpha 
                \left[
                       \fr{1}{2J}
                       \left( B^{0\alpha}_{mi} \eta_\BB{j}{n} +
                             B^{0\alpha}_{jm} \eta_\BB{i}{n}
                        \right)
                        \hat{\pi}^\BBS{m}{n} \delta
                \right] \label{3.2 releventan deo} \\
             && -\p_\alpha 
                \left[ 
                        \fr{1}{4J}
                        \left(
                                B^{0\alpha}_{ij} +
                                B^{0\alpha}_{mi} \delta^\B{m}_\B{j} +
                                 B^{0\alpha}_{jm} \delta^\B{m}_\B{i}
                        \right)
                       \hat{\pi}^\B{n}{}_\B{n} \delta
                \right] \label{3.2 nula}
\end{eqnarray}
\end{subequations}
It should be noted that (\ref{3.2 nula}) is vanishing because of~(\ref{G nula}), so (\ref{3.2 releventan deo}) is the only nontrivial part that remains. We further simplify it using (\ref{B=2H kompaktno}):

\begin{eqnarray}
\{ a B_{ij}, \fr{1}{2} {P^2}' \}
            &=& -\p_\alpha 
                \left[
                       \fr{1}{J}
                       \left( H^{0\alpha}_{mi} \eta_{jn} +
                             H^{0\alpha}_{jm} \eta_{in}
                        \right)
                        \hat{\pi}^\BBS{m}{n} \delta
                \right] \NN \\
            &=&  -\p_\alpha 
                \left[N
                \left(
                        -\eta_{jn} h_i{}^0 h_m{}^\alpha +
                        \eta_{in} h_j{}^0 h_m{}^\alpha 
                \right) \hat{\pi}^\BBS{m}{n} \delta
                \right] \NN \\
            &=& \p_\alpha 
                \left[
                \left(
                        n_i \hat{\pi}_\BBS{m}{j} - n_j \hat{\pi}_\BBS{m}{i}
                \right)
                h^{\B{m}\alpha } \delta
                \right]
\end{eqnarray}

\boldmath
\subsection*{$\{ a B_{ij}, (J \B{\mc{L}}_T)' \}$}
\unboldmath
Since $J \B{\mc{L}}_T$ does not depend on momenta, we can immediately conclude that:
\begin{equation}
    \{ a B_{ij}, (J \B{\mc{L}}_T)' \} = 0
\end{equation}
If we now take into account all the results that we derived in this section, we get:
\begin{equation}
\{ a B_{ij}, \mc{H}'_\perp \} 
            =  \p_\alpha 
                \left[
                \left(
                        n_i \hat{\pi}_\BBS{m}{j} - n_j \hat{\pi}_\BBS{m}{i}
                \right)
                h^{\B{m}\alpha } \delta
                \right] 
                -
                a\p_\alpha 
                \left[
                        n^k \delta(x-x')
                 \varepsilon^{\alpha 0 \beta \gamma  }_{ijmk} T^m{}_{  \beta \gamma  }
                \right] \label{H sa B rezultat}
\end{equation}

\boldmath
\subsection*{$\text{Poisson Brackets Involving}\;  \mc{H}_{ij}$}
\unboldmath
In this section, we will examine Poisson brackets in (\ref{PS za H}).
 Since $\{ \mc{H}_{ij}, \delta^{\B{k}'}_\B{r} \}$ is an important side result, it is useful to first calculate $\{ \mc{H}_{ij}, n_k' \}$: 

\begin{eqnarray}
\{ \mc{H}_{ij}, n_k' \}
            &=& b_{j\alpha} \{ \pi_i{}^\alpha, n_k' \} -
                b_{i\alpha} \{ \pi_j{}^\alpha, n_k' \} \NN \\
            &=& b_{j\alpha} n_i h_\B{k}{}^\alpha \delta(x-x') -
                b_{i\alpha} n_j h_\B{k}{}^\alpha \delta(x-x') \NN \\
            &=& \left(
                        \eta_\BB{j}{k} n_i - \eta_\BB{i}{k} n_j
                \right) \delta(x-x') \NN \\
            &=& \left(
                        \eta_{jk} n_i - \eta_{ik} n_j
                \right) \delta(x-x')  \label{H sa n}
\end{eqnarray}
where we used $\eta_\BB{i}{j} = \eta_{ij} - n_i n_j$. Now it is clear that:

\begin{eqnarray}
\{ \mc{H}_{ij}, \delta^{\B{k}'}_\B{r} \}
            = \{ \mc{H}_{ij}, \delta^k_r - {n^k}' n'_r \}
            = -\left(
                    \delta^k_j n_r + \eta_{jr} n^k
              \right) n_i \delta(x-x') - (i \leftrightarrow j) \label{H sa delta}
\end{eqnarray}
Later, we will need to calculate $\{ \mc{H}_{ij}, (J \B{\mc{L}}_T)' \}$, so it is necessary to know both $\{\mc{H}_{ij}, T^{'}_{k\BB{m}{n}} \}$ and $\{ \mc{H}_{ij}, J'\}$. Since $T_{k\BB{m}{n}}= h_\B{m}{}^\beta h_\B{n}{}^\gamma \left( \p_\beta b^k{}_\gamma - \p_\gamma b^k{}_\beta \right)$ we obtain here a few side results that will make our work easier. Firstly:

\begin{eqnarray}
\{ \mc{H}_{ij}, b_{k\beta}' \}
            &=&  b_{j\alpha} \{ \pi_i{}^\alpha, b_{k\beta}' \} -
                 b_{i\alpha} \{ \pi_j{}^\alpha, b_{k\beta}' \} \NN \\
            &=& \left(
                        \eta_{ik} b_{i\beta} - \eta_{ik} b_{j\beta}
                \right) \delta(x-x') \label{Hij sa b}
\end{eqnarray}
Secondly:
\begin{eqnarray}
\{ \mc{H}_{ij}, h_\B{k}{}^{\beta '} \}
            &=& \{ \mc{H}_{ij}, \delta^{\B{p}'}_\B{k} \} h_p{}^\beta +
                \{ \mc{H}_{ij}, h_p{}^{\beta '} \} \delta^{\B{p}'}_\B{k} \NN \\
            &=& - \left(
                        h_j{}^\beta n_k + \eta_{jk} \fr{g^{0\beta}}{\sqrt{g^{00}}}
                  \right) n_i \delta +
                  \left(
                        h_i{}^\beta n_k + \eta_{ik} \fr{g^{0\beta}}{\sqrt{g^{00}}}
                  \right) n_j \delta \NN \\
             && + b_{j\alpha} h_\B{k}{}^\alpha h_i{}^\beta \delta -
                  b_{i\alpha} h_\B{k}{}^\alpha h_j{}^\beta \delta \NN \\
            &=& \eta_{jk} 
                \left(
                        h_i{}^\beta - \fr{g^{0\beta}}{\sqrt{g^{00}}} n_i
                \right) \delta -
                 \eta_{ik} 
                \left(
                        h_j{}^\beta - \fr{g^{0\beta}}{\sqrt{g^{00}}} n_j
                \right) \delta \NN \\
            &=& \left(
                        \eta_{jk} h_\B{i}{}^\beta  - \eta_{ik} h_\B{j}{}^\beta 
                \right) \delta
\end{eqnarray}
Thirdly:
\begin{equation}
\{ \mc{H}_{ij}, h_\B{m}{}^{\beta '} h_\B{h}{}^{\gamma '} \}
            = \left(
                    \eta_{jm} h_\B{i}{}^\beta h_\B{n}{}^\gamma +
                    \eta_{jn} h_\B{i}{}^\gamma h_\B{m}{}^\beta
              \right) \delta - (i \leftrightarrow j) \label{H sa hh}
\end{equation}
The last thing we derive here is:
\begin{eqnarray}
\{ \mc{H}_{ij}, J' \}
            &=& b_{j\alpha} \{ \pi_i{}^\alpha, J' \} -
                b_{i\alpha} \{ \pi_j{}^\alpha, J' \} \NN \\
            &=& \left(
                - b_{j\alpha} J h_\B{i}{}^\alpha
                + b_{i\alpha} J h_\B{j}{}^\alpha
                \right) \delta \NN \\
            &=& \left(
                        -\eta_\BB{j}{i} + \eta_\BB{i}{j}
                \right) J \delta = 0
\end{eqnarray}
Before calculating $\{ \mc{H}_{ij}, (J \B{\mc{L}}_T)' \}$, we will first find the other two terms in  (\ref{PS za H}), because it turns out they are much easier but equally relevant:

\boldmath
\subsection*{$\{ \mc{H}_{ij}, -(n^{k} \p_\beta \pi_k{}^\beta)' \}$}
\unboldmath

\begin{eqnarray}
\{ \mc{H}_{ij}, -(n^{k} \p_\beta \pi_k{}^\beta)' \}
            &=& -b_{j\alpha} \p^{'}_\beta \pi_k{}^{\beta'} \{\pi_i{}^\alpha, n^{k'}\} 
                -\pi_i{}^\alpha n^{k'} \{ b_{j\alpha}, \p^{'}_\beta \pi_k{}^{\beta'}\}
                - (i \leftrightarrow j) \NN \\
            &=& -b_{j\alpha} (\p_\beta \pi_k{}^\beta) n_i h^{\B{k}\alpha} \delta +
                \pi_i{}^\beta n_j^{'} \p_\beta \delta  - (i \leftrightarrow j) \NN \\
            &=& -(\p_\beta \pi_j{}^\beta) n_i \delta +
                (\p_\beta \pi_k{}^\beta) n_i n^k n_j \delta +
                \pi_i{}^\beta \p_\beta \left[ n_j \delta \right] 
                - (i \leftrightarrow j) \NN \\
            &=&  -(\p_\beta \pi_j{}^\beta) n_i \delta
                 -\pi_j{}^\beta \p_\beta \left[ n_i \delta \right]
                 +(\p_\beta \pi_i{}^\beta) n_j \delta
                 +\pi_i{}^\beta \p_\beta \left[ n_j \delta \right] \NN \\
            &=& -\p_\beta
                \left[
                \left(
                        \pi_j{}^\beta n_i - \pi_i{}^\beta  n_j 
                \right) \delta
                \right] \NN \\
            &=& -\p_\beta
                \left[
                \left(
                        n_i \hat{\pi}_{j\B{k}} - n_j \hat{\pi}_{i\B{k}}
                \right) h^{\B{k} \beta} \delta
                \right]
\end{eqnarray}

\boldmath
\subsection*{$\{ \mc{H}_{ij}, \fr{1}{2} {P^2}' \}$}
\unboldmath

\begin{equation}
\{ \mc{H}_{ij}, \fr{1}{2} {P^2}' \}
            = \fr{1}{2aJ} \{ \mc{H}_{ij}, \hat{\pi}^{'}_\BB{m}{n} \}\hat{\pi}^\BBS{m}{n}
            -\fr{1}{4aJ} \eta^{mn}  \{ \mc{H}_{ij}, \hat{\pi}^{'}_\BB{m}{n} \} \hat{\pi}^\B{p}{}_\B{p} \label{H sa P2}
\end{equation}
\begin{subequations}
\begin{eqnarray}
\{ \mc{H}_{ij}, \pi_{m\B{n}}' \}
            &=& \pi_i{}^\alpha  b_{n\beta} \{ b_{j\alpha}, \pi_m{}^{\beta '} \} +
                b_{\alpha j} \pi_m{}^\beta \{ \pi_i{}^\alpha, b_{n\beta}' \}
                - (i \leftrightarrow j) \NN \\
            &=& \pi_i{}^\alpha b_{n\alpha } \eta_{jm} -
                b_{j \alpha} \pi_m{}^\alpha \eta_{in}
                -(i \leftrightarrow j) \NN \\
            &=& \left(
                        \eta_{jm} \hat{\pi}_{i \B{n}}
                      - \eta_{in} \hat{\pi}_{m \B{j}}
                \right) \delta -
                \left(
                        \eta_{im} \hat{\pi}_{j\B{n}}-
                        \eta_{jn} \hat{\pi}_{m\B{i}}
                \right) \delta 
\end{eqnarray}

\begin{eqnarray}
\{ \mc{H}_{ij}, \hat{\pi}_{\BB{m}{n}}' \}
            &=&  \{ \mc{H}_{ij}, \hat{\pi}_{k\B{n}}' \} \delta^\B{k}_\B{m} +
                 \{ \mc{H}_{ij}, \delta^{\B{k}'}_\B{m} \} \hat{\pi}_{k\B{n}} \NN \\
            &=& \left(
                        \eta_\BB{j}{m} \hat{\pi}_{i\B{n}} -
                        \eta_{in} \hat{\pi}_\BB{m}{j}
                \right) \delta -
                \left(
                        \eta_\BB{i}{m} \hat{\pi}_{j\B{n}} -
                        \eta_{jn} \hat{\pi}_\BB{m}{i}
                \right) \delta  \NN \\
            && +\left(
                    -n_m n_i \hat{\pi}_\BB{j}{n} + n_m n_j \hat{\pi}_\BB{i}{n}
                    -\eta_\BB{j}{m} n_i \hat{\pi}_{\perp \B{k}}
                    +\eta_\BB{i}{m} n_j \hat{\pi}_{\perp \B{n}}
                \right) \delta \NN \\
            &=& \left(
                        \eta_\BB{j}{m} \hat{\pi}_{i\B{n}} +
                        n_m n_j \hat{\pi}_\BB{i}{n} -
                        \eta_\BB{j}{m} n_i \hat{\pi}_{\perp \B{k}}
                \right) \delta  \NN \\
            &&  + \left(
                        -\eta_\BB{i}{m} \hat{\pi}_{j\B{n}} -
                        n_m n_i \hat{\pi}_\BB{j}{n} +
                        \eta_\BB{i}{m} n_j \hat{\pi}_{\perp \B{n}}
                \right) \delta \NN \\
            && +\left( 
                \eta_{jn} \hat{\pi}_\BB{m}{i} - \eta_{in} \hat{\pi}_\BB{m}{j}
                \right) \delta  \NN \\
            &=& \left(
                        \eta_{jm} \hat{\pi}_\BB{i}{n} -
                        \eta_{in} \hat{\pi}_\BB{m}{j} +
                        \eta_{jn} \hat{\pi}_\BB{m}{i} -
                        \eta_{im} \hat{\pi}_\BB{j}{n}
                \right) \delta \label{H sa piBB}
\end{eqnarray}
Now, it is straightforward to see that:
\begin{eqnarray}
\{ \mc{H}_{ij}, \hat{\pi}^{'}_\BB{m}{n} \}\hat{\pi}^\BBS{m}{n} = 0
\qquad \qquad
\eta^{mn}  \{ \mc{H}_{ij}, \hat{\pi}^{'}_\BB{m}{n} \} = 0
\end{eqnarray}
\end{subequations}
Substituting these into (\ref{H sa P2}) gives:
\begin{equation}
    \{ \mc{H}_{ij}, \fr{1}{2} {P^2}' \} = 0
\end{equation}

\boldmath
\subsection*{$\{ \mc{H}_{ij}, (J \B{\mc{L}}_T)' \}$}
\unboldmath

\begin{equation}
\{ \mc{H}_{ij}, (J \B{\mc{L}}_T)' \}
            = \fr{aJ'}{2} \{\mc{H}_{ij},T^{'}_{k\BB{m}{n}} \} T^{'k\BB{m}{n}}
            +a J' \{\mc{H}_{ij}, T^{'}_\BBB{r}{m}{n} \} T^{'\BBB{m}{r}{n}}
            - 2aJ' \{\mc{H}_{ij}, \B{T}^{'}_\B{k} \} \B{T}^{'\B{k}} \label{H sa L}
\end{equation}
where $\B{T}_\B{k} = T^\B{m}{}_{\B{m}\B{k}}$. We can see that there are three terms that we need to calculate. Using \eqref{Hij sa b} and (\ref{H sa hh}):
\begin{eqnarray}
\{ \mc{H}_{ij}, T'_{k\beta \gamma} \}
            &=& b_{j\alpha} \{ \pi_i{}^\alpha,
                \p^{'}_\beta b'_{k\gamma } - \p^{'}_\gamma b'_{k\beta} \}
                - (i \leftrightarrow j) \NN \\
            &=& -\eta_{ik} 
                \left[ 
                        b_{j\gamma} \p^{'}_\beta \delta -
                        b_{j\beta} \p^{'}_\gamma \delta 
                \right] - (i \leftrightarrow j) \label{H sa T obicno}
\end{eqnarray}

\begin{eqnarray}
\{ \mc{H}_{ij}, T'_{k\BB{m}{n}} \}
            &=& \{ \mc{H}_{ij}, {h'}_\B{m}{}^\beta {h'}_\B{n}{}^\gamma \}
                T'_{k\beta \gamma} +
                \{ \mc{H}_{ij}, T'_{k\beta \gamma} \} 
                {h'}_\B{m}{}^\beta {h'}_\B{n}{}^\gamma \NN \\
            &=& \left[
                \left(
                        \eta_{jm} T_{k\BB{i}{n}} - 
                        \eta_{jn} T_{k\BB{i}{m}}
                \right) \delta -\eta_{ik}
                \left(
                        b_{j\gamma} \p^{'}_\beta \delta - 
                        b_{j\beta} \p^{'}_\gamma \delta 
                \right) {h'}_\B{m}{}^\beta {h'}_\B{n}{}^\gamma
                \right] \NN \\ 
            &&    - (i \leftrightarrow j)
\end{eqnarray}
Hence, the first term in (\ref{H sa L}) is given by:
\begin{subequations}
\begin{eqnarray}
\fr{aJ'}{2} \{\mc{H}_{ij},T^{'}_{k\BB{m}{n}} \} T^{'k\BB{m}{n}}
            &=& \fr{aJ'}{2} 
                \left[ 
                        -T_{k\BB{i}{n}} T^{k\B{n}}{}_\B{j}
                        -T_{k\BB{i}{m}} T^{k\B{m}}{}_\B{j}
                \right] \delta \NN \\
            && -\fr{aJ'}{2} \eta_{ik} 
                \left(
                        b_{j\gamma} \p^{'}_\beta \delta
                        -b_{j\beta} \p^{'}_\gamma \delta 
                \right) 
                {h'}_\B{m}{}^\beta {h'}_\B{n}{}^\gamma  T^{'k\BB{m}{n}} \\
            &&
                - (i \leftrightarrow j)
\end{eqnarray}
\end{subequations}
Note that the first line gives no contribution to the end result because of the symmetry  $ (i \leftrightarrow j)$, whereas the two terms in the second line are equal to each other. Then, we apply theorem \ref{Teorema sa deltom}:

\begin{eqnarray}
\fr{aJ'}{2} \{\mc{H}_{ij},T^{'}_{k\BB{m}{n}} \} T^{'k\BB{m}{n}}
            &=&  a\eta_{ik}b_{j\gamma} \p_\beta
                \left(
                        JT^{k\BB{m}{n}} h_\B{m}{}^\beta  h_\B{n}{}^\gamma \delta
                \right) - (i \leftrightarrow j) \NN \\
            &=& a\p_\beta 
                \left(
                \eta_{ik}b_{j\gamma}  JT^{k\BB{m}{n}} h_\B{m}{}^\beta  
                h_\B{n}{}^\gamma \delta
                \right)  \NN \\
            && - a\eta_{ik} JT^{k\BB{m}{n}} h_\B{m}{}^\beta  
                h_\B{n}{}^\gamma \delta \p_\beta b_{j\gamma}
                -(i \leftrightarrow j) \NN \\
            &=& a \p_\beta 
                \left(
                        JT_{i\BB{m}{j}} h^{\B{m}\beta } \delta 
                \right) -
                \fr{aJ}{2} T_i{}^\BB{m}{n} T_{j\BB{m}{n}} \delta 
                -(i \leftrightarrow j) \NN \\
            &=&a  \p_\beta 
                \left[ J
                \left(
                        T_{i\BB{m}{j}} - T_{j\BB{m}{i}}
                \right)
                h^{\B{m}\beta } \delta
                \right]
\end{eqnarray}
The second term in (\ref{H sa L}) is calculated as follows:
\begin{eqnarray}
\{\mc{H}_{ij}, T^{'}_\BBB{r}{m}{n} \}
            &=& \{\mc{H}_{ij}, \delta^{'\B{k}}_\B{r} \} T^{'}_{k\BB{m}{n}}
                + \{\mc{H}_{ij}, T^{'}_{k\BB{m}{n}} \} \delta^{'\B{k}}_\B{r} \NN \\
            &=&
            \left[
                \left(
                        \eta_{jm} T_\BBB{r}{i}{n} - \eta_{jn} T_\BBB{r}{i}{m}
                \right) \delta 
                -\eta^{'}_\BB{i}{r}
                \left(
                        b_{j\gamma} \p^{'}_\beta \delta 
                        -b_{j\beta} \p^{'}_\gamma  \delta
                \right) {h'}_\B{m}{}^\beta  {h'}_\B{n}{}^\gamma 
                \right] \NN \\
            && -n_i \left( \delta^k_j n_r + \eta_{jr} n^k \right) 
                T_{k\BB{m}{n}} \delta 
                 - (i \leftrightarrow j)       
\end{eqnarray}

\begin{eqnarray}
J' \{\mc{H}_{ij}, T^{'}_\BBB{r}{m}{n} \} T^{'\BBB{m}{r}{n}}
            &=& - n_i \left( T_{j \BB{m}{n}} T^\BBB{m}{r}{n} n_r
                +n^k T_{k \BB{m}{n}} T^\B{m}{}_\B{j}{}^\B{n} \right) J\delta \NN \\
            && + \left( \eta_{jm} T_\BBB{r}{i}{n} T^\BBB{m}{r}{n} 
                -\eta_{jn} T_\BBB{r}{i}{m} T^\BBB{m}{r}{n} \right) J \delta \NN \\
            && + \eta_\BB{i}{r} 
                \left(
                        b_{j\gamma} \p_\beta \delta - b_{j \beta} \p_\gamma \delta 
                \right) h_\B{m}{}^\beta h_\B{n}{}^\gamma T^\BBB{m}{r}{n} J \NN \\
            &&+ b_{j\gamma} \delta \p_\beta 
                \left[
                    \eta_\BB{i}{r} h_\B{m}{}^\beta h_\B{n}{}^\gamma T^\BBB{m}{r}{n} J
                \right] \NN \\
            &&
                -b_{j\beta } \delta \p_\gamma  
                \left[
                    \eta_\BB{i}{r} h_\B{m}{}^\beta h_\B{n}{}^\gamma T^\BBB{m}{r}{n} J
                \right] -
                (i \leftrightarrow j) \NN \\
            &=& n_i T_{\perp \BB{m}{n}} T^{\BB{m}{n}}{}_\B{j} J \delta +
                \left( T_\BBB{r}{i}{n} T_\B{j}{}^\BB{r}{n} 
                    - T_\BBB{r}{i}{m} T^\BB{m}{r}{}_\B{j} \right) J \delta \NN \\
            &&-\p_\beta  \left[ \delta b_{j\gamma} h_\B{m}{}^\beta h_\B{n}{}^\gamma 
                        T^\BB{m}{n}{}_\B{i} J \right]
                +\p_\gamma  \left[ \delta b_{j\beta } h_\B{m}{}^\beta h_\B{n}{}^\gamma 
                        T^\BB{m}{n}{}_\B{i} J \right] \NN \\
            && + (\p_\beta b_{j\gamma})  h_\B{m}{}^\beta h_\B{n}{}^\gamma 
                T^\BB{m}{n}{}_\B{i} J \delta
               - (\p_\gamma  b_{j\beta }) h_\B{m}{}^\beta h_\B{n}{}^\gamma 
                T^\BB{m}{n}{}_\B{i} J \delta \NN \\
            &&    - (i \leftrightarrow j) \NN \\
            &=& n_i T_{\perp \BB{m}{n}} T^{\BB{m}{n}}{}_\B{j} J \delta +
                \left( T_\BBB{r}{i}{n} T_\B{j}{}^\BB{r}{n} 
                    - T_\BBB{r}{i}{m} T^\BB{m}{r}{}_\B{j} \right) J \delta \NN \\
            &&    -\p_\beta  \left[ J \delta h^{\B{m}\beta} \left( T_\BBB{m}{j}{i} +
                    T_\BBB{j}{i}{m} \right) \right] + 
                    T_{j \BB{m}{n}} T^\BB{m}{n}{}_\B{i} J \delta 
                    - (i \leftrightarrow j) \NN \\
            &=& \left(
                        n_i T_{\perp \BB{m}{n}} T^\BB{m}{n}{}_\B{j}
                        +n_j T_{\perp \BB{m}{n}} T^\BB{m}{n}{}_\B{i}
                \right) J \delta 
                - T_\BBB{r}{i}{m} T^\BB{m}{r}{}_\B{j} J \delta  \NN \\
            && -\p_\beta  \left[ J \delta h^{\B{m}\beta} \left( T_\BBB{m}{j}{i} +
                    T_\BBB{j}{i}{m} \right) \right]
                     - (i \leftrightarrow j) \NN \\
            &=& -\p_\beta  \left[ J \delta h^{\B{m}\beta} \left( T_\BBB{m}{j}{i} +
                    T_\BBB{j}{i}{m} \right) \right]
                     - (i \leftrightarrow j)
\end{eqnarray}
The third term in (\ref{H sa L}) can be obtained similarly:
\begin{eqnarray}
\{\mc{H}_{ij}, \B{T}^{'}_\B{n} \}
            &=& \eta^{rm} \{\mc{H}_{ij}, T^{'}_\BBB{r}{m}{n} \} \NB
            &=& \left( T_\BBB{j}{i}{n} - \eta_{jn} T^\B{m}{}_\BB{i}{m}
                \right) \delta - \eta^{'}_\BB{i}{r}
                \left(
                        b_{j\gamma} \p^{'}_\beta \delta -
                        b_{j\beta } \p^{'}_\gamma \delta
                \right) h^{'\B{r}\beta } h^{'}_\B{n}{}^\gamma \NB
            && - n_i \left[
                            \delta^k_j n^m + \delta^m_j n^k
                    \right] T_{k\BB{m}{n}} \delta  - (i \leftrightarrow j)
\end{eqnarray}

\begin{eqnarray}
\{\mc{H}_{ij}, \B{T}^{'}_\B{n} \} J' \B{T}^{'\B{n}}
            &=&  T_\BBB{j}{i}{n} \B{T}^\B{n} J \delta
             +  b_{j\gamma} \p_\beta 
              \left[
                    \B{T}^\B{n} h_\B{i}{}^\beta h_\B{n}{}^\gamma J \delta
                \right] 
            -  b_{j\beta } \p_\gamma 
              \left[
                    \B{T}^\B{n} h_\B{i}{}^\beta h_\B{n}{}^\gamma J \delta
                \right] \NB 
            && -n^i T_{\perp \BB{j}{n}} \B{T}^\B{n} J \delta 
                - (i \leftrightarrow j) \NB
        &=& T_\BBB{j}{i}{n} \B{T}^\B{n} J \delta
            +\p_\beta \left[ \B{T}_\B{j} h_\B{i}{}^\beta J \delta \right]
            -\p_\gamma \left[ \eta_\BB{j}{i} \B{T}^\B{n} h_\B{n}{}^\gamma J \delta 
                        \right] \NB
        &&
            -(\p_\beta b_{j\gamma}) \B{T}^\B{n}  h_\B{i}{}^\beta h_\B{n}{}^\gamma J \delta
            +(\p_\gamma  b_{j\beta }) \B{T}^\B{n}  h_\B{i}{}^\beta h_\B{n}{}^\gamma J \delta \NB
        && -n^i T_{\perp \BB{j}{n}} \B{T}^\B{n} J \delta 
                - (i \leftrightarrow j)
\end{eqnarray}
The third term in the first line of the last equality gives no contribution to the result because of the symmetry  $(i \leftrightarrow j)$. Two terms in the second line of the same equality are proportional to the torsion tensor, which is written as $-T_{j\BB{i}{n}}=-T_\BBB{j}{i}{n} - n_j T_{\perp \BB{i}{n}}$:
\begin{eqnarray}
\{\mc{H}_{ij}, \B{T}^{'}_\B{n} \} J' \B{T}^{'\B{n}}
            &=& \p_\beta \left[ \B{T}_\B{j} h_\B{i}{}^\beta J \delta \right] -
                \left(
                        n_i T_{\perp \BB{j}{n}} +  n_j T_{\perp \BB{i}{n}}
                \right) \B{T}^\B{n} J \delta - (i \leftrightarrow j) \NB
            &=& \p_\beta \left[ \B{T}_\B{j} h_\B{i}{}^\beta J \delta \right]
                - (i \leftrightarrow j)
\end{eqnarray}
Now, we combine all of the previously obtained results:
\begin{eqnarray}
\{ \mc{H}_{ij}, (J \B{\mc{L}}_T)' \}
            &=&  a\p_\beta 
                \left[ J
                \left(
                        T_{i\BB{m}{j}} - T_{j\BB{m}{i}}
                \right)
                h^{\B{m}\beta } \delta
                \right] \\
               && -a\p_\beta  \left[ J \delta h^{\B{m}\beta} \left( T_\BBB{m}{j}{i} +
                T_\BBB{j}{i}{m} - T_\BBB{m}{i}{j} -T_\BBB{i}{j}{m} \right) \right] \NB
             && -2a\p_\beta \left[ \B{T}_\B{j} h_\B{i}{}^\beta J \delta \right]
                 +2\p_\beta \left[ \B{T}_\B{i} h_\B{j}{}^\beta J \delta \right] \NB
            &=& -a\p_\beta  \left[ J \left( n_i T_{\perp \BB{j}{m}}
                                        -n_j T_{\perp \BB{i}{m}} \right)
                                        h^{\B{m}\beta} \delta 
                            \right] \NB
            && -2 a\p_\beta \left[ J \left( T_\BBB{m}{j}{i}
                                    - \eta_\BB{m}{j} \B{T}_\B{i}
                                    + \eta_\BB{m}{i} \B{T}_\B{j}
                                      \right)  h^{\B{m}\beta} \delta 
                            \right]
\end{eqnarray}
where we used $T_{i\BB{m}{j}}=T_\BBB{i}{m}{j} + n_i T_{\perp \BB{m}{j}}$. Finally:
\begin{align}
 \{ \mc{H}_{ij}, \mc{H}'_\perp \}
            =&a  \p_\beta  \left[ J \left( n_i T_{\perp \BB{j}{m}}
                                        -n_j T_{\perp \BB{i}{m}} \right)
                                        h^{\B{m}\beta} \delta 
                            \right] \NB
            +& 2a \p_\beta \left[ J \left( T_\BBB{m}{j}{i}
                                    - \eta_\BB{m}{j} \B{T}_\B{i}
                                    + \eta_\BB{m}{i} \B{T}_\B{j}
                                      \right)  h^{\B{m}\beta} \delta 
                            \right] \NB
            -&                a \p_\beta
                \left[
                \left(
                        n_i \hat{\pi}_{j\B{k}} - n_j \hat{\pi}_{i\B{k}}
                \right) h^{\B{k} \beta} \delta
                \right]
                            \label{H sa L rezultat}
\end{align}

\boldmath
\subsection*{$\{  \phi_{ij}, \mathcal{H^{'}}_\perp\}$}
\unboldmath
Let us now combine the results from (\ref{H sa B rezultat}) and (\ref{H sa L rezultat}):

\begin{eqnarray}
\{ \phi_{ij}, \mc{H}^{'}_\perp \}
            &=& \{a B_{ij} , \mc{H}^{'}_\perp \} +
                \{ \mc{H}_{ij}, \mc{H}^{'}_\perp \} \NB
            &=&  \p_\alpha 
                \left[
                \left(
                        n_i \hat{\pi}_\BBS{m}{j} - n_j \hat{\pi}_\BBS{m}{i}
                \right)
                h^{\B{m}\alpha } \delta
                \right] 
                -
                a\p_\alpha 
                \left[
                        n^k \delta(x-x')
                \varepsilon^{\alpha 0 \beta \gamma  }_{ijmk} T^m{}_{  \beta \gamma  }
                \right] \NB
            && +a\p_\beta  \left[ J \left( n_i T_{\perp \BB{j}{m}}
                                        -n_j T_{\perp \BB{i}{m}} \right)
                                        h^{\B{m}\beta} \delta 
                            \right] \NB
            && +2a \p_\beta \left[ J \left( T_\BBB{m}{j}{i}
                                    - \eta_\BB{m}{j} \B{T}_\B{i}
                                    + \eta_\BB{m}{i} \B{T}_\B{j}
                                      \right)  h^{\B{m}\beta} \delta 
                            \right] \NB
            && -\p_\beta
                \left[
                \left(
                        n_i \hat{\pi}_{j\B{k}} - n_j \hat{\pi}_{i\B{k}}
                \right) h^{\B{k} \beta} \delta
                \right] \label{rez 1}
\end{eqnarray}
This result can be further simplified. By combining the first and fifth term in (\ref{rez 1}) one finds:
\begin{multline}
 \p_\alpha 
                \left[
                \left(
                        n_i \hat{\pi}_\BBS{m}{j} - n_j \hat{\pi}_\BBS{m}{i}
                \right)
                h^{\B{m}\alpha } \delta
                \right] 
                -\p_\beta
                \left[
                \left(
                        n_i \hat{\pi}_{j\B{k}} - n_j \hat{\pi}_{i\B{k}}
                \right) h^{\B{k} \beta} \delta
                \right] = \\
        = -\fr{1}{2} \p_\beta 
            \left[ 
                    h^{\B{k}\beta}\delta \left( n_i \mc{H}_{jk} -
                                                n_j \mc{H}_{ik}
                                        \right)
            \right]
\end{multline}
On the other hand, the second and fourth term in (\ref{rez 1}) cancel each other out because of (\ref{Posledica}), while the third term can be rewritten using (\ref{PomRel6}):
\begin{equation}
 a\p_\beta  \left[ J \left( n_i T_{\perp \BB{j}{m}}
                                        -n_j T_{\perp \BB{i}{m}} \right)
                                        h^{\B{m}\beta} \delta 
                            \right]
            =-a\p_\beta  \left[ 
            \fr{1}{2} \left( n_i B_{jk} - n_j B_{ik} \right) h^{\B{k}\beta  }
               \delta \right]
\end{equation}
Combining everything yields:
\begin{eqnarray}
\{ \phi_{ij}, \mc{H}^{'}_\perp \}
            &=& -\fr{1}{2} \p_\beta \left[ h^{\B{k}\beta} \delta \left[
                    n_i \left(a  B_{jk} +\mc{H}_{jk} \right) -
                    n_j \left( aB_{ik} +\mc{H}_{ik} \right) \right] \right] \NB
            &=& -\fr{1}{2} \p_\beta \left[ 
                \left(
                        n_i \phi_{jk} - n_j \phi_{ik}
                \right) h^{\B{k}\beta} \delta 
                \right] \label{treca puasonova zagrada spec slucaj}
\end{eqnarray}
\subsection*{Constraint Algebra}
Here are all the Poisson brackets that include if-constraints:

\begin{subequations} \label{algebra veza dodatne veze}
\begin{empheq}[box=\widefbox]{align}
\{\phi_{ij},{\phi'}_{kl} \} &= 
 \left( \eta_{ik} \phi_{lj} + \eta_{jk}\phi_{il} \right) \delta - (k \leftrightarrow l) \\
  \{ \phi_{ij}, \mc{H'}_\beta \} &= \p_\beta \left[ \phi_{ij} \delta(x-x') \right] \\
 \{ \phi_{ij}, \mc{H}^{'}_\perp \}   &= -\fr{1}{2} \p_\beta \left[ 
                \left(
                        n_i \phi_{jk} - n_j \phi_{ik}
                \right) h^{\B{k}\beta} \delta 
                \right]
\end{empheq}
\end{subequations}
Once again,  we emphasize that the constraint algebra \eqref{algebra veza genericki} of TG is still valid here. Thus, we found all six nontrivial Poisson brackets in TEGR. One can see that each Poisson bracket is a linear combination of constraints. Therefore, all 14 constraints $\pi_i{}^0, \mc{H}_\alpha, \mc{H}_\perp, \phi_{ij}$ are first-class quantities. Since the total Hamiltonian is also a constraint\footnote{Up to a total divergence term, which is irrelevant here, as we noted earlier.}, it follows that the consistency conditions for $\phi_{ij}$ are identically fulfilled.
\subsection*{Number of Degrees of Freedom}

Using \eqref{Broj stepeni slobode}, and the fact that $b^k{}_\mu$ has $16$ independent components, one can conclude that the number of degrees of freedom is:

\begin{equation} 
    \text{number of degrees of freedom} = 2\cdot 16 - 2 \cdot 14 = 4
\end{equation}
That is the number of degrees of freedom in the phase space, whereas the corresponding number in the configurational space is  $4/2=2$. This is in agreement with the GR. In the next chapter, we will examine the construction of the generator of gauge symmetries, and find the gauge transformations of the canonical variables.

\chapter{Generators of Gauge Symmetries}
\label{glava konstrukcija generatora}
In this chapter, we will construct the generators of gauge symmetries in the case of TG and TEGR. After that, we will examine the gauge transformations of canonical variables $b^k{}_\mu$ and $\pi_k{}^\mu$.

\section{Gauge Generator in the Case of TG}
\subsection*{Constructing the Generators}
As we saw in~\S \ref{Kastelani}, there is a bijective correspondence between the primary first-class constraints and gauge generators. In this case, the only primary first-class constraints are $\pi_i{}^0$, so we use them as a starting point in Castellani's algorithm\footnote{ $G_{i}^k = \pi_i{}^0$ is also a valid starting point.}

\begin{equation}
    G_{i}^k = -\pi_i{}^0
\end{equation}
Index $i$ runs over primary first-class constraints, whereas $k$ represents the number of steps in Castellani's algorithm, which is so far unknown. The next step is calculating the Poisson bracket between $G^k_i$ and the total Hamiltonian. One can simplify this a little bit by noticing that $\{ \pi_i{}^0, H_T \} = \{ \pi_i{}^0, H_c\}$, up to a first-class constraint\footnote{This is sufficient for Castellani's algorithm \eqref{Kastelanijev algoritam}.}

\begin{equation} \label{generator medjukorak 1}
\{  G_{i}^k , \mc{H}_T \} = n_i \mc{H}_\perp +h_\B{i}{}^\beta \mc{H}_\beta \equiv \mc{H}_i
\end{equation}
where we defined $\mc{H}_i$. This is not a primary first-class constraint, so we will continue the algorithm by calculating $G^{k-1}_i$:
\begin{equation}
G^{k-1}_i = V_{\text{PFC}} -  \{  G_{i}^k , \mc{H}_T \}   
\end{equation}
Using the fact that every primary first-class constraint can be written as a linear combination of $\pi_i{}^0$, we can write the previous expression as:
\begin{equation}
G_i^{k-1} (x)= -\int dy \;\alpha_i{}^m (x,y) \pi_m{}^0 (y) - \mc{H}_i(x)
\end{equation}
where $\alpha_i{}^m$ are yet to be determined.\par\null\par

Now, we need to see if the Poisson bracket between $G_i^{k-1}$ and the total Hamiltonian is a primary first-class constraint. As before, it is justified to use the canonical instead of the total Hamiltonian:
\begin{equation}\label{generator medjukorak 2}
\{G_i^{k-1}, \mc{H}_c \} = \alpha_i{}^m \mc{H}_m - \{\mc{H}_i, \mc{H}_c \}
\end{equation}
From the definition of  $\mc{H}_i$ \eqref{generator medjukorak 1} it is easy to prove that $\mc{H}_c = b^m{}_0 \mc{H}_m$. Using that in the previous expression gives:
\begin{equation} \label{generator medjukorak 4}
\{G_i^{k-1}, \mc{H}_c \} = \alpha_i{}^m \mc{H}_m -   b^m{}_0 \{ \mc{H}_i, \mc{H}_m \}  
\end{equation}
In the second term, we use the definition of $\mc{H}_i$:
\begin{equation} \label{generator medjukorak 3}
\{ \mc{H}_i, \mc{H}_m \}=  \{ n_i \mc{H}_\perp +h_\B{i}{}^\alpha \mc{H}_\alpha, n_m \mc{H}_\perp + h_\B{m}{}^\beta \mc{H}_\beta \} 
\end{equation}
Expanding this expression gives us 16 new terms. Some of them are vanishing $\{n_i, n_m \} = \{n_i, h_\B{m}{}^\beta \} = \{ h_\B{i}{}^\alpha, h_\B{m}{}^\beta \} = 0$. Others can be obtained using \eqref{algebra veza genericki} and:
\begin{subequations}
\begin{eqnarray}
\{n_i, \mc{H'}_\beta \} &=& - \left( \p_\beta b^n{}_\alpha \right) n_n h_\B{i}{}^\alpha\delta \\ 
\{ h_\B{i}{}^\alpha, \mc{H'}_\beta \} &=& \left(\p_\beta b^n{}_\gamma \right) \left(n_n n_i {}^3g^{\alpha\gamma} - h_\B{i}{}^\gamma h_\B{n}{}^\alpha \right) \delta - \delta^\alpha_\beta h_\B{i}{}^\gamma \p_\gamma \delta \\
\{ n_i, \mc{H'}_\perp \} &=& - \left(T_{\perp \B{i}\perp}+ h_\B{i}{}^\alpha \p_\alpha \right) \delta \\
\{h_\B{i}{}^\alpha, \mc{H'}_\perp \} &=& -h^{\B{m}\alpha} \left( h_\B{i}{}^\beta \p_\beta n_m +T_{\BB{m}{i}\perp} - n_i T_{\perp \B{m}\perp }\right) \delta + {}^3g^{\alpha\beta} n_i \p_\beta \delta 
\end{eqnarray}
\end{subequations}
Hence, \eqref{generator medjukorak 3} can be rewritten as:
\begin{equation}
  \{ \mc{H}_i, \mc{H}_m \}= T^n{}_{im} \mc{H}_n \delta  
\end{equation}
Substituting this result in \eqref{generator medjukorak 4} gives:
\begin{equation}
    \{G^{k-1}_i, \mc{H}_c \} = \alpha_i{}^n \mc{H}_n - b^m{}_0 T^n{}_{im} \mc{H}_n \delta
\end{equation}
We can now see that it is possible to end Castellani's algorithm if we choose  $\alpha_i{}^n =  b^m{}_0 T^n{}_{im}$. Thus, we conclude that $k=1$, so the corresponding gauge generator can be written as:
\begin{empheq}[box=\widefbox]{align} \label{GENERATOR 1}
    G = - \int d^3 x \left[   \Dot{\varepsilon}^i \pi_i{}^0 + \varepsilon^i \left(\mc{H}_i + b^m{}_0 T^n{}_{im} \pi_n{}^0  \right)   \right]
\end{empheq}

\subsection*{Gauge Transformations of $\boldsymbol{b^k{}_\mu}$ и $\boldsymbol{\pi_k{}^\mu}$}
It turns out that it is useful to replace $\varepsilon^k$ in \eqref{GENERATOR 1} with the new  gauge parameter $\xi^\mu$,  defined as:
\begin{equation}
    \varepsilon^k = \xi^\mu b^k{}_\mu
\end{equation}
Hence, the gauge generator \eqref{GENERATOR 1} can be written as:
\begin{eqnarray}
G &=& -\int d^3 x \; \left[ \p_0 \left( \xi^\mu b^k{}_\mu \right) \pi_k{}^0 + \xi^\mu \left( b^k{}_\mu \mc{H}_k + b^k{}_\mu b^m{}_0 T^i{}_{km} \pi_i{}^0 \right)\right] \NB
&=&
    -\int d^3 x \; \left[ \Dot{\xi}^\mu b^k{}_\mu \pi_k{}^0 + \xi^\mu P_\mu                                  \right] \label{generator drugi oblik}
\end{eqnarray}
where
\begin{equation}
P_\mu := b^k{}_\mu \mc{H}_k + b^k{}_\mu b^m{}_0 \pi_i{}^0 T^i{}_{km} + \pi_k{}^0 \p_0 b^k{}_\mu = b^k{}_\mu \mc{H}_k + \pi_i{}^0 \p_\mu b^i{}_0
\end{equation}
On shell, $P_0$ is equal to the total Hamiltonian up to a total divergence:
\begin{equation} \label{velicina P0}
    P_0 = b^k{}_0 \mc{H}_k + \pi_i{}^0 \p_0 b^i{}_0 = \mc{H}_c + \pi_i{}^0 \p_0 b^i{}_0 = \mc{H}_T - \p_\alpha D^\alpha
\end{equation}
which follows from:
\begin{equation*}
    \p_0 b^i{}_0 = \{ b^i{}_0, \mc{H}_T \} = \{ b^i{}_0, \mc{H}_c + u^k \pi_k{}^0 \} = u^i
\end{equation*}
On the other hand:
\begin{equation}
P_\alpha = b^k{}_\alpha \mc{H}_k + \pi_i{}^0 \p_\alpha b^i{}_0 = \mc{H}_\alpha + \pi_i{}^0 \p_\alpha b^i{}_0 = \pi_i{}^\mu \p_\alpha b^i{}_\mu - \p_\beta \left( \pi_i{}^\beta b^i{}_\alpha \right)
\end{equation}
Now it is straightforward to derive the gauge transformations of $b^k{}_\mu$:
\begin{eqnarray}
\delta_0 b^k{}_\mu &=& \{ b^k{}_\mu, G  \} = -\{ b^k{}_\mu, \int dx' \; \left[ \Dot{\xi'}^\nu {b'}^i{}_\nu {\pi'}_i{}^0 + {\xi'}^\nu {P'}_\nu \right] \} \NB
&=&
    -\Dot{\xi}^\nu b^k{}_\nu \delta^0_\mu - \int dx' \; {\xi'}^0 \{ b^k{}_\mu, {P'}_0 \} - \int dx' \;   {\xi'}^\alpha \{ b^k{}_\mu, {P'}_\alpha \} \label{generator medjukorak 5}
\end{eqnarray}
The second and the third term are given by:
\begin{subequations} \label{generator medjukorak 6}
\begin{equation}
 - \int dx' \; {\xi'}^0 \{ b^k{}_\mu, {P'}_0 \} = - \int dx' \; {\xi'}^0 \{ b^k{}_\mu, \mc{H'}_T - {\p'}_\alpha {D'}^\alpha \} = -{\xi}^0 \Dot{b}^k{}_\mu - b^k{}_0 \delta^\beta_\mu \p_\beta \xi^0
\end{equation}
\begin{eqnarray}
 - \int dx' \;   {\xi'}^\alpha \{ b^k{}_\mu, {P'}_\alpha \} &=&  - \int dx' \;   {\xi'}^\alpha \{ b^k{}_\mu, {\pi'}_i{}^\mu \p^{'}_\alpha {b'}^i{}_\mu - \p^{'}_\beta \left( {\pi'}_i{}^\beta {b'}^i{}_\alpha \right) \} \NB
 &=&
    -\xi^\alpha \p_\alpha b^k{}_\mu  - b^k{}_\alpha \delta^\beta_\mu \p_\beta \xi^\alpha
\end{eqnarray}
\end{subequations}
Gauge transformation of tetrads is thus obtained by substituting \eqref{generator medjukorak 6} into \eqref{generator medjukorak 5}:
\begin{empheq}[box=\widefbox]{align}
 \delta_0 b^k{}_\mu = -b^k{}_\rho \p_\mu \xi^\rho - \xi^\rho \p_\rho b^k{}_\mu   \label{kalibraciona transf b}
\end{empheq}
The procedure for obtaining the gauge transformations of the momenta $\pi_k{}^\mu$ is very similar:
\begin{eqnarray}
 \delta_0 \pi_k{}^\mu &=& 
 -\int dx' \; \{ \pi_k{}^\mu, \Dot{\xi'}^\nu {b'}^i{}_\nu {\pi'}_i{}^0 \} - \int dx' \; {\xi'}^0 \{  \pi_k{}^\mu, {P'}_0 \} - \int dx' \;   {\xi'}^\alpha \{  \pi_k{}^\mu, {P'}_\alpha \} \NB
 &=&
    \Dot{\xi}^\mu \pi_k{}^0 - \xi^0\p_0 \pi_k{}^\mu + \delta^\mu_0 \left( \p_\alpha \xi^0 \right) \pi_k{}^\alpha - \p_\alpha\left( \xi^\alpha \pi_k{}^\mu \right) + \left(\p_\beta \xi^\alpha \right) \pi_k{}^\beta \delta^\mu_\alpha \NB
&=&
    \left( \p_\nu \xi^\mu \right) \pi_k{}^\nu - \xi^\nu \p_\nu \pi_k{}^\mu - \left(\p_\alpha\xi^\alpha \right) \pi_k{}^\mu
\end{eqnarray}
Hence:
\begin{empheq}[box=\widefbox]{align}
 \delta_0 \pi_k{}^\mu = \left( \p_\nu \xi^\mu \right) \pi_k{}^\nu - \xi^\nu \p_\nu \pi_k{}^\mu - \left(\p_\alpha\xi^\alpha \right) \pi_k{}^\mu \label{kalibraciona transf pi}
\end{empheq}
It is very important to note that gauge transformations \eqref{kalibraciona transf b} are actually local translations. Thus, we conclude that TG can be interpreted as a gauge theory of translations. In the next section, we will see that TEGR has additional gauge symmetries, which unfortunately do not have such a clear geometric interpretation.

\section{Gauge Generators in the Case of TEGR}
As noted earlier, beside the sure constraints, TEGR also has if-constraints. From \eqref{algebra veza dodatne veze} we concluded that those constraints are first-class quantities. According to Castellani's algorithm, there should also exist the corresponding gauge symmetries. Before we start constructing these new generators, it is necessary to clarify one important issue that arises with the emergence of these new primary constraints.\par\null\par

Procedure for constructing the gauge generator in TG is no longer valid in TEGR, even though $\pi_i{}^0$ are still primary first-class constraints. TG and TEGR look similar, but these theories do not even have the same total Hamiltonians. This is in itself a sufficient indicator that should make us doubt the validity of the previously obtained generator in this new case. Luckily, there is no need to repeat the whole procedure from the previous section because we know that the gauge transformations  \eqref{kalibraciona transf b} and \eqref{kalibraciona transf pi} are still valid\footnote{This can be directly checked regardless of the construction procedure.}. Thus, one can easily guess the form of the generator which gives the correct gauge transformations. From the derivation in the previous section, we can see that such a generator is given by \eqref{generator drugi oblik}, where\footnote{Here, $\mc{H}_T$ is the total Hamiltonian of TEGR, not TG! } $P_0 = \mc{H}_T - \p_\alpha D^\alpha$. In other words, a new gauge generator (in the case of TEGR) corresponding to $\pi_i{}^0$ can be obtained from the old gauge generator (in the case of TG) by formally substituting the old total Hamiltonian with the new one. Let us now examine a gauge generator that corresponds to  $\phi_{ij}$:

\subsection*{Castellani's Algorithm for $\phi_{ij}$}
The starting point in Castellani's algorithm is:
\begin{equation}
    G_{ij}^k = \phi_{ij}
\end{equation}
Now, we have to check if the Poisson bracket between $ G_{ij}^k$ and the total Hamiltonian $\mc{H}_T = \mc{H}_c + u^k \pi_k{}^0 + \frac{1}{2} u^{kl} \phi_{kl} $ is a primary first-class constraint. In this case:  $\{ \phi_{ij}, \mc{H}_T \} = \{ \phi_{ij}, \mc{H}_c \} = \{ \phi_{ij}, N\mc{H}_\perp \} + \{ \phi_{ij}. N^\beta \mc{H}_\beta \} $. We already calculated $\{ \phi_{ij}, \mc{H}_\perp \}$ in \eqref{treca puasonova zagrada spec slucaj}. Similarly, we can obtain:

\begin{equation*}
    \{ \phi_{ij}, N \mc{H}_\perp \} = -\frac{1}{2} \p_\beta \left[ N \left( n_i \phi_{jk} - n_j \phi_{ik} \right) h^{\B{k}\beta} \delta \right]
\end{equation*}
Also,  $\{ \phi_{ij}, \mc{H}_\beta \}$ was calculated in \eqref{druga puasonova zagrada spec slucaj}. Similarly:
\begin{equation*}
    \{\phi_{ij}, N^\beta \mc{H}_\beta \} = \p_\beta \left(N^\beta \phi_{ij} \delta \right)
\end{equation*}
Hence:
\begin{equation}
 \{ \phi_{ij}, \mc{H}_T \} =     -\frac{1}{2} \p_\beta \left[ N \left( n_i \phi_{jk} - n_j \phi_{ik} \right) h^{\B{k}\beta} \delta \right] +  \p_\beta \left(N^\beta \phi_{ij} \delta \right)
\end{equation}
The right-hand side represents the primary first-class constraint. As a consequence, Castellani's procedure immediately ends. Thus, the gauge generator is given by:
\begin{empheq}[box=\widefbox]{align}
    G =\frac{1}{2} \int d^3 x \;  \varepsilon^{ij} \phi_{ij}
\end{empheq}
\subsection*{Gauge Transformations of $\boldsymbol{b^k{}_\mu}$ и $\boldsymbol{\pi_k{}^\mu}$}
Gauge transformations of tetrads are obtained straightforwardly:
\begin{eqnarray}
\delta_0 b^k{}_\mu &=&
\{ b^k{}_\mu, G'\} = \fr{1}{2}\int dx' \; \varepsilon^{ij} \{ b^k{}_\mu, {\pi'}_i{}^\alpha {b'}_{j\alpha} - {\pi'}_j{}^\alpha {b'}_{i\alpha} \}= \varepsilon^{ki} b_{i\alpha} \delta^\alpha_\mu
\end{eqnarray}
This can be written as:
\begin{empheq}[box=\widefbox]{align}
    \delta_0 b^k{}_0 = 0; \quad \delta_0 b^k{}_\alpha = \varepsilon^{ki}b_{i\alpha} \label{delovanje generatora spec b}
\end{empheq}
Let us now examine gauge transformations of momenta:
\begin{equation}
    \delta_0 \pi_i{}^\mu  = \int dx' \; \{ \pi_k{}^\mu, a{B'}_{ij} + \mc{H'}_{ij} \}\fr{1}{2}{\varepsilon'}^{ij} \equiv \delta_0^{(B)}\pi_k{}^\mu + \delta_0{}^{(H)}\pi_k{}^\mu \label{generator poslednji medjurezultat}
\end{equation}
where 
\begin{eqnarray}
\delta_0^{(B)}\pi_k{}^\mu 
    &=&
        \fr{1}{2}\int dx' \; \{ \pi_k{}^\mu,a {B'}_{ij} \} {\varepsilon'}^{ij} \NB
    &=&
         \fr{a}{2}\int dx' \;{\varepsilon'}^{ij} \varepsilon^{0\alpha\beta\gamma}_{ijlm}\p^{'}_\alpha \{ \pi_k{}^\mu, {b'}^l{}_\beta {b'}^m{}_\gamma \} \NB
    &=&
        a\varepsilon^{0\alpha\beta\gamma}_{ijkm} b^m{}_\gamma \p_\alpha \varepsilon^{ij}
\end{eqnarray}

\begin{eqnarray}
\delta_0^{(H)}\pi_k{}^\mu
    &=&
        \fr{1}{2}\int dx' \; {\varepsilon'}^{ij} \{ \pi_k{}^\mu, {\mc{H'}}_{ij}\}\NB
    &=&
         \fr{1}{2}\int dx' \; {\varepsilon'}^{ij} \{ \pi_k{}^\mu,{\pi'}_i{}^\alpha{b'}_{j\alpha} - {\pi'}_j{}^\alpha{b'}_{i\alpha} \}\NB
    &=&
        -\delta^\mu_\alpha \varepsilon^i{}_k \pi_i{}^\alpha
\end{eqnarray}
Substituting these results in \eqref{generator poslednji medjurezultat} gives:
\begin{equation}
    \delta_0 \pi_k{}^\mu = a\varepsilon^{0\alpha\mu\gamma}_{ijkm} b^m{}_\gamma \p_\alpha \varepsilon^{ij} -\delta^\mu_\alpha \varepsilon^i{}_k \pi_i{}^\alpha
\end{equation}
This result can be rewritten as:
\begin{empheq}[box=\widefbox]{align}
    \delta_0 \pi_k{}^0 = 0; \quad \delta_0 \pi_k{}^\beta = a\varepsilon^{0\alpha\beta\gamma}_{ijkm} b^m{}_\gamma \p_\alpha\varepsilon^{ij} - \varepsilon^i{}_k \pi_i{}^\beta 
    \label{aaaa}
\end{empheq}
Gauge transformations of tetrads~\eqref{delovanje generatora spec b} look like Lorentz transformations,\newline but (\ref{aaaa}) shows clear differences.\par\null\par

We know that gauge transformations should preserve the constraints, $\delta_0 \phi_{ij} \approx 0$. In other words, gauge transformations map the points from the hypersurface $\Gamma$ (picture \ref{hip}) into (not necessarily the same) points on $\Gamma$. This property can be used as a test (necessary but not sufficient) to check our results.

\begin{equation}
\delta_0 \phi_{ij} = \{\phi_{ij}, {G'} \} =\fr{1}{2} \int dx'\; \{\phi_{ij}, {\phi'}_{kl} \} \varepsilon^{kl} \approx 0 
\end{equation}
where we used the fact that $\phi_{ij}$ are first-class constraints. A similar analysis can be repeated for the other generator (and other constraints). This completes the canonical analysis of this paper. We could expand it by adding the calculation of conserved charges, but that is beyond the scope of this paper.

\chapter*{Conclusion}
\addcontentsline{toc}{chapter}{Conclusion}

In this paper, we have:

\begin{itemize}
    \item derived many identities in ADM basis, which we have heavily used throughout this paper.
    \item proved that TEGR is equivalent to GR.
   \item reviewed the canonical analysis of constrained systems and then applied it to the case of electrodynamics. 
    \item applied the canonical formalism in the case of TG and TEGR. In particular, we've constructed the corresponding canonical Hamiltonians, found all the constraints that exist in the theories, and then calculated the Poisson brackets between them. Thus obtained constraint algebra was used to classify the constraints themselves and to calculate the gauge symmetry generator using Castellani's algorithm. Within the framework of this analysis, we've also obtained that the number of degrees of freedom in the configuration space of TG is 8, while in the case of TEGR it is 2, as expected, having in mind its equivalence to GR.
    \item found the gauge transformations of $b^k{}_\mu$ and $\pi_k{}^\mu$. We've concluded that local translations are symmetries of both TG and TEGR. Besides, TEGR has additional symmetries (and constraints) compared to TG, and consequently fewer degrees of freedom than TG. Although these additional symmetries are very reminiscent of Lorentz transformations, they do not have a standard geometrical interpretation. The additional symmetries could have gone unnoticed within Lagrange's formalism. This is one of the benefits of the canonical formalism. Moreover, we are sure that we have found all gauge symmetries because the formalism itself requires finding all the constraints, and by applying Castellani's algorithm we can obtain all gauge symmetries.
\end{itemize}
Thus, we have given a complete analysis of the canonical structure of TG and TEGR. In the future, this paper can be extended by adding the calculation of conserved charges. Besides, the obtained results provide a step forward towards a better understanding of $f(\mc{L}_T)$ theories of gravity.
 
\subsection*{Acknowledgment}
I would like to thank Dr Milutin Blagojević for carefully reading this text. His useful suggestions have made this paper much better and clearer.
\newpage

\chapter*{}
 \nocite{*} 
\clearpage
\appendix
\renewcommand{\thechapter}{}
\titleformat{\chapter}[display]
   {\normalfont\huge\bfseries\centering}{\centering\chaptertitlename\ \thechapter}{20pt}{\Huge}
\titlespacing*{\chapter} 
   {0pt}{20pt}{20pt}
\chapter*{Appendix} 
\markboth{APPENDIX}{}
\addcontentsline{toc}{chapter}{Appendix}
\renewcommand{\thesection}{A} 
\renewcommand{\theequation}{\thesection.\arabic{equation}}
\newtheorem{LemaD}{Lemma}[section]
\newtheorem{PosledicaD}{Consequence}[section]
\section{The dependence of the ADM variables on \texorpdfstring{$\boldsymbol{b^k{}_0}$}{bk0}} \label{DodatakB}

The main goal of this section is to prove:

\begin{Teorema} \label{TeoremaZavisnosti} \normalfont
 $\quad$
 \begin{itemize}
     \item  $n^k, h_{\Bar{k}}{}^{ \mu}, \mathcal{H}_\alpha,\mathcal{H}_\perp,J=\frac{b}{N}$ do not depend on $b^k{}_{ 0}$
     \item  $N,N^\alpha,b$ linearly depend on $b^k{}_{0}$.
 \end{itemize}
 \end{Teorema}
 \noindent
 We will divide the proof into lemmas in order to make it clearer.
  \begin{LemaD} \normalfont
 $^3 h_a{}^{ \alpha} \equiv h_a{}^{ \alpha}-\frac{h_a{}^{ 0} h_0{}^{ \alpha}}{h_0{}^{ 0}}$ and $b^a{}_{ \beta}$ are inverses of each other. In other words $^3 h_a{}^{ \alpha} b^a{}_{\beta}=\delta^\alpha _\beta$ 
 \end{LemaD}
 
 \begin{Dokaz} 
  \normalfont
 \begin{equation}
     ^3 h_a{}^{ \alpha} b^a{}_{\beta}=(h_a{}^{ \alpha}-\frac{h_a{}^{ 0} h_0{}^{ \alpha}}{h_0{}^{ 0}})b^a{}_{ \beta}=
     h_k{}^{\alpha}b^k{}_{\beta}-h_0{}^{\alpha} b^0{}_{ \beta}-\frac{h_0{}^{\alpha}}{h_0^0}(h_k{}^{ 0} b^k{}_{\beta} -h_0{}^{ 0}b^0{}_{\beta})=\delta^\alpha_\beta
 \end{equation}
  \flushright
 \qedsymbol
 \end{Dokaz}
 
 \begin{LemaD} \normalfont
If we define a new quantity $h_a$ as $h_a \equiv \frac{h_a{}^{ 0}}{h_0{}^{ 0}}$, then: $h_a=-^3h_a{}^{\alpha}b^0{}_{\alpha}$
 \end{LemaD}
 
 \begin{Dokaz} 
 \normalfont
 \begin{equation}
     ^3h_a{}^{\alpha}b^0{}_{\alpha}=
     (h_a{}^{ \alpha}-\frac{h_a{}^{ 0} h_0{}^{ \alpha}}{h_0{}^{ 0}})b^0{}_{\alpha}=
     h_a{}^{ \mu}b^0{}_{\mu}-h_a{}^{ 0}b^0{}_{ 0}-\frac{h_a{}^{0}}{h_0{}^{ 0}}(h_0{}^{\mu} b^0{}_{\mu}-h_0{}^{ 0}b^0{}_{ 0})=
     -\frac{h_a{}^{ 0}}{h_0{}^{0 }}
 \end{equation}
 \noindent
  \flushright
 \qedsymbol
 \end{Dokaz}
 \noindent
 Using these two lemmas, it is now obvious that $^3h_a{}^{\alpha} \text{and } h_b$ do not depend on $b^k{}_{ 0}$
 \begin{PosledicaD}
 \normalfont
 From $n_k \propto h_k{}^{ 0}\propto \frac{h_k{}^{ 0}}{h_0{}^{0}}$ we conclude that $n\propto (1,h_b)\equiv l$. Using the fact that $n$ is a unit vector we get $n=\frac{l}{\sqrt{l \cdot l}}$. We proved that $h_b$ does not depend on  $b^k{}_{ 0}$. That means that $n^k$ also cannot depend on $b^k{}_{ 0}$.
 
 \end{PosledicaD}
 
 \begin{LemaD} \normalfont
 $h_{\Bar{k}}{}^{ \mu}$ does not depend on $b^k{}_{ 0}$
 \end{LemaD}
 \begin{Dokaz}
 \normalfont
 Let us define a new quantity $^3 h_l {}^{ \mu}=h_l{}^{ \mu}-\frac{h_l{}^{ 0}h_0{}^{ \mu}}{h_0{}^{0}}$. This quantity is  equal to ${}^3 h_a{}^\alpha$ for $l=a$ and $\mu=\alpha$. Otherwise, (for  $l=0$ or $\mu=0$) it is vanishing. Taking into account that we already proved that ${}^3 h_a{}^\alpha$ is independent of $b^k{}_{ 0}$, we can draw the same conclusion for  $^3 h_l {}^{ \mu}$. Further:

 \begin{eqnarray}
 (\delta^l_k- n^l n_k) ^3 h_l{}^{\mu}&=&
 h_k{}^{ \mu}-\frac{h_k{}^{0} h_0{}^{\mu}}{h_0{}^{0}}-n_k n^l h_l{}^{ \mu}+n^l n_k \frac{h_l{}^{ 0} h_0{}^{\mu}}{h_0{}^{0}} \NB
 &=& h_k{}^{ \mu}-n_k n^l h_l {}^{\mu}=\delta^{\Bar{l}}_k h_l{}^{\mu}=h_{\Bar{k}}{}^{ l}
 \end{eqnarray}
 Using the fact that the left-hand side is independent of $b^k{}_{ 0}$ we can draw the same conclusion for the right-hand side $h_{\Bar{k}}{}^{ l}$.
 \flushright
 \qedsymbol
 \end{Dokaz}
 
 \begin{PosledicaD} \normalfont
 Since $N=n_k b^k {}_{ 0};\; N^\alpha =h_{\Bar{k}}{}^{ \alpha}b^k{}_{ 0}$ we conclude that  $N$ and $N^\alpha$ linearly depend on $b^k{}_{ 0}$.
 \end{PosledicaD}
 
 \begin{LemaD} \label{B: lema 4}
 $b:=detb^k{}_{\mu}=\frac{N}{n_0}detb^a{}_{\alpha} \equiv J N$
 \end{LemaD}
 \begin{Dokaz}
 \normalfont
 Let us introduce a new notation in which $\overline{A}$ represents the matrix which is obtained by erasing the first row and the first column from the matrix A. Inverse matrix ($A^{-1}$) elements will be denoted as $c_{ij}$. Then, from the inverse matrix theorem, we know that $c_{11}=\frac{det \overline{A}}{detA}$. It turns out to be more useful to write that result as $detA=\frac{det\overline{A}}{c_{11}}$. In the case of matrix $b^k{}_{\mu}$ the previous formula gives us  $b=\frac{det b^a{}_{\alpha}}{h_0{}^{ 0}}$. If we now take into account that $N=\frac{1}{\sqrt{g^{00}}},\quad n_0=\frac{h_0{}^{ 0}}{\sqrt{g^{00}}}$ it is easy to see that lemma \ref{B: lema 4} holds.
 
  \flushright
  \qedsymbol
 \end{Dokaz}
 
 \begin{PosledicaD}
 \normalfont
 We can now see that  $b$ linearly depends on $b^k{}_{ 0}$. That means that $J$ is independent of $b^k{}_{ 0}$ because it can be written as 
  $J=\frac{det b^a{}_{\alpha}}{n_0}$. Using all of the results obtained so far, as well as (\ref{ H beta}) it becomes obvious that $\mathcal{H}_\beta$ does not depend on $b^k{}_{ 0}$. The same conclusion can be made for $\mathcal{H}_\perp$ using:

 \begin{equation}
     \mathcal{H}_\perp=\pi_i{}^{\Bar{m}}T^i{}_{\perp \Bar{m}}-n^i \partial_\alpha \pi_i{}^{\alpha}-J\mathcal{L}_{T}
 \end{equation}
 \end{PosledicaD}
\noindent
Thus, the proof of the theorem is completed.
\renewcommand{\thesection}{B}
\section{Jacobi's Theorem and Its Consequences} \label{DodatakD}
\renewcommand{\theequation}{\thesection.\arabic{equation}}

\begin{Teorema} \label{determinanta}
\normalfont
Let $A \in \mathbb{C}^{n \times n}$ be a nonsingular matrix. Let $I$ and $J$ be two subsets of the set  $N = \{1,2...n\}$, such that $|I|=|J|$. Then:
\begin{equation} \label{teorema G1}
    \mathrm{det}\left( A[I,J] \right) = (-1)^{\sum I + \sum J} \cdot \mathrm{det}A \cdot \mathrm{det} \left( (A)^{-1}[J_c,I_c] \right)
\end{equation}
where:
\begin{itemize}
   \item $I_c = N / I$
   \item  $A[I,J]$ is a  $|I| \times |J|$  matrix, which can be obtained by erasing all rows indexed by $I_c$ and by erasing all columns indexed by $J_c$.
    \item  $\sum I$ is a sum of all elements in $I$.
    
\end{itemize}
\end{Teorema}
\noindent
This is sometimes referred to as Jacobi's theorem. We will first prove it in the special case $I=J=\{1...k\}$ and then we will extend our proof to the general case:

\begin{LemaD} \label{ G 1 lema} \normalfont
Theorem \eqref{determinanta} is valid for $I=J=\{1...k\}$, $k<n$.
\end{LemaD}
\begin{Dokaz}
\normalfont
We will introduce a new notation in which $A_i$ denotes $i-$th column of the matrix A, while $e_i$ denotes the column whose elements are all zeros except the $i$th element which is 1. Then it is obvious that:
\begin{equation*}
    A \cdot [e_1  \; e_2\; ... \;e_k\; (A^{-1})_{k+1}\; ... \;(A^{-1})_n ] = [A_1 \;... \;A_k\; e_{k+1}\; ... \;e_n]
\end{equation*}
Taking the determinant of each side gives us $\mathrm{det} A \cdot \mathrm{det} \left(A^{-1}[J_c,I_c] \right) = \mathrm{det} A[I,J]$. Thus, the lemma is proved.
\flushright
\qedsymbol
\end{Dokaz}
Now, we will start proving the general case. It is important to notice that there always exist permutation matrices $C,R$, such that $A[I,J] = (CAR)[K,K]$, $K=\{1...k\}$, where $k=|I|=|J|$. Lemma \ref{ G 1 lema} is applicable to $(CAR)[K,K]$, so:
\begin{eqnarray*}
\mathrm{det}\left(A[I,J]\right) &=& \mathrm{det}\left(CAR[K,K]\right) = \mathrm{det}\left(CAR\right) \cdot   \mathrm{det}\left( \left(CAR\right)^{-1}[K_c,K_c]\right) \\
&=&(-1)^{\sigma(C) + \sigma(R)} \cdot \mathrm{det}A \cdot \mathrm{det}\left( \left(R^{-1}A^{-1}C^{-1}\right)[K_c,K_c]\right) \\
&=& (-1)^{\sigma(C) + \sigma(R)} \cdot \mathrm{det}A \cdot \mathrm{det} \left( A^{-1} [J_c,I_c] \right)
\end{eqnarray*}
where $\sigma(C)$ is a parity of the permutation  $C$. It is easy to see that $(-1)^{\sigma(C)+\sigma(R)} = (-1)^{\sum I + \sum J}$. That completes proof of theorem \ref{determinanta}.\par\null\par

\subsubsection*{Determinant of a submatrix}

Formula $\mathrm{det} A \equiv a = -\frac{1}{n!} \varepsilon^{j_1...j_n}\varepsilon_{i_1...i_n}a^{i_1}{}_{j_1} ... a^{i_n}{}_{j_n}$ is valid for every matrix $A \in  \mathbb{C}^{n \times n}$. It is easy to show that a similar formula is also valid for the determinant of a submatrix\footnote{The summation is not performed over $m_i$ and $n_j$.}

\begin{equation} \label{determinanta podmatrice}
a^{m_1...m_k}_{n_1...n_k} = -\frac{(-1)^{\sum I + \sum J} \cdot (-1)^{\sigma(m_1 ... m_k)}\cdot (-1)^{\sigma(n_1 ... n_k)}}{(n-k)!} \varepsilon^{m_1...m_k j_1 ...} \varepsilon_{n_1 ... n_k i_1 ...} a^{i_1}{}_{j_1} a^{i_2}{}_{j_2} ...   
\end{equation}
where  $a^{m_1...m_k}_{n_1...n_k}$  is the determinant\footnote{Determinant with a sign! We take into account the permutations of both upper and lower indices.} of a matrix obtained from $A$ by erasing rows $m_1...m_k$  and columns $ n_1 ... n_k$. Now, using \eqref{teorema G1} and \eqref{determinanta podmatrice} we obtain:
\begin{equation}
 -\frac{ (-1)^{\sigma(m_1 ... m_k) + \sigma(n_1 ... n_k)}}{(n-k)!} \varepsilon^{m_1...m_k j_1 ...} \varepsilon_{n_1 ... n_k i_1 ...} a^{i_1}{}_{j_1} a^{i_2}{}_{j_2} ...   = \mathrm{det}A \cdot \mathrm{det} \left( A^{-1} [J_c, I_c] \right)   \label{G: relacija}
\end{equation}
where $I_c = \{ m_1, ... m_k\},\quad J_c = \{ n_1, ... n_k\}$. \par\null\par

\noindent
Even though  \eqref{G: relacija} looks cumbersome, consequences of that relation are especially important to us:
\subsubsection*{Consequences of \eqref{G: relacija}}
\begin{enumerate}
    \item If we use $b^k{}_\mu$ as our matrix $A$ in \eqref{G: relacija} then its inverse of transpose is $h_k{}^\mu$, while its determinant is $b$. Using $I_c = \{ i, j\},\quad J_c = \{ \mu, \nu \}$, relation \eqref{G: relacija} becomes:
    \begin{equation}
        -\fr{1}{2} \varepsilon^{\mu\nu\rho\sigma}_{ijmn} b^m{}_\rho b^n{}_\sigma = b \left( h_i{}^\mu h_j{}^\nu - h_i{}^\nu h_j{}^\mu \right) \label{B=2H}
    \end{equation}
    Sometimes it is useful to introduce a new notation $B^{\mu\nu}_{ij} =  \varepsilon^{\mu\nu\rho\sigma}_{ijmn} b^m{}_\rho b^n{}_\sigma$, $H_{ij}^{\mu \nu} = b \left( h_i{}^\mu h_j{}^\nu - h_i{}^\nu h_j{}^\mu \right)$. Now, \eqref{B=2H} can be written as:
    \begin{equation} \label{B=2H kompaktno}
        2H_{ij}^{\mu\nu} = -B^{\mu\nu}_{ij}
    \end{equation}
    
    \item Equation \eqref{G: relacija}  is very easy to apply up to a sign. However, determining the sign can be tiresome. That is why we will now derive another useful result using a different method to obtain the sign. \newline 
    As earlier, we will apply \eqref{G: relacija} to  $b^k{}_\mu$, but in this case, we will take the submatrix to be $3 \times 3$ (not $2 \times 2$ as earlier):
    \begin{equation}
        \fr{\kappa}{6b} \epsilon^{\mu\nu\rho\sigma}_{ikmn} b^k{}_\nu b^m{}_\rho b^n{}_\sigma =   h_i{}^\mu
    \end{equation}
    where $|\kappa|=1$, but for now we don't know if $k$ is $+1$ or $-1$. By imposing a condition that  $h_i{}^\mu b^j{}_\nu = \delta^j_i$ we see that $\kappa = -1$:
    \begin{equation}
         h_i{}^\mu=-\fr{1}{6b} \epsilon^{\mu\nu\rho\sigma}_{ikmn} b^k{}_\nu b^m{}_\rho b^n{}_\sigma 
    \end{equation}
    
    \item Using the same reasoning as earlier we get:
    \begin{equation} \label{b preko h}
        b^i{}_\mu = -\fr{b}{6} \varepsilon^{ijkl}_{\mu\nu\rho\sigma}h_j{}^\nu h_k{}^\rho h_l{}^\sigma
    \end{equation}
\end{enumerate}

\renewcommand{\thesection}{C}
\section{Useful relations for calculating Poisson brackets} 
\label{DodatakC}
\renewcommand{\theequation}{\thesection.\arabic{equation}}

This section has two parts. The first part contains relations that are used in TG, whereas the second part contains relations that are used in TEGR. 
\subsection*{Useful relations in TG}
The proof of the next few lemmas will rely on the following relation:
\begin{equation}
\tau^\mu{}_{\nu}=bT^\mu{}_{\nu}+\Delta^\mu{}_{\nu}= H_k{}^{ \mu\gamma} T^k{}_{\nu\gamma}-NJ\mathcal{L_T}\delta^\mu_\nu-b^k{}_{\nu}\partial_\gamma H_k{}^{ \mu \gamma } \label{V1}
\end{equation}
which represents an equivalent form of~\eqref{TAU}.
\begin{LemaD}
\begin{equation}
    \tau^0{}_{ \alpha }=\mathcal{H}_\alpha =JT_{\perp \alpha }+\Delta^0{}_{\alpha }
\end{equation}
\end{LemaD}

\begin{Dokaz}
\normalfont
From~\eqref{V1} we can see that:
    \begin{equation}
        \tau^0{}_{\alpha }=\pi_k{}^{\gamma}T^k{}_{\alpha \gamma }-b^k{}_{\alpha }\partial_\gamma \pi_k{}^{\gamma}=\mathcal{H}_\alpha  \nonumber
    \end{equation}
where we used that $H_k{}^{0\nu}=\pi_k{}^\nu$. Returning to~\eqref{V1} again:
    \begin{equation}
        \mathcal{H}_\alpha =\tau^0{}_{\alpha }=bT^0{}_{\alpha }+\Delta^0{}_{\alpha }=\frac{b}{N}T_{\perp \alpha }+\Delta^0{}_{\alpha }=JT_{\perp \alpha }+\Delta^0{}_{\alpha }  \nonumber
    \end{equation}
    \flushright
    \qedsymbol
\end{Dokaz}

\begin{LemaD}
\begin{equation}
    \tau^0{}_{\perp}=\mathcal{H}_\perp=J T_{\perp \perp}+\Delta^0{}_{\perp}
\end{equation}
\end{LemaD}

\begin{Dokaz}
\normalfont
Using~\eqref{V1} gives us:
\begin{eqnarray}
\tau^0{}_{\perp}&=&\pi_k{}^{\gamma }T^k{}_{\perp \gamma }-NJg^{0\perp}\mathcal{L_T}-b^{k\perp}\partial_\gamma \pi_k{}^{\gamma } \NB
&=&\hat{\pi}_k{}^{\Bar{m}}T^k{}_{\perp \Bar{m}}-J\mathcal{L_T}-n^k\partial_\gamma \pi_k{}^{\gamma} \NB
&=&
JT_{\perp \perp}+\Delta^0{}_{\perp}=\mathcal{H}_\perp \nonumber
\end{eqnarray}
If we now use~\eqref{V1} again:
 \begin{equation}
     \mathcal{H}_\perp=\tau^0{}_{\perp}=bT^{0\perp}+\Delta^0{}_{\perp}=\frac{b}{N}T_{\perp\perp}+\Delta^{0}{}_{\perp}=JT_{\perp\perp}+\Delta^0{}_{\perp}  \nonumber
 \end{equation}
 \flushright
 \qedsymbol
\end{Dokaz}

\begin{LemaD}
\begin{equation}
   T^k{}_{\perp \Bar{l}}=J\frac{\partial T_{\perp \perp}}{\partial \hat{\pi}_k{}^{ \Bar{l}}}
\end{equation}
\end{LemaD}

\begin{Dokaz}
\normalfont
From the definition of $T^\mu{}_\nu$ we get:
\begin{equation} \label{T ort ort}
    T_{\perp \perp}=\frac{1}{b}H_k{}^{\perp \gamma }T^k{}_{\perp \gamma }-g^{\perp \perp}\mathcal{L}=\frac{1}{J}\pi_k{}^{\gamma }T^k{}_{\perp \gamma }-\mathcal{L_T}  
    \end{equation}
    \begin{eqnarray}
      \frac{\partial T_{\perp \perp}}{\partial \hat{\pi}_p{}^{\Bar{l}}}&=&\frac{1}{J}\frac{\partial \pi_k{}^{\gamma}}{\partial \hat{\pi}_p{}^{\Bar{l}}}T^k{}_{\perp \gamma }+\frac{1}{J} \pi_k{}^{\gamma}\frac{\partial T^k{}_{\perp \gamma }}{\partial\hat{\pi}_p{}^{\Bar{l}}}-\frac{\partial\mathcal{L_T}}{\partial\hat{\pi}_p{}^{\Bar{l}}} \NB
      &=&\frac{1}{J}T^p{}_{\perp \Bar{l}}+\frac{1}{J}{\pi}_k{}^{\gamma } \frac{\partial T^k{}_{\perp \gamma}}{\partial \hat{\pi}_p{}^{\Bar{l}}}-\frac{\partial \mathcal{L_T}}{\partial T^k{}_{\perp \gamma}} \frac{\partial T^k{}_{\perp\gamma}}{\partial\hat{\pi}_p{}^{\Bar{l}}}\NB
      &=& \frac{1}{J} T^p{}_{\perp \Bar{l}}+\frac{1}{J}\pi_k{}^{\gamma}
      \frac{\partial T^k{}_{\perp\gamma}}{\partial\hat{\pi}_p{}^{\Bar{l}}}-
      \frac{1}{J}\pi_k{}^{\gamma}\frac{\partial T^k{}_{\perp\gamma}}{\partial\hat{\pi}_p{}^{\Bar{l}}}=\frac{1}{J} T^p{}_{\perp \Bar{l}}  \nonumber
      \end{eqnarray} 
      \flushright
      \qedsymbol
\end{Dokaz}

\begin{LemaD} \normalfont
The Legendre identities hold:
\begin{equation}
    \frac{\partial \mathcal{L}}{\partial T^i{}_{ \Bar{k}\Bar{m}}}=-\frac{\partial T_{\perp \perp}}{\partial T^i{}_{ \Bar{k}\Bar{m}} } \quad
\end{equation}
\end{LemaD}

\begin{Dokaz}
\normalfont
Using (\ref{T ort ort}) to express  $T_{\perp \perp}$:
      \begin{equation}
          \frac{\partial T_{\perp\perp}}{\partial T^i{}_{\Bar{p}\Bar{s}}}=\frac{1}{J}\pi_k{}^{\gamma}\frac{\partial T^k{}_{\perp \gamma}}{\partial T^i{}_{ \Bar{p}\Bar{s}}}-\frac{\partial\mathcal{L_T}}{\partial T^i{}_{\Bar{p}\Bar{s}}}=-\frac{\partial\mathcal{L_T}}{\partial T^i{}_{\Bar{p}\Bar{s}}}
      \end{equation}
      because $\frac{\partial T^k{}_{\perp\gamma}}{\partial T^i{}_{\Bar{p}\Bar{s}}}\sim \delta^{\Bar{p}}_{\perp}=\delta^k_l n^l \delta^{\Bar{p}}_k=\delta^k_l n^l (\delta^p_k-n^p n_k)=n^p-n^p=0$
      \flushright
      \qedsymbol
\end{Dokaz}
\subsection*{Useful relations in TEGR}
\begin{LemaD}
\begin{equation} \label{G: lema 5}
\varepsilon_{imnp}^{\mu\lambda \omega\theta} b^i{}_\mu = \fr{1}{2} \left[ B_{mn}^{\lambda \omega}h_p{}^\theta + B_{pm}^{\lambda\omega}h_n{}^\theta + B_{np}{}^{\lambda \omega} h_m{}^\theta \right] \equiv \fr{1}{2} f_{mnp}^{\lambda\omega \theta}
\end{equation}\end{LemaD}

\begin{Dokaz}
\normalfont
Relation \eqref{b preko h}  can be equivalently written as:
\begin{equation*} 
    b^i{}_\mu = \fr{1}{24} \varepsilon^{ijkl}_{\mu\nu\rho\sigma} B^{\nu\rho}_{jk}h_l{}^\sigma
\end{equation*}
Now it is straightforward to obtain that:
\begin{equation*} 
    \varepsilon_{imnp} b^i{}_\mu = \fr{1}{24} \varepsilon_{imnp}\varepsilon^{ijkl} \varepsilon_{\mu\nu \rho\sigma} B^{\nu\rho}_{jk} h_l{}^\sigma = -\fr{1}{12} \varepsilon_{\mu\nu \rho\sigma}f^{\nu\rho\sigma}_{mnp}
\end{equation*}
If we notice that $f^{\nu\rho\sigma}_{mnp}$ is antisymmetric with respect to both upper and lower indices, we can complete the proof simply by multiplying each side of the previous equation by $\varepsilon^{\nu\lambda\omega\theta}$.
\flushright
\qedsymbol
\end{Dokaz}
\noindent
Using \eqref{G: lema 5} one can now obtain a few additional relations: 

\begin{eqnarray} \label{G: relacija 1}
\varepsilon^{0\alpha\beta \gamma}_{ijkl} b^l{}_\gamma b_{n\beta} &=& \fr{1}{2}\left( \eta_{kn}B^{0\alpha}_{ij} + \eta_{jn}B^{0\alpha}_{ki} + \eta_{in}B^{0\alpha}_{jk}\right) \\
\varepsilon^{0\alpha\beta \gamma}_{ijkl} b^l{}_\gamma b^k{}_\delta &=& \fr{1}{2}\left( \delta^\beta_\delta B^{0\alpha}_{ij} + \delta^\alpha_\delta B^{\beta 0 }_{ij} \right) \\
0&=&  B^{0\alpha}_{ij} + 
    B^{0\alpha}_{ki} \delta^\B{k}_\B{j} + 
    B^{0\alpha}_{jk}\delta^\B{k}_\B{i}  \label{G nula}
\end{eqnarray}
\begin{LemaD} \label{PomRel6}
\normalfont
\begin{equation}
P_{ij}{}^\alpha :=
    \fr{1}{2}N
            \left(
                     n_i B_{jk} -n_j B_{ik}
            \right)
    h^{\B{k}\alpha} = 
    -b \left(
             n_i T_{\perp \BB{j}{k}} - n_j T_{\perp \BB{i}{k}}
        \right)
    h^{\B{k}\alpha}
\end{equation}
where $B_{ij}:= \p_\alpha B^{0\alpha }_{ij}$.
\end{LemaD}
\begin{Dokaz}
\normalfont
Using lemma \ref{Lema 1.5}, as well as the equation \eqref{B=2H kompaktno}, it is easy to see that:
\begin{equation}
 B_{ij}= -2J \left(T_{\perp \BB{i}{j}} - n_i \B{T}_\B{j} + n_j \B{T}_\B{i}\right)   \label{G: medjurezultat}
\end{equation}

Using \eqref{G: medjurezultat} it is now straightforward to calculate:
\begin{eqnarray}
P_{ij}{}^\alpha &=&
    -JN \left(
                n_i T_{\perp \BB{j}{k}} - n_i n_j \B{T}_\B{k} + n_k n_i \B{T}_\B{j} - 
                n_j T_{\perp \BB{i}{k}} + n_i n_j \B{T}_\B{k} - n_j n_k \B{T}_\B{i}
        \right) h^{\B{k}\alpha} \NN \\
                &=& -b
        \left(
                n_i T_{\perp \BB{j}{k}}- 
                n_j T_{\perp \BB{i}{k}}
        \right) h^{\B{k}\alpha}
\end{eqnarray}
\flushright
\qedsymbol
\end{Dokaz}

\begin{LemaD}
\begin{equation}
\varepsilon^{\alpha 0\beta \gamma}_{ijmk} T^m{}_{\beta \gamma}
=
h_i{}^\alpha B_{kj} - h_j{}^\alpha B_{ki} + h_k{}^\alpha B_{ji} -
2J h_m{}^\alpha \left(
                        n_i T^m{}_\BB{j}{k} + 
                        n_k T^m{}_\BB{i}{j} +
                        n_j T^m{}_\BB{k}{i}
                \right)
\end{equation}
\end{LemaD}
\begin{Dokaz}
\normalfont

There are many ways to express a determinant of any matrix. We will use one of those ways in the case of inverse tetrad matrix:
\begin{eqnarray}
h \varepsilon^{\alpha 0\beta \gamma}_{ijmk} 
&=&
-2h_i{}^\alpha \left(
                        h_j{}^0h_{[m}{}^\beta h_{k]}{}^\gamma +
                        h_m{}^0h_{[k}{}^\beta h_{j]}{}^\gamma +
                        h_k{}^0h_{[j}{}^\beta h_{m]}{}^\gamma 
                \right) \NN \\
&&
+2h_j{}^\alpha \left(
                        h_i{}^0h_{[m}{}^\beta h_{k]}{}^\gamma +
                        h_m{}^0h_{[k}{}^\beta h_{i]}{}^\gamma +
                        h_k{}^0h_{[i}{}^\beta h_{m]}{}^\gamma 
                \right) \NN \\
&&
+2h_k{}^\alpha \left(
                        h_i{}^0h_{[j}{}^\beta h_{m]}{}^\gamma +
                        h_j{}^0h_{[m}{}^\beta h_{i]}{}^\gamma +
                        h_m{}^0h_{[i}{}^\beta h_{j]}{}^\gamma 
                \right)  \NN \\
&&
-2h_m{}^\alpha \left(
                        h_i{}^0h_{[j}{}^\beta h_{k]}{}^\gamma +
                        h_j{}^0h_{[k}{}^\beta h_{i]}{}^\gamma +
                        h_k{}^0h_{[i}{}^\beta h_{j]}{}^\gamma 
                \right)
\end{eqnarray}
Let us multiply each side by $T^m{}_{\beta \gamma}$:
\begin{subequations}
\begin{eqnarray}
\varepsilon^{\alpha 0\beta \gamma}_{ijmk}  T^m{}_{\beta \gamma}
&=&
-2J h_i{}^\alpha \left(
                        n_j T^m{}_{mk} + n_m T^m{}_{kj} + n_k T^m{}_{jm}
                \right) \label{DL3 1red} \\
&&
+ 2J h_j{}^\alpha \left(
                        n_i T^m{}_{mk} + n_m T^m{}_{ki} + n_k T^m{}_{im}
                \right) \label{DL3 2red} \\
&&
+ 2J h_k{}^\alpha \left(
                        n_i T^m{}_{jm} + n_j T^m{}_{mi} + n_m T^m{}_{ij}
                \right) \label{DL3 3red} \\
&&
- 2J h_m{}^\alpha \left(
                        n_i T^m{}_{jk} + n_j T^m{}_{ki} + n_k T^m{}_{ij}
                \right) \label{DL3 4red}
\end{eqnarray}
\end{subequations}
The second line can be obtained from the first line, up to a sign, by substituting $(i \leftrightarrow j)$. Similarly, the third line can be obtained from the first line, up to a sign, by substituting $(i \leftrightarrow k) $. Therefore, to calculate the first three rows it is enough to calculate only the first, while the last row must be considered separately. \newline 
Let us now calculate (\ref{DL3 1red}), taking into account that $T_{\perp kj} = T_{\perp \BB{k}{j}} + n_k T_{\perp \perp \B{j}} - n_j T_{\perp \perp \B{k}}$, as well as $n_j T^m{}_{m \B{k}} = n_j T_{\perp \perp \B{k}} + n_j T^\B{m}{}_{\B{m} \B{k}}$:

\begin{eqnarray}
-2J h_i{}^\alpha \left(
                        n_j T^m{}_{mk} + n_m T^m{}_{kj} + n_k T^m{}_{jm}
                \right)
&=&
-2J h_i{}^\alpha \left(
                        T_{\perp \BB{k}{j}} +
                        n_j T^\B{m}{}_\BB{m}{k} -
                        n_k T^\B{m}{}_\BB{m}{j}
                \right) \NN \\
&=&
h_i{}^\alpha B_{kj} \label{DL3 Rez1}
\end{eqnarray}
where we used (\ref{G: medjurezultat}). Now, we immediately conclude that (\ref{DL3 2red}) and (\ref{DL3 3red}) are equal to $-h_j{}^\alpha B_{ki}$ and $h_k{}^\alpha B_{ji}$ respectively. The only thing that remains is to calculate (\ref{DL3 4red}):

\begin{equation}
- 2J h_m{}^\alpha \left(
                        n_i T^m{}_{jk} + n_j T^m{}_{ki} + n_k T^m{}_{ij}
                \right)
=
- 2J h_m{}^\alpha \left(
                        n_i T^m{}_\BB{j}{k} + 
                        n_k T^m{}_\BB{i}{j} + 
                        n_j T^m{}_\BB{k}{i}
                  \right)
\end{equation}
That completes the proof of our lemma.
\flushright
\qedsymbol
\end{Dokaz}

\begin{PosledicaD}
\normalfont
\begin{eqnarray}
\varepsilon^{\alpha 0\beta \gamma}_{ijmk}  T^m{}_{\beta \gamma} n^k N
&=& -2b h_m{}^\alpha T^m{}_\BB{i}{j} - Nh_\perp{}^\alpha B_{ij}
    +2bh_i{}^\alpha \B{T}_\B{j} - 2bh_j{}^\alpha \B{T}_\B{i} \NN \\
&=&
    -2bh_m{}^\alpha \left(
                            T^m{}_\BB{i}{j} -
                            \delta^m_i \B{T}_\B{j} +
                            \delta^m_j \B{T}_\B{i}
                    \right)
    -Nh_\perp{}^\alpha B_{ij} \NN \\
&=&
    -2bh_m{}^\alpha \left(
                            T^\B{m}{}_\BB{i}{j} -
                            \delta^\B{m}_\B{i} \B{T}_\B{j} +
                            \delta^\B{m}_\B{j} \B{T}_\B{i}
                    \right) \NN \\
&&
    -2bh_\perp{}^\alpha \left(
                                T_{\perp \BB{i}{j}} -
                                n_i \B{T}_\B{j} +
                                n_j \B{T}_\B{i}
                        \right)
    -Nh_\perp{}^\alpha B_{ij}       \NN \\
&=&
    -2bh_\B{m}{}^\alpha \left(
                            T^\B{m}{}_\BB{i}{j} -
                            \delta^\B{m}_\B{i} \B{T}_\B{j} +
                            \delta^\B{m}_\B{j} \B{T}_\B{i}
                    \right)
\end{eqnarray}
\noindent
We conclude that:

\begin{equation}
    \varepsilon^{\alpha 0\beta \gamma}_{ijmk}  T^m{}_{\beta \gamma} n^k N
=   -2bh_\B{m}{}^\alpha \left(
                            T^\B{m}{}_\BB{i}{j} -
                            \delta^\B{m}_\B{i} \B{T}_\B{j} +
                            \delta^\B{m}_\B{j} \B{T}_\B{i}
                    \right) \label{Posledica}
\end{equation}
\end{PosledicaD}
\renewcommand\bibname{References}
\bibliographystyle{nar}
\bibliography{Literatura.bib}
\end{document}